\documentclass{JINST}

\usepackage{url}
\usepackage{textcomp}
\usepackage{wrapfig}
\usepackage{nth}
\usepackage{xspace}
\usepackage{afterpage}
\usepackage{wasysym}

\graphicspath{ {./} {./pix/} }

\newcommand{\txtpm}{\(\pm\)}

\newcommand{\cf}{{c.f.}\xspace}
\newcommand{\pe}{{p.e.}\xspace\ }

\input{library.def}

\title{Calibration and performance of the photon sensor response of FACT --\\The First G-APD Cherenkov telescope}

\newcommand{\ethz}{$^a$}
\newcommand{\tudo}{$^b$}
\newcommand{\unige}{$^c$}
\newcommand{\uniw}{$^e$}

\newcommand{\aut}{$^*$}

\author{
A.~Biland\ethz,
T.~Bretz\ethz\thanks{Corresponding authors: {\tt thomas.bretz@phys.ethz.ch, dorner@astro.uni-wuerzburg.de}},
J.~Bu\ss\tudo,
V.~Commichau\ethz,
L.~Djambazov\ethz,
D.~Dorner\uniw\aut,
S.~Einecke\tudo,
D.~Eisenacher\uniw,
J.~Freiwald\tudo,
O.~Grimm\ethz,
H.~von Gunten\ethz,
C.~Haller\ethz,
C.~Hempfling\uniw,
D.~Hildebrand\ethz,
G.~Hughes\ethz,
U.~Horisberger\ethz,
M.~L.~Knoetig\ethz,
T.~Kr\"ahenb\"uhl\ethz,
W.~Lustermann\ethz,
E.~Lyard\unige,
K.~Mannheim\uniw,
K.~Meier\uniw,
S.~Mueller\tudo,
D.~Neise\tudo,
A.-K.~Overkemping\tudo,
A.~Paravac\uniw,
F.~Pauss\ethz,
W.~Rhode\tudo,
U.~R\"oser\ethz,
J.-P.~Stucki\ethz,
T.~Steinbring\uniw,
F.~Temme\tudo,
J.~Thaele\tudo,
P.~Vogler\ethz,
R.~Walter\unige,
Q.~Weitzel\ethz\\
\llap{\ethz}ETH Zurich, Switzerland~~\,--\,~~Institute for Particle Physics, Otto-Stern-Weg 5, 8093 Zurich\\
\llap{\tudo}TU Dortmund, Germany~~\,--\,~~Experimental Physics 5, Otto-Hahn-Str.~4, 44221 Dortmund\\
\llap{\unige}University of Geneva, Switzerland\\
ISDC Data Center for Astrophysics, Chemin d'Ecogia~16, 1290 Versoix\\
\llap{\uniw}Universit\"at W\"urzburg, Germany \\
Institute for Theoretical Physics and Astrophysics, Emil-Fischer-Str.~31, 97074 W\"urzburg\\
\begin{center}{\bf\sc\center Dedicated to the memory of Eckart Lorenz}\end{center}
}

\renewcommand{\[}{\begin{equation}}
\renewcommand{\]}{\end{equation}}

\abstract{
The First G-APD Cherenkov Telescope (FACT) is the first in-operation
test of the performance of silicon photo detectors in Cherenkov
Astronomy. For more than two years it is operated on La Palma,
Canary Islands (Spain), for the purpose of long-term monitoring of
astrophysical sources. For this, the performance of the photo detectors
is crucial and therefore has been studied in great detail. Special care
has been taken for their temperature and voltage dependence
implementing a correction method to keep their properties stable.
Several measurements have been carried out to monitor the performance.
The measurements and their results are shown, demonstrating the
stability of the gain below the percent level. The resulting stability
of the whole system is discussed, nicely demonstrating that silicon
photo detectors are perfectly suited for the usage in Cherenkov
telescopes, especially for long-term monitoring purpose. }


\begin{document}

\newpage
\section{Introduction}

The First G-APD Cherenkov telescope (FACT) is the first Imaging
air-Cherenkov telescope to use silicon based sensors for photo
detection. Commencing operation in October 2011, the telescope is
operated remotely and automatic. A complete introduction and detailed
description of its hardware and software can be found in~\cite{Design}.

\subsection{General introduction}

The telescope is dedicated to the monitoring of the brightest known
gamma-ray sources, mainly Active Galactic Nuclei (AGN) with their
highly variable flux. Combining the observed energy spectrum with the
spectral information obtained at other wavelengths (e.g.\ radio or
X-ray data) gives an insight on cosmic particle acceleration. To
understand the highly variable flaring behaviour on all time scales
continuous monitoring for several months or years is necessary.
Although other, much more sensitive instruments are currently available
as the H.E.S.S., MAGIC and VERITAS telescopes, their high discovery
potential is best used for the detection of new sources at TeV energies
and precision studies of known sources. For  long-term and especially
continous monitoring the observation time of those instruments is too
expensive which suggests the construction of an inexpensive instrument
with an adapted sensitivity, c.f.~\cite{bretz08}. In addition, the
application of silicon based photo sensors promised the increase of
available observation time due to their robustness against light
exposure as string moon light.

The presented study is one of the first long-term tests of these
sensors under real environmental conditions. The Imaging Air-Cherenkov
Technique is an indirect measurement method where Cherenkov light
flashes emitted by atmospheric particle cascades induced from cosmic
ray particles are imaged. The camera comprises 1440 photo sensors each
read out individually. As photo sensors, Geiger-mode avalanche photo
diodes (G-APDs) are used. These silicon based devices are fast and
sensitive photo sensors with a high potential and future impact.
Consequently, they are an ideal alternative to photo multiplier-tubes
for all new projects such as the Cherenkov Telescope Array
(CTA,~\cite{CTA, CTA2}). In contrast to photo multipliers, silicon
photo sensors can easily be operated under bright light conditions as
moon lit nights allowing a significant increase in duty cycle
especially important for long-term monitoring. Although current sensors
usually have a sensitive area of not more than 60\,mm\(^2\), the
rapidly decreasing prices allow for several sensors to be put together
in a single channel. This enables their application also in large scale
detectors such as the large-size telescopes planned for CTA as shown
in~\cite{Ribordy}. 

A primary goal of building the FACT camera was to prove the
applicability of G-APDs under real environmental conditions. It had to
be shown that the properties of G-APDs can be kept under control,
despite their strong dependence on temperature and applied
voltage. Equally important is the proof that their intrinsic
properties, such as their internal crosstalk behavior, does not negatively
affect the data quality in Cherenkov telescopes. Initial data analysis
(see~\cite{Gamma}) show that previous conclusions drawn about their
applicability in Cherenkov astronomy hold. The following
chapters will discuss measurement techniques and results demonstrating
the excellent performance achieved.

\subsection{Overview}
As of today there are already significantly improved sensors available
on the market, still, either a temperature stabilization or an adequate
voltage correction is necessary. While a temperature stabilization
usually involves mechanical parts like fans, a correction can be
implemented as well by an adjustment of the support voltage. To
simplify maintenance, mechanical parts in or at the camera should be
avoided in general. Consequently, no active temperature control was
implemented for the FACT camera. A general overview of the properties
of G-APD sensors is given later in section~\ref{sec:gapd} including a
discussion of the applicability of these sensors in Cherenkov
telescopes. A summary of the camera hardware and a detailed description
of the implemented feedback system are given in section~\ref{sec:feedback}.

To understand the system in details, three different methods have been
applied: the analysis of dark count spectra, measurements of 
the amplitude of an external light pulser and the dependency of
the trigger rate from the applied threshold. Due to the detailedness, 
each method and the derived results is described individually in the
following. After two  applications made possible by the obtained
stability of the system, common conclusions are drawn at the end of the
paper.

\paragraph{Dark count spectra}
Dark count spectra at different temperatures and voltages have been
measured. To calibrate and monitor the system, Dark count spectra
histogram the discharge induced from thermal excitation without the
background of impinging photons. Although G-APD sensors consist of many
individual diodes, each of them issues nearly identical signals.
Therefore, dark count spectra are a direct measurement of the gain.
Crosstalk between individual diodes can induce higher order
multiplicities which renders the determination of the gain independent
from the precise knowledge of the baseline. From these measurements
taken under varying conditions the correlation of the sensor properties
as a function of temperature and voltage is derived. This dependency is
used for a fine tuning of the voltage calibration allowing to push the
precision of the voltage setting to the limit defined by the hardware.
Merging the results from all measurements, an average dark count
spectrum extending to high multiplicities is extracted with very high
precision. In addition, the exact pulse shape can be deduced. The
method used and the obtained results are presented in
chapter~\ref{sec:DCS}. The function which describes the distribution of
higher order multiplicities very precisely, is derived in
appendix~\ref{sec:appendix}.

\paragraph{Light pulser measurements}
Although measurements of dark count spectra already allow a precise
determination of many of the sensor properties, they are not suited as
a crosscheck for measurements during bright light conditions where the
noise introduced from background photons dominates. To prove the
stability of the system for measurements during varying light
conditions, an external light pulser is used. The method used and the
obtained result are discussed in chapter~\ref{sec:lightpulser}.

\paragraph{Ratescans}
While light pulser measurements rely on the stability and precision of
the light pulser, a source completely independent of the
instrumentation is available: the dominating background of cosmic ray
induced air showers. Using the trigger system of the telescope with
varying thresholds allows for the determination of the trigger rate
as a function of the trigger threshold. As the cosmic ray background is
independent of temperature and sky brightness, it provides a direct
measurement of the independence of the system from these variables.
Furthermore, any deviation from the expected rates is an indication
for additional attenuation in the atmosphere and facilitates a direct
measurement of the atmospheric conditions. Ratescans are discussed in
more detail in chapter~\ref{sec:ratescans}.

\paragraph{Applications}
A parametrization of the rates as a function of the current, i.e.\ a
measure for the sky brightness, permits to derive the ideal trigger
threshold directly from the measurement. This is shown in
chapter~\ref{sec:threshold}. Once a correlation between current and
threshold is available, a prediction of the current derived from sky
properties enables a more efficient observation scheduling. In
addition, the comparison between the predicted and the eventually
measured current provides an additional tool to assess the sky quality.
The current prediction is introduced in chapter~\ref{sec:prediction}.

\subsection{Geiger-mode avalanche photo diodes}\label{sec:gapd}

To avoid confusion due to different naming conventions used by
different communities or experiments, the following paragraphs
summarize the most important properties of Geiger-mode avalanche
photo diode (G-APD) based sensors and introduce the notations used in
this paper.

\paragraph{The sensors} 

When an avalanche photo-diode operates above the {\em
breakdown voltage}, an incident photon induces a complete discharge.
The probability for this process is the Geiger-probability in the
following simply called {\em photo detection efficiency}. Random
discharges induced by thermal excitation of electron-hole pairs are
called {\em dark counts}. The discharge occurs in form of an avalanche
originating from the primary electron-hole pair. To stop the cascade, an
internal quenching resistor will decrease the voltage below the
breakdown voltage. While the breakdown voltage is temperature
dependent, the released charge depends on the physical properties of
the cell and the voltage difference between the applied voltage and the
breakdown voltage: the {\em overvoltage}. This process makes the
released charge a unique property of each cell and independent of the
angle and the energy of the impacting photon. In the following, the
charge, released by a single breakdown, will also be called {\em photon
equivalent} (p.e.). After a breakdown, for a short time which is on the order of
nanoseconds, no further cascade can be induced in the diode,
commonly known as {\em dead time}. After this short time, the cell is
re-charged, also known as {\em recovery time}. During the re-charging
process, the charge released by another potential breakdown is
decreased accordingly.

In the literature, several definitions of the breakdown voltage are
suggested. Hereafter, the voltage at which the extrapolated gain is
zero will be referred to as the breakdown voltage.\\

In some cases, trapped charges left over from an avalanche can be
released later and potentially trigger another discharge in the same
cell. The probability for such a delayed release decreases in the first
order exponentially with a decay time on the order of several times the
recovery time. These spurious signals are called {\em afterpulses}. A
detailed measurement of their properties is discussed
in~\cite{Retiere}.\\

In a photo sensor, many G-APD cells are combined. The industrial
production and the high precision of silicon processing ensures a small
parameter spread between them. During the discharge of a single cell,
every avalanche process itself emits a random number of photons
proportional to the charge released in the avalanche. Some of these photons can
directly or indirectly trigger a discharge in neighboring cells. This
process is known as {\em optical crosstalk}, often simply called {\em
crosstalk}. While afterpulses are delayed, crosstalk induced signals
are usually prompt. The fraction of signals with at least two
synchronous avalanches out of the total number of signals is called
{\em crosstalk probability}. A detailed characterization of the single
cell response and the detailed properties of optical crosstalk can be
found in~\cite{Eckert}.\\

The common voltage applied to the sensor, hereafter {\em bias voltage},
is usually distributed using a passive filter network. To apply the correct bias
voltage, the voltage drop at the serial resistance needs to be known. 
Any breakdown induces also a voltage drop at the serial resistance. Due
to AC coupling, high enough rates induce a DC current. If the rate of
breakdowns is changing, a direct or indirect measurement of the voltage
drop is mandatory to correct this and keep the overvoltage stable. The
necessary ability to change the bias voltage accordingly also enables
the possibility to compensate for the change of the breakdown voltage
due to temperature.

\paragraph{Application in Cherenkov astronomy}

In Cherenkov astronomy, flashes of Cherenkov light emitted by particle
cascades, which are induced in the atmosphere by very high energy
cosmic ray particles, are observed with pixelized photon detectors.
While the light flashes are of nano-second duration, the number of
recorded photons in an image can range from a few to a few thousand.
The limiting factor for a trigger and a good image reconstruction is
the contrast of the image to the night-sky background. While on clear
moonless nights, the rate of detected photons per pixel is on the order
of tens of MHz, it can exceed ten GHz during full moon. As long as the
rate of random signals induced by dark counts in each channel is well
below the photon rate from the night-sky background, dark counts have
no significant influence on the data. In most of today's analysis
methods, image reconstruction relies on a mainly statistical analysis
of the measured light distribution. Consequently, for a reasonably low
statistical error on the obtained parameters, at least a few tens of
photons need to be detected. Optical crosstalk just increases this
signal statistically. The increase in fluctuation can be neglected
compared to the intrinsic fluctuations in the shower and the Poisson
fluctuation of the signal. The influence of afterpulses can be
eliminated completely by choosing the charge integration time small
enough. This is discussed in more detail in~\cite{IEEE}.  
yet 

\subsection{The feedback system}\label{sec:feedback}

\paragraph{Motivation} In Cherenkov astronomy, an estimate of the
energy of the primary particle can only be obtained from detailed Monte
Carlo simulations of the shower image. Therefore, a good agreement of
the simulation with the data is necessary. In turn, a very detailed
understanding of the detector is needed. In view of the sensors, this
requires a systematic assessment of the gain and the probability for
optical crosstalk. While the quality of simulations only influences the
quality of the analysis result, the stability of the gain directly
influences the quality of the recorded data. 

An inhomogeneous gain over
the camera requires a local adaption of the trigger threshold to keep
the physics response homogeneous throughout the camera. This results
in a complicated system difficult to calibrate and operate, or in an
effort required for reasonable simulations which exceed the available
resources significantly. To avoid this, best effort is made to keep the
threshold as homogeneous as possible over the camera and stable within
reasonable time intervals. To suppress the remaining effect of
instabilities of the gain, the threshold is increased artificially in
the analysis. This achieves a homogeneous and stable response which is
easy to simulate. Consequently, a more stable gain directly translates
to a lower energy threshold.

As a rough estimate, a change in gain at constant trigger threshold
converts to a change in energy threshold linearly to quadratically. In
order to avoid a significant influence on the energy threshold,
a stability of the gain on a percent level is required.

%


\paragraph{Concept} The gain of G-APDs depends directly on the sensor
temperature and indirectly on the background light level. For the
sensors in use, an uncorrected temperature difference of
10\,\textcelsius{} as well as the typical current during a
three-quarter moon night, would reduce the gain by \(\sim\)50\%. To
correct the effect of both, a real-time feedback system was implemented.
As feedback values, temperature sensors in the sensor compartment and
the current readout of each bias voltage channel are available.

Since the temperature effect on the breakdown voltage is linear, well
defined and to first order identical for all channels, the bias
voltage can be adapted with a unique coefficient of 55\,mV/K. This is
done every 15\,s being small compared to the temperature gradient
induced from changes in the ambient temperature. For a more precise
correction, the temperature for each bias voltage patch is interpolated
or extrapolated from the available sensor readout. For details on the
interpolation algorithm, see~\cite{IEEE}.

To compensate for the voltage drop induced by varying background
light, the current readout of each bias voltage channel is used. From
the current measured at a rate of 1\,Hz and averaged over three
seconds, the voltage drop is calculated and the voltage adapted
accordingly. This time interval is still short compared to the expected
change induced from bright stars and the rotating star field. A
detailed description of the feedback algorithm can be found
in~\cite{IEEE} and a sketch in~\cite{IEEERT} as well.


\paragraph{Hardware}

In the following paragraphs, the main characteristics of the hardware 
are repeated briefly for completeness. A detailed overview of the
hardware is given in~\cite{Design}. 

The camera uses sensors from Hamamatsu (MPPC
S10362-33-50C~\cite{DataSheet}). The sensors have an active area of
3\,mm\,x\,3\,mm and a total number of 3600 Geiger-mode avalanche photo
diodes each 50\,\(\mu\)m\,x\,50\,\(\mu\)m in size. They are supposed to
be operated at their nominal {\em operation voltage} \(U_{op}\) as
provided individually for each sensor by the manufacturer. Whether this
is the ideal operation voltage for application in a Cherenkov telescope
is out of the scope of this paper. The dark count rate of these sensors
does not exceed 1\,MHz/mm\(^2\), even during the summer months when the
temperature can rise up to 30\,\textcelsius{} at night. This rate is
negligible, compared to a count rate of more than 30\,MHz per sensor
from the diffuse night-sky background light.

The data acquisition system is based on the domino
ring-sampler~\cite{DRS4appl} (DRS\,4) with its nine readout channels.
It facilitates one readout channel per sensor. The bias voltage is
provided by 320 bias voltage channels of which 160 serve four and the
other 160 five sensors at a time.  To avoid large discrepancies between
the sensors in one bias channel, the sensors for each channel were
selected to have very similar operation voltages. The voltage can be
set up to 90\,V in 4096 steps corresponding to a resolution of
\(\sim\)22\,mV. It is created using an op-amp OPA454 as programmable
voltage source, fed by a 12\,bit current source. Each bias voltage
channel has its own current readout with a resolution of 12\,bit in the
range up to 5\,mA. The trigger system comprises 160 trigger channels
where each is the discriminated sum of nine signal channels
corresponding to two bias voltage channels.

To measure the temperature, 31 PT-1000 temperature sensors have been
distributed homogeneously in the sensor compartment close to the photo
sensors. Since three temperature sensors showed problems after
assembly, only 28 are available for readout. Their signals are digitized
with a precision of 1\,mV which corresponds to 0.26\,\textcelsius.

Although not a single readout channel has failed since assembly, in
total, 12 readout channels show diverse problems~\cite{Design}. These
channels, and their corresponding bias voltage channels, are excluded
from the following considerations and studies, so that only 1428  will
be refered in the following.


\paragraph{Systematic error}

The precision of the gain control is ultimately limited by harware
specifications, more precisely: the resolution of the temperature
measurement, the current readout, the bias voltage setting, the
operation voltage as supplied by the manufacturer and the precision of
the temperature coefficient. Numbers for these are given in
table~\ref{tab:precision} converted into a corresponding bias voltage
setting. 

\begin{table}[h]
\centering
\begin{tabular} {|l|ccc|}\cline{2-4}
\multicolumn{1}{c|}{}         & \(\Delta U\) & \(U_{ov}\)=1.1\,V & \(U_{ov}\)=1.4\,V\\\hline
Temperature sensor            &  14\,mV & 1.3\%             & 1.0\%           \\
Current measurement           & 3.5\,mV & 0.4\%             & 0.3\%           \\
Operation voltage \(U_{op}\)  &  10\,mV & 0.9\%             & 0.7\%           \\
Bias voltage setting          &  22\,mV & 2.0\%             & 1.6\%           \\
Temperature coefficient       & 1\,mV/K & 0.9\%/10\,K & 0.7\%/10\,K           \\
Resistors                     &         & \multicolumn{2}{c|}{O(1\%)} \\\hline
Pre-amplifier                 &         & \multicolumn{2}{c|}{O(1\%)} \\\hline
\end{tabular}
\caption{Precision of voltage correction originating from different sources.
Numbers are given as absolute values \(\Delta U\) and as the
corresponding relative error in gain assuming operation at an
over-voltage of 1.1\,V and 1.4\,V. Not directly related to the voltage
supply, but still influencing the charge measurement is the
pre-amplifier.}  
\label{tab:precision}  
\end{table}

The values for the temperature sensors, current measurement and voltage
settings do not include calibration accuracy. Even if all erros were added
linearly, a maximum systmatic error of not more than \(\sim\)5\% is
expected.


\section{Calibration and monitoring with dark count spectra}

To achieve a maximum precision around the breakdown voltage, an
accurate calibration of the absolute voltage and the current readout of
each bias voltage channel is important. The details of the calibration
procedure are given in~\cite{IEEE}. Applying the calibration procedure 
described therein, it turned out that the very high precision of
\(\sim\)22\,mV at \(\sim\)70\,V corresponding to 0.3\textperthousand{}
was still not achieved. A possible reason could be that the temperature
of the room where the power supply is located during calibration and
operation was different, see also section~\ref{sec:temp}.
Therefore, another voltage offset was introduced derived from an
indirect measurement of the average breakdown voltage of the sensors in
each channel. This calibration is obtained from gain measurements as a
function of voltage. For gain determination, dark count spectra are
used, as well as to monitor the sensor properties. Measurement and
analyis of the dark count spectra and the obtained results are
discussed hereafter.

\subsection{Method}

\subsubsection{Dark count spectrum}\label{sec:DCS}

Due to the excellent single-\pe resolution of G-APDs, most of their
properties can be extracted from their dark count spectrum. To obtain a
dark count spectrum, randomly triggered data is recorded and the
signals are extracted. To avoid a bias on the extraction from
overlapping pulses, the camera lid is kept closed.
Consequently, only dark counts with their crosstalk and afterpulse
signals are recorded. Dark count spectra recorded with open
lid during dark time would show an order of magnitude higher
count rates. Due to the higher probability for random coincidences,
this complicates the analysis significantly and reduced the quality of
the result. It is therefore avoided. While the effect of afterpulses is
suppressed by the way the signals are extracted, coincident crosstalk
events create signals with higher multiplicities. In this context, a
reasonable amount of crosstalk is essential for calibration purpose. 

Although the current increases with dark count rate, i.e.\ with rising
temperature, the induced voltage drop is still negligible and below the
resolution of the voltage setting. Dark count spectra are therefore
ideally suited to measure the dependency of the sensor properties from
the temperature without a significant bias from the induced voltage
drop.

\paragraph{Measurement}
Dark count spectra are recorded as single runs at the beginning and end
of the night or during bad weather periods when no observations can
be scheduled. Since the ambient temperature can not be controlled, the
measurement is limited to the temperature range provided by the
environment on site. For each run, 10,000 events are triggered at a
rate of 77\,Hz reading out the maximum sampling depth of 1024 samples
per channel at a rate of 2\,Gsample/s. The trigger rate is chosen close
to the saturation limit of the data acquisition system to keep the run
short limiting ambient temperature changes to a minimum. The number of
events was chosen to have good statistics at multiplicities \(N>2\)
even at low temperatures. Before further processing, the data is
calibrated to correct for properties of the DRS\,4 ring sampler and
convert the digital values to physical units, see~\cite{DRS4appl}.

\subsubsection{Signal extraction}

To extract charges from the recorded data, for each run and channel, a
baseline is determined and subtracted, and a pulse extraction algorithm
is applied. The baseline is determined, from a histogram filled with
the calibrated ADC values. Calculating the maximum of a second order
polynomial from the logarithm of the three maximum bins, provides a
refined position for the position of maximum. This is identical to
fitting a Gaussian to these three bins.

\begin{figure}[t]
 \centering
 \includegraphics*[width=0.99\textwidth,angle=0,clip]{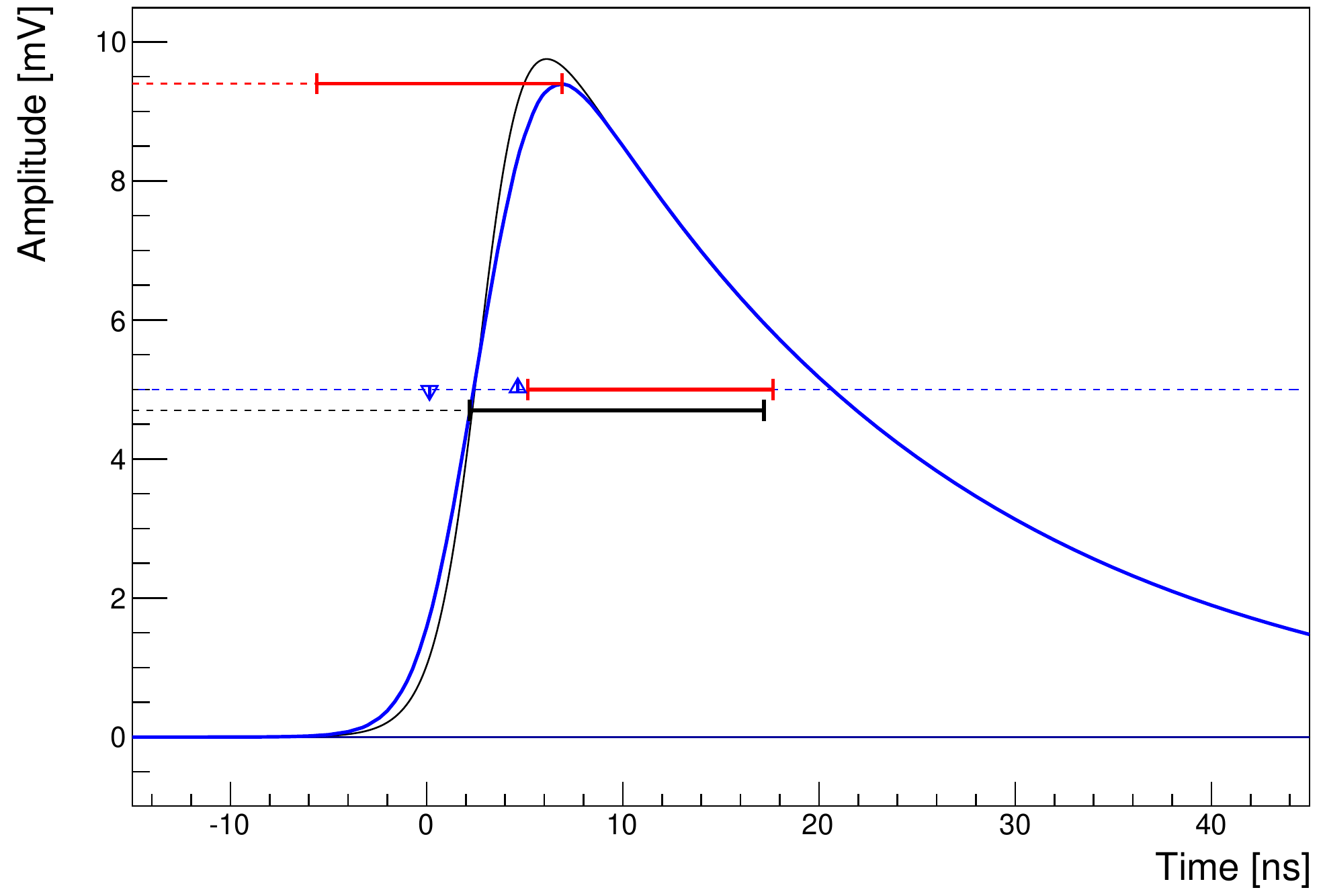}
\caption{Illustration of the signal extraction algorithm using the
shown pulse (black) as example. The algorithm, based on the a sliding
average filtered pulse (blue) and the reference lines are
described in details in the text.
}
\label{fig:pulse-extraction}
\end{figure}

To illustrate the pulse extraction algorithm, an example pulse for a
single discharge is shown in figure~\ref{fig:pulse-extraction} as a
thin black line. 

The first twenty and last ten samples of each event are discarded
because they suffer from additional noise. Before pulse extraction a
sliding average filter of ten samples is applied, which is possible due
to oversampling. This improves the signal-to-noise ratio and eliminates
some occasional noise with a corresponding frequency. The resulting
pulse shape is shown in blue. To identify the pulses, the samples are
scanned for a leading edge defined as a threshold crossing between two
consecutive samples. Studies have shown that good results are obtained
with a threshold of 5\,mV (dashed blue line). In addition, it is
required that the amplitude four samples before and after the threshold
crossing, is still below, respectively above the threshold, indicated
by the two blue arrows. A local maximum is searched between five and 35
samples after the threshold crossing (bottom red line). Within a range
of 30 samples before the determined maximum (top, red line), the last
sample which does not fall below 50\% of that maximum is sought. The
position of this sample is called {\em arrival time} hereafter. If
the distance between the maximum and the arrival time exceeds 14
samples (7\,ns), the pulse is discarded as wrong identification. Starting
from the arrival time, 30 samples of the raw signal are integrated,
hereafter called {\em extracted signal}. The next search starts at the
end of the integration window. To allow for a proper calculation of the
dark count rate, the number of searched samples is computed as well.

An example of a spectrum obtained from a single channel in a single run
is shown in figure~\ref{fig:pixelspectrum}.

\subsubsection{Parameter extraction}

In the obtained spectrum, the distance between two consecutive peaks
represents a measurement for the charge released in a single breakdown,
i.e.\ the gain. Higher order multiplicities reflect the process in which
optical crosstalk spreads from a primary breakdown. The total number of
detected signals is a measure of the dark count rate.

To get a more precise estimate of these properties, the spectrum of
each individual pixel is fit to a corresponding distribution
function. Empirically, it has been found that the best fit is obtained
with a slightly modified Erlang distribution. This distribution
describes the probability to measure a multiplicity of \(N\)
synchronous signals induced by a single primary breakdown. A more
detailed discussion can be found in appendix~\ref{sec:appendix}.

\paragraph{Spectrum function}

While the modified Erlang distribution describes the distribution of
the multiplicity \(N\), for the proper description of a real
measurement additional noise components have to be taken into account.
Measuring a real sensor, the fluctuations on the released charge and
electronics noise smear out the distribution. While electronics noise
can be considered independent of the number of breakdowns, the
fluctuation on the released charge scales with the multiplicity. Both
types of noise are assumed to be Gaussian with width \(\sigma_{el}\)
for the constant noise and \(\sigma_{pe}\) for the amplitude dependent
noise.

The resulting distribution can be expressed as a sum of Gaussian
functions for multiplicity~\(N\), each with the gain \(g\) and a
baseline shift \(x_0\). The gain \(g\) corresponds to charge extracted
from a single avalanche. Therefore, in the following gain and extracted
charge will be used interchangeably.

\[ f(x) = A_1\cdot a_1\sum\limits_{n=1}^{n=\infty}P_n\,\frac{ e^{
-\frac{1}{2}\,\left[\frac{x-x_n}{\sigma_n}\right]^2 }  }{a_n}\label{eq:spectrum}  \]
with the offsets \(x_n\), the width \(\sigma_n\) and the normalization \(a_n\)
\[ x_n = x_0+ng\,,\qquad \sigma_n=\sqrt {n\,\sigma_{pe}^2+\sigma_{el}^2}\qquad\mbox{and}\qquad a_n = \sigma_n\,\sqrt{2\pi} \]
and the modified Erlang distribution (see equation~\ref{eq:Erlang}) as the distribution function \(P_n\)
\[ P_n = \frac{(nq)^{n-1}}{\left[(n-1)!\right]^{\nu}}\qquad\mbox{with}\qquad q = p\cdot e^{-p}\,. \]

For easier determination of start values in a fit, the function is
written such that \(A_1\) denotes the amplitude of the single-\pe peak.
The normalized amplitude \(A_1'\) independent of the bin-width \(w\) of the
fitted histogram is then given as \(A_1'=A_1/w\).

Knowing the effective on-time \(T_{eff}\) from the pulse-extraction,
the dark count rate \(R\) can be estimated as

\[ R = \frac{A_1'\cdot a_1\sum\limits_{n=1}^{n=\infty}P_n}{ T_{eff} }\,.\]

The crosstalk probability \(p_{xt}\) is

\[ p_{xt} = \frac{\sum\limits_{n=2}^{n=\infty}P_n}{\sum\limits_{n=1}^{n=\infty}P_n}\,.\label{eq:crosstalk}\]

\paragraph{Fit procedure}

\begin{figure}[tb]
 \centering
 \includegraphics*[width=\textwidth,angle=0,clip]{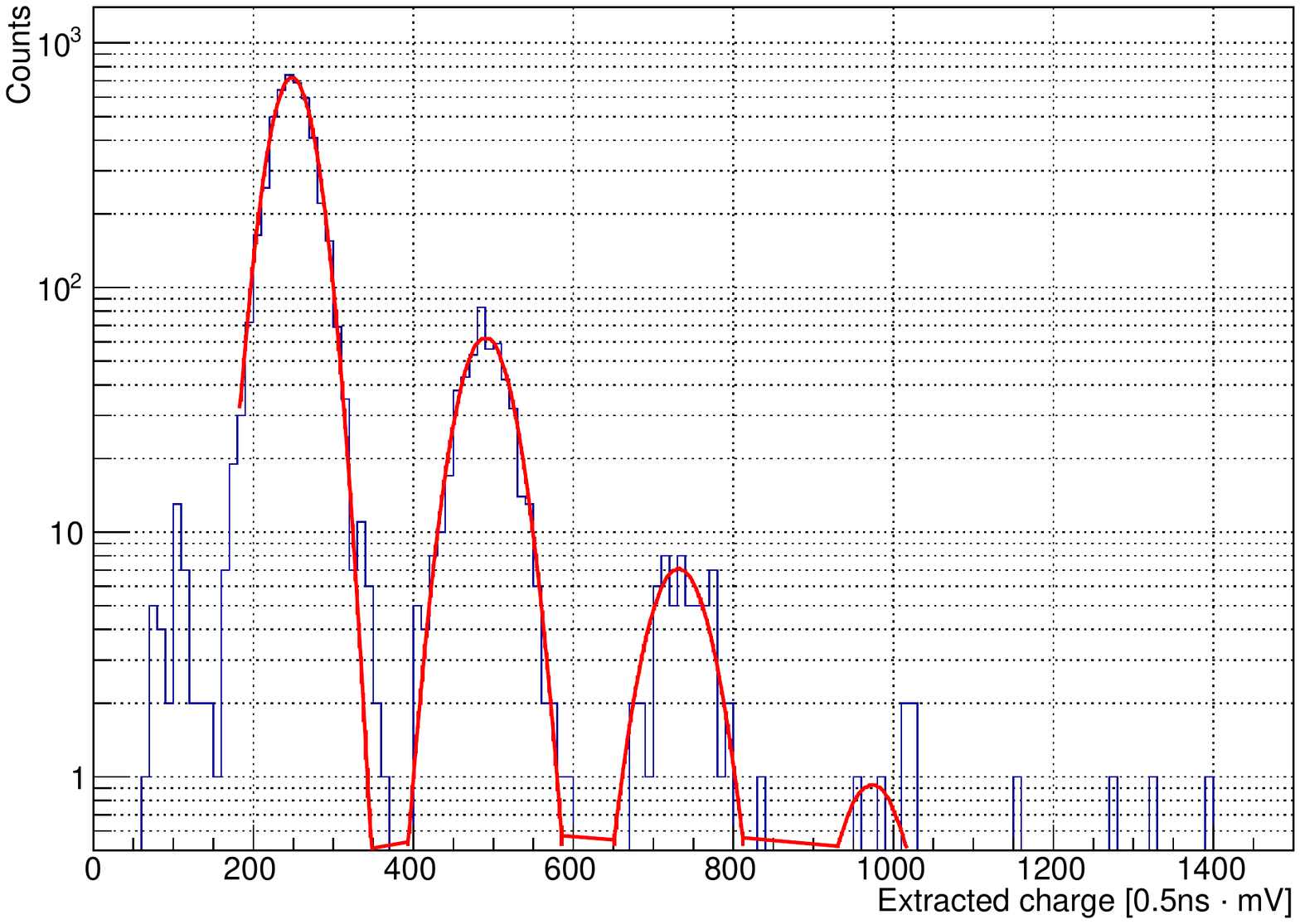}
\caption{
Example for the spectrum of charges extracted from a single
run and one pixel fitted with a modified Erlang distribution (red line). The
plateaus close to the x-axis are artifacts introduced by the
used plotting tool.
}
\label{fig:pixelspectrum} 
\end{figure}

To obtain reasonable fit results, meaningful start values need to be
obtained for all channels. This is achieved by fitting a high statistics
spectrum, combined from all channels. An example of a randomly selected
single pixel spectrum with the corresponding fit is shown in
figure~\ref{fig:pixelspectrum}. 

\subsubsection{Offset calibration}\label{sec:calibration}

Although during measurement, the feedback system keeps the gain stable,
it does not yet calibrate its absolute value. Such a systematic offset
can, for example, originate from inaccurate knowledge of the serial
resistances or other systematic errors as mentioned in
table~\ref{tab:precision}. Since the sum of all systematic errors can
easily exceed the precision of the voltage setting, an additional
absolute calibration is required. Strictly speaking, a correction not
only for offset but also for slope is necessary, for both, temperature
and current coefficient. The influence of an imprecise slope can be
neglected for the operational range of only a few volts. A simple
voltage offset per channel is already enough to push the systematic
error below the precision of the voltage setting.


To calculate these offsets, the gain is determined from data taken at
different voltages. An example of such a measurement
is shown in figure~\ref{fig:ovtest} (top left).
The data represents the average obtained from all 
channels and measurements taken at the same voltage, the error
bars the standard deviation.
From a linear fit to this average, the nominal gain at \(\Delta
U=0\)\,V is obtained. To obtain a reference for each individual bias voltage
channel, the readout channels belonging to one bias channel are
averaged and fit linearly as well. From these fit and the
nominal gain, a voltage offset for each channel is calculated and
applied.

\subsection{Results}

With the determined offsets applied, measurements were taken to
determine also the dependency of the variables on voltage.
Dark count spectra obtained from runs taken at varying ambient
temperatures allow for the determination of the dependency on the
sensor temperature. 

\subsubsection{Voltage dependency}

\begin{figure}[p]
 \centering
 \includegraphics*[width=0.495\textwidth,angle=0,clip]{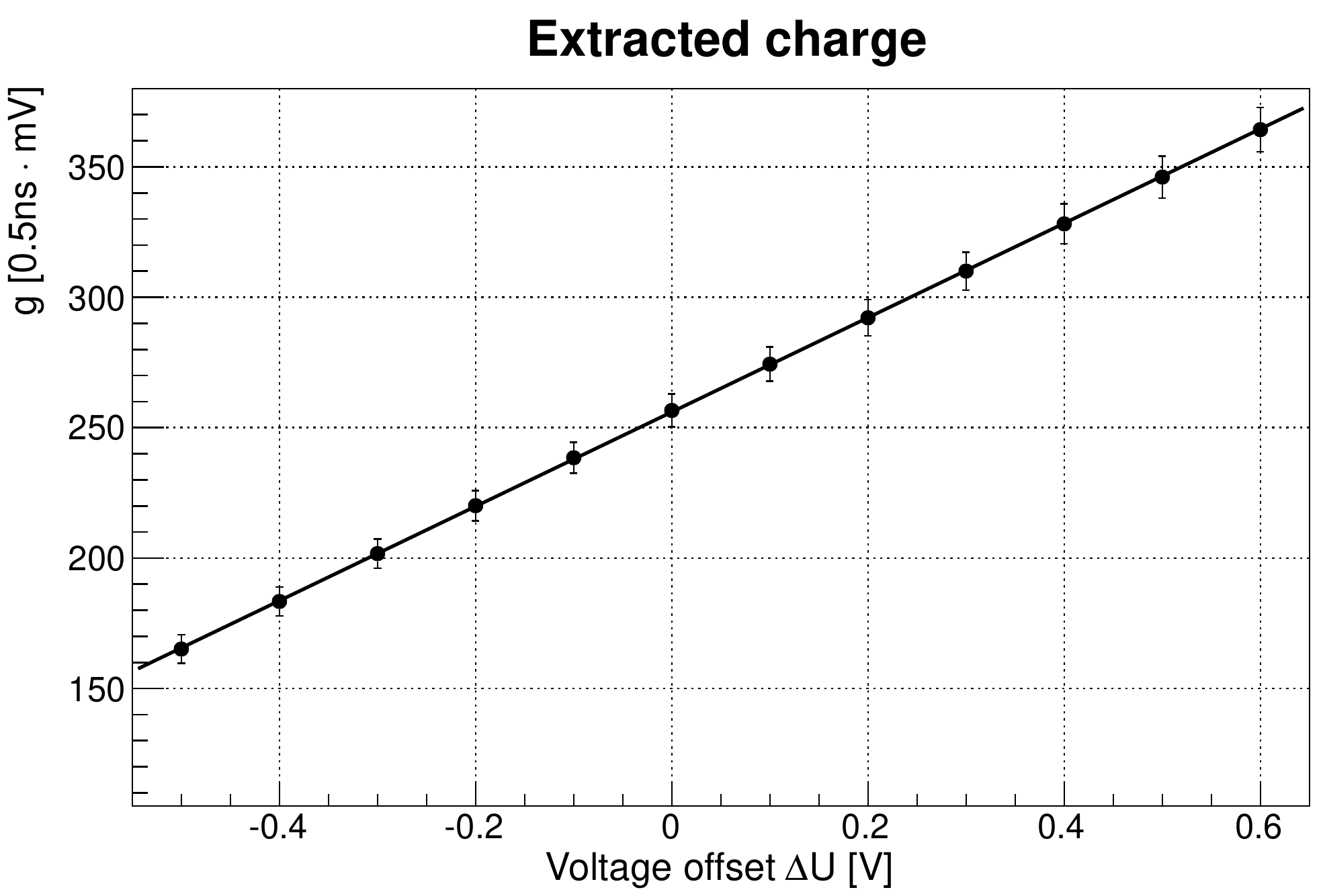}\hfill
 \includegraphics*[width=0.495\textwidth,angle=0,clip]{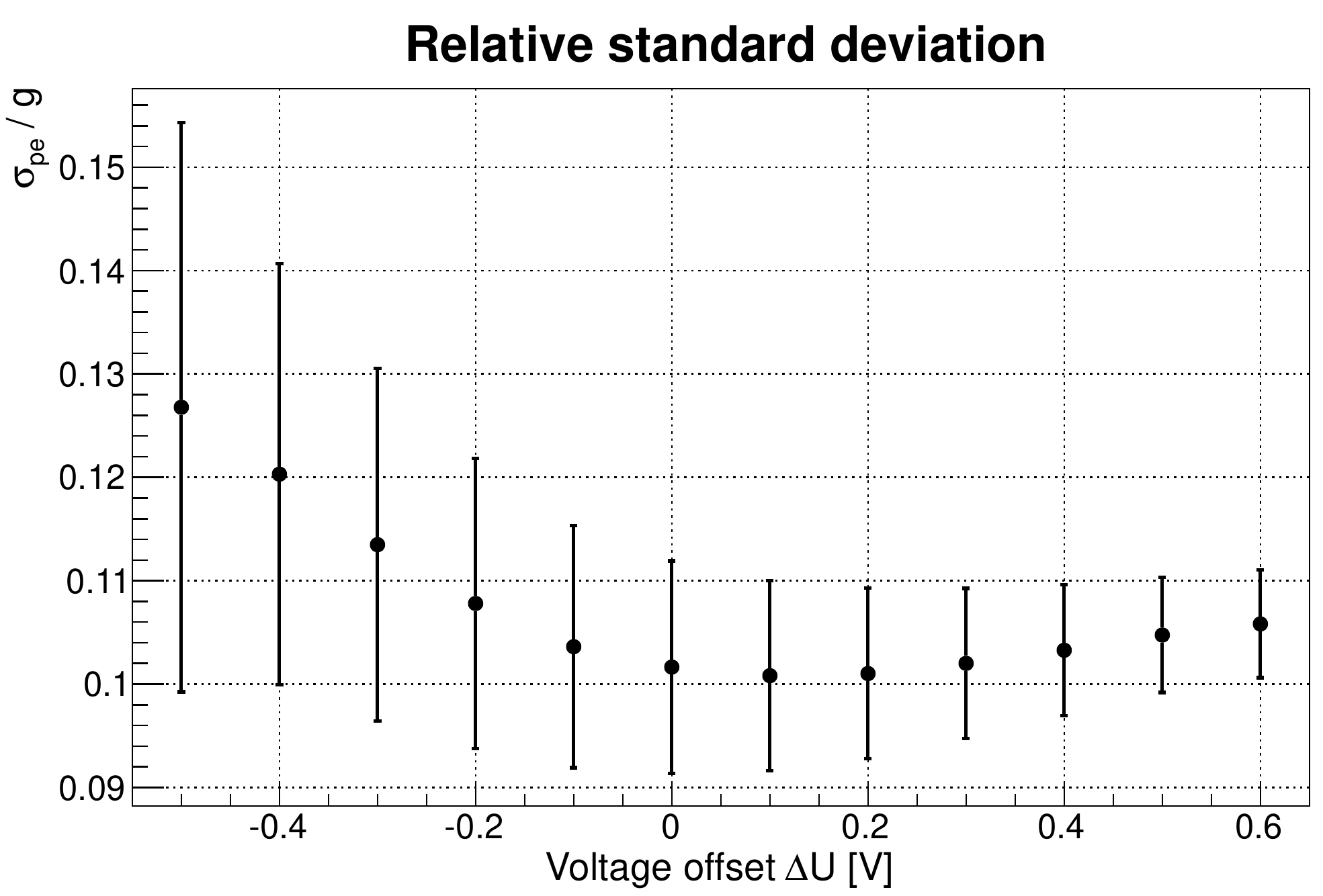}\\[2em]
 \includegraphics*[width=0.495\textwidth,angle=0,clip]{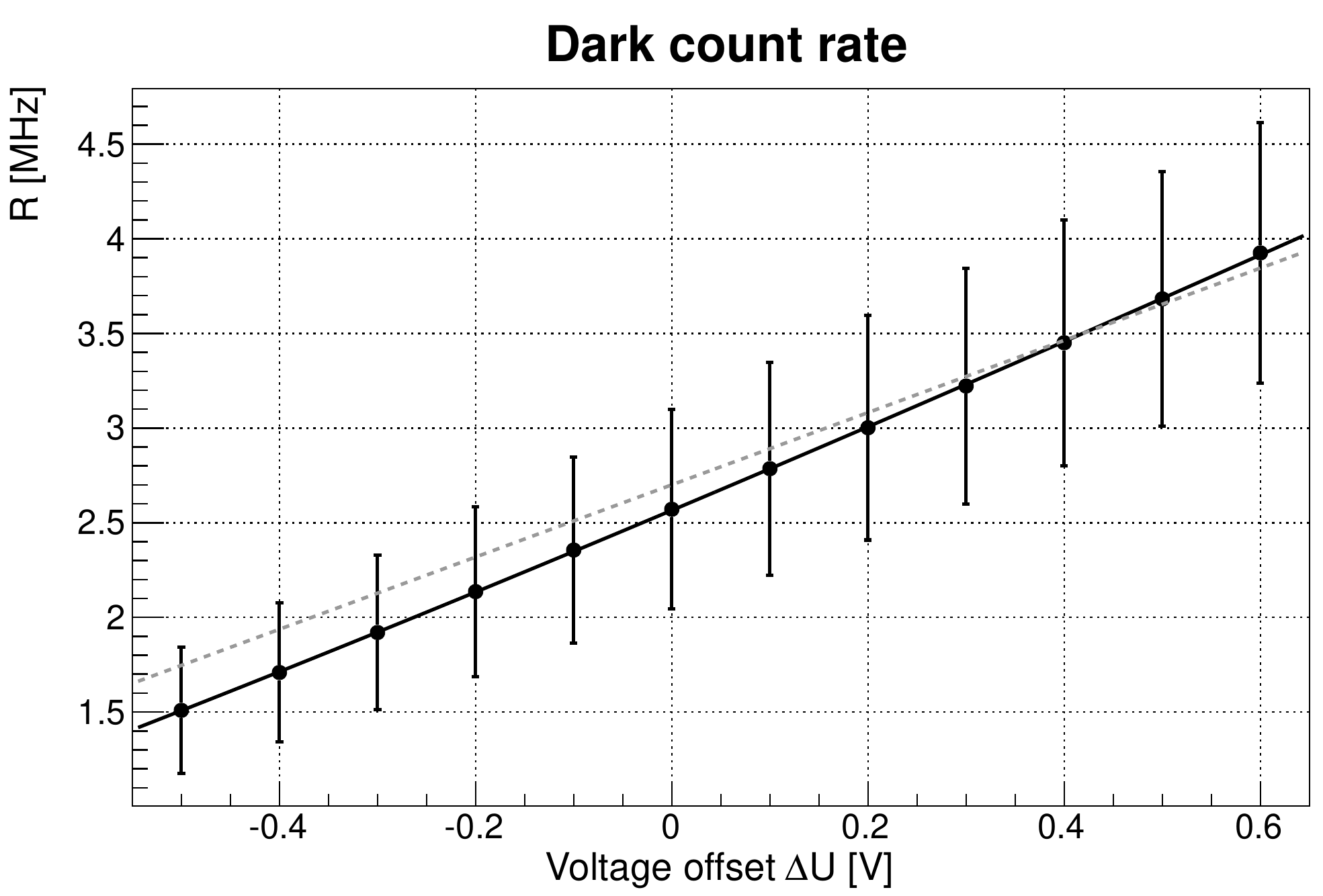}\hfill
 \includegraphics*[width=0.495\textwidth,angle=0,clip]{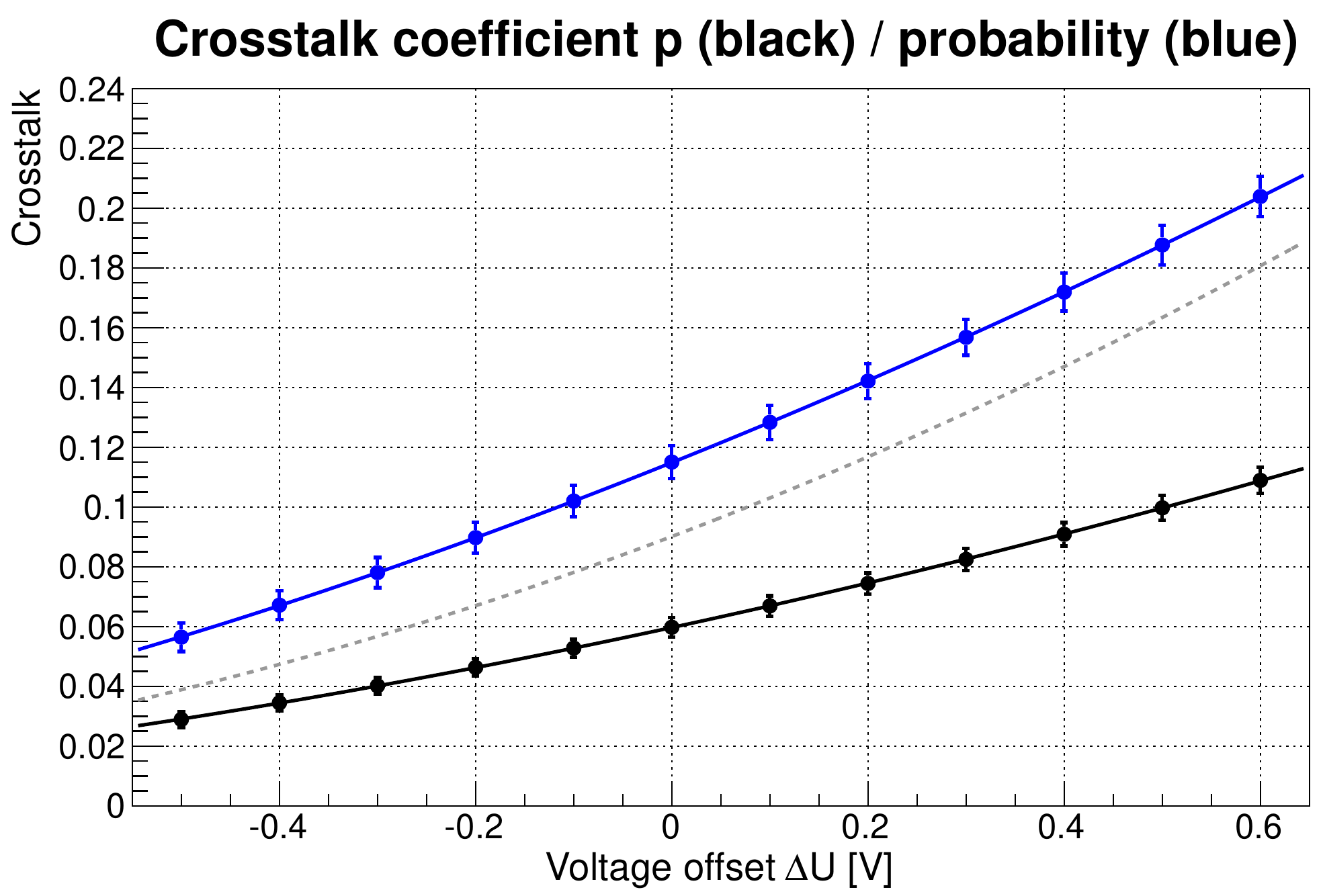}\\[2em]
 \includegraphics*[width=0.495\textwidth,angle=0,clip]{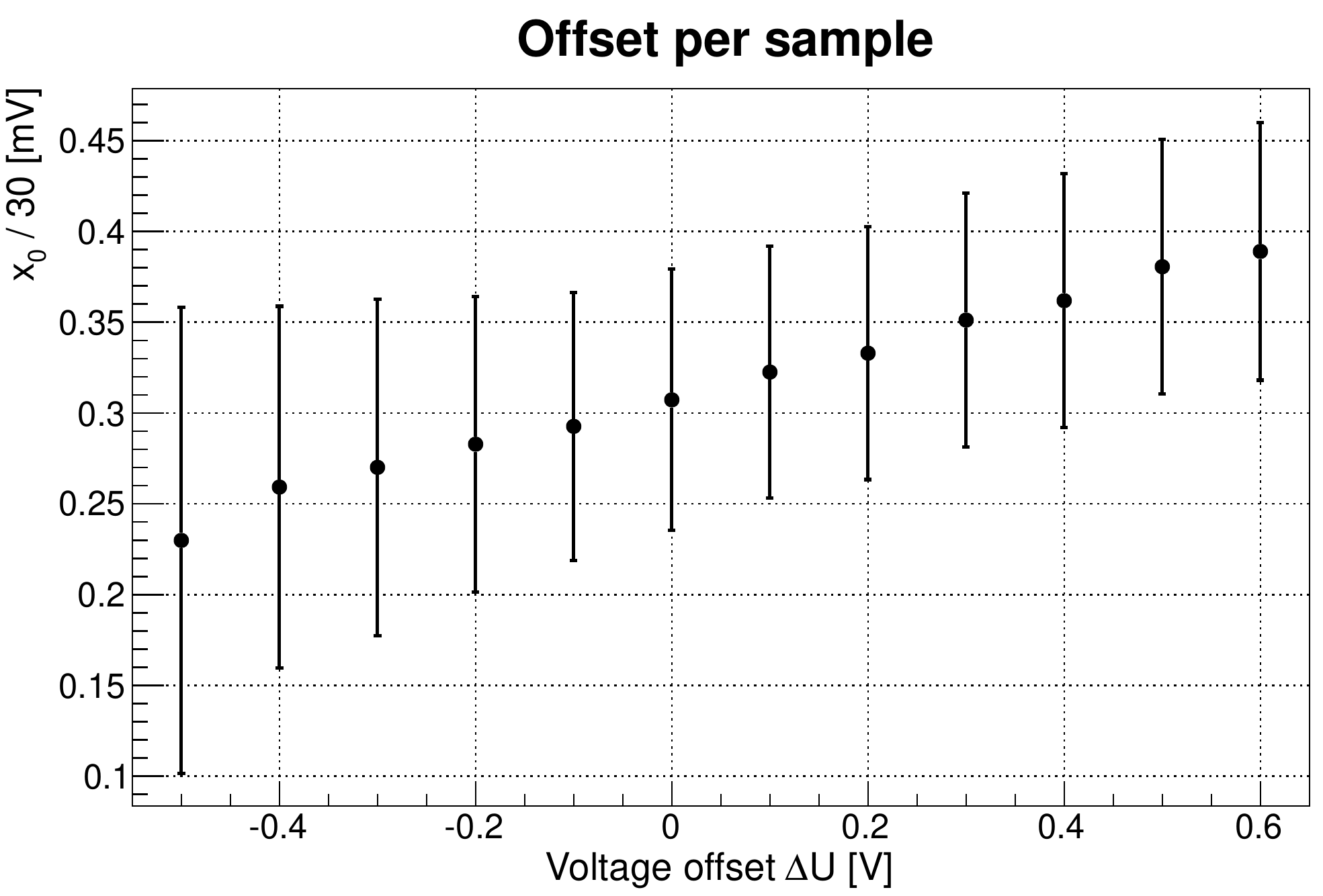}\hfill
 \includegraphics*[width=0.495\textwidth,angle=0,clip]{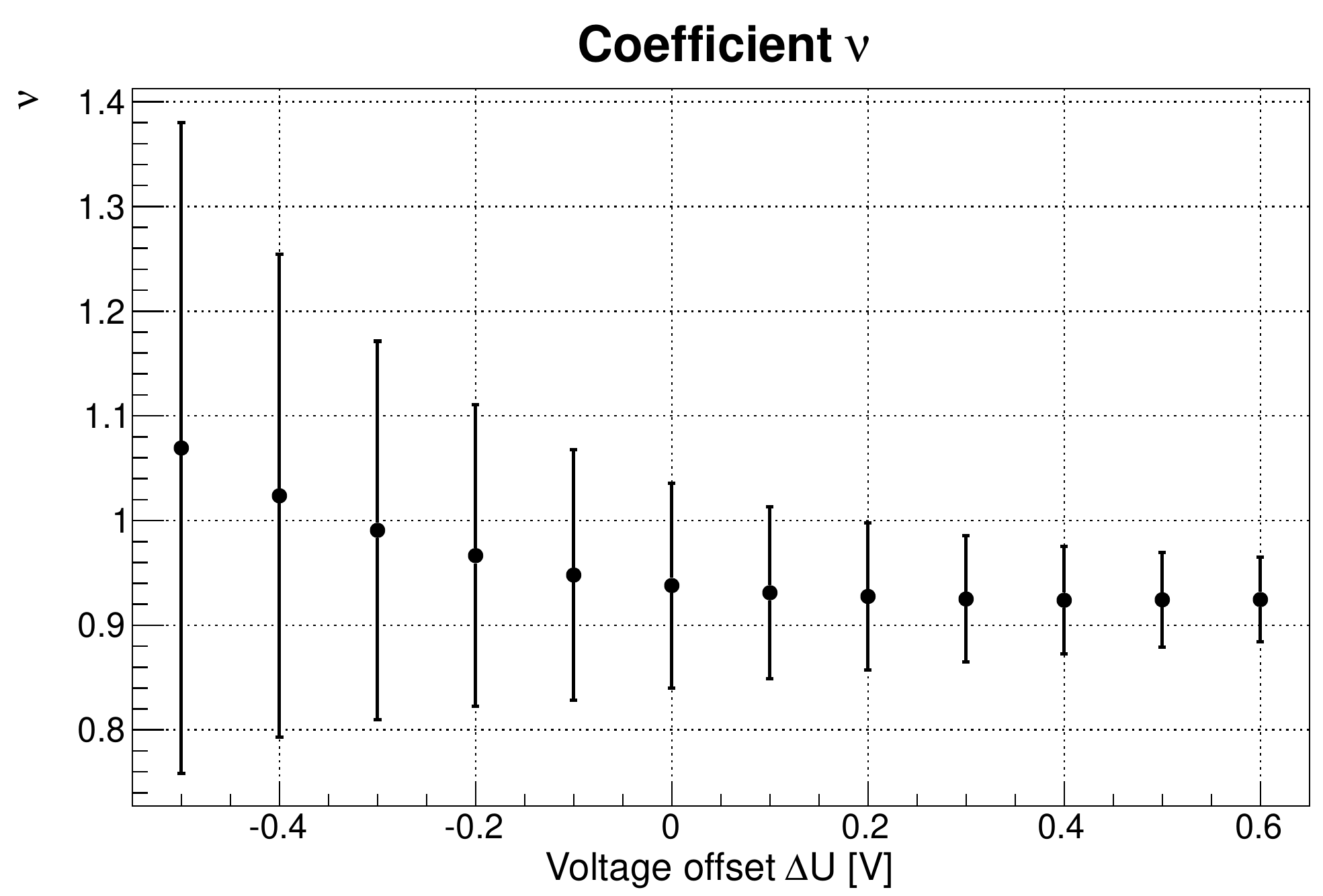}\\[1em]
\caption{Average of the fit parameters of all channels obtained from
several measurements. Each measurement contains data taken at 14
different voltages. An offset of 0\,V corresponds to the operation
voltage of the sensors. Error bars show the standard deviation of the
obtained parameters. Data was taken at sensor temperatues between
8.4\,\textdegree{}C and 10.7\,\textdegree{}C. The gray dashed lines are measurements taken
by~\cite{Retiere} for a similar sensor type. Their dark count rate
measured for a 1\,mm\(^2\) sensor was scaled linearly to 9\,mm\(^2\)
for an easier comparison. The crosstalk values refer to the probability
\(p\) as obtained from the fit (black) and the crosstalk probability
(blue).}  \label{fig:ovtest}  

\end{figure}

To derive the voltage dependency, in total 10 measurements where taken
at 14 voltages, each between 14/03/2014 and 17/03/2014 at average
sensor temperatures between 8.4\,\textdegree{}C and
10.7\,\textdegree{}C. From the parameters obtained by the fit, their
average for each voltage is calculated. The results are shown in
figure~\ref{fig:ovtest}. The error bars represent the standard
deviation of the parameters obtained from all measurements and
channels. The larger ones are dominated by the change of the parameters
with sensor temperature, see section~\ref{sec:temp}.  The increasing
standard deviation and fluctuations towards lower
voltages are due to a significantly decreasing fit quality introduced
by a worsening in separation of the individual peaks in the spectrum.

To describe the voltage dependency of all variables, the following
function has been fit to the dark count rate, the gain, the noise
relative to the gain, the crosstalk coefficient and the crosstalk
probability as shown in figure~\ref{fig:ovtest}. The baseline and the
coefficient \(\nu\) of the modified Erlang distribution can be
considered constant within the obtained statistical errors.

\begin{table}[th]
\centering
\begin{tabular}{|l||c|c|c|}\hline
Par(\(\Delta U\))                 &  \(c_0\)                  &  \(c_1\)            & \(c_2\)                 \\\hline\hline
Dark count rate \(R\)             &  (1.8\,\txtpm\,4.7)\,MHz  &  1.4\,\txtpm\,2.2    &  1.1\,\txtpm\,2.0      \\\hline
Extracted charge \(g\)            &  181\,\txtpm\,6           &  1.42\,\txtpm\,0.04  &  1 (fixed)             \\\hline
Crosstalk coefficient \(p\)       &  0.03\,\txtpm\,0.03       &  1.5\,\txtpm\,0.6    &  1.8\,\txtpm\,0.7      \\\hline
Crosstalk probability \(p_{xt}\)  &  0.07\,\txtpm\,0.05       &  1.4\,\txtpm\,0.5    &  1.62\,\txtpm\,0.6     \\\hline
\end{tabular}
\caption{Coefficients obtained from the fits of the function given in equation~\protect\ref{eq:parameters}
to the measurements as shown in figure~\protect\ref{fig:ovtest}.}
\label{tab:coefficients}
\end{table}

%
\afterpage{\clearpage}

\[Par(\Delta U) = c_0 \cdot \left[\frac{\Delta U}{\mbox{V}} + c_1\right]^{c_2}
\label{eq:parameters}\]

The resulting coefficients are summarized in table~\ref{tab:coefficients}.\\

For comparison, measurements obtained in~\cite{Retiere} for the dark count
rate and the crosstalk probability are shown as gray dashed line. As
their measurements have been carried out for a slightly different type
of sensor (50\,\(\mu\)m cell size, 400 cells) a perfect match is not
expected. To achieve comparability, their dark count rates have been
scaled up linearly to an area of 9\,mm\(^2\). It is known, that dark
count rates for the same sensor type can change already from production
to production due to the purity of the silicon. Taking also into
account that both measurements have most probably been taken at
different temperatures, their slopes and even their absolute values fit
surprisingly well. 

\paragraph{Overvoltage}

From the extrapolated voltage offset at which dark count rate, gain and
crosstalk vanish, the overvoltage can be derived. While dark count
rate and crosstalk depend on absolute temperature as well, the effect
on the gain is fully compensated by the feedback system. This is
illustrated by the small error bars demonstrating the small variation
between the measurements. Therefore, the gain is the best measurement
for the overvoltage consistent with the results obtained from the other
parameters. Using the definition of overvoltage
from section~\ref{sec:gapd}, the fit shown in figure~\ref{fig:ovtest}
(top left) and the corresponding fit as presented in
table~\ref{tab:coefficients} suggest that the operation voltage of the
sensors correspond to an overvoltage of 1.4\,V which is consistent
with the datasheet of the devices.

\subsubsection{Temperature dependency}\label{sec:temp}

The temperature dependency of the coefficients is derived from 295 runs
taken between 11/01/2014 and 21/03/2014 at average sensor temperatures
ranging from 4\,\textdegree{}C to 19\,\textdegree{}C. From the fit
results to all measurements,  average and standard deviation of all
channels are calculated.

\begin{figure}[t]
 \centering
 \includegraphics*[width=\textwidth,angle=0,clip]{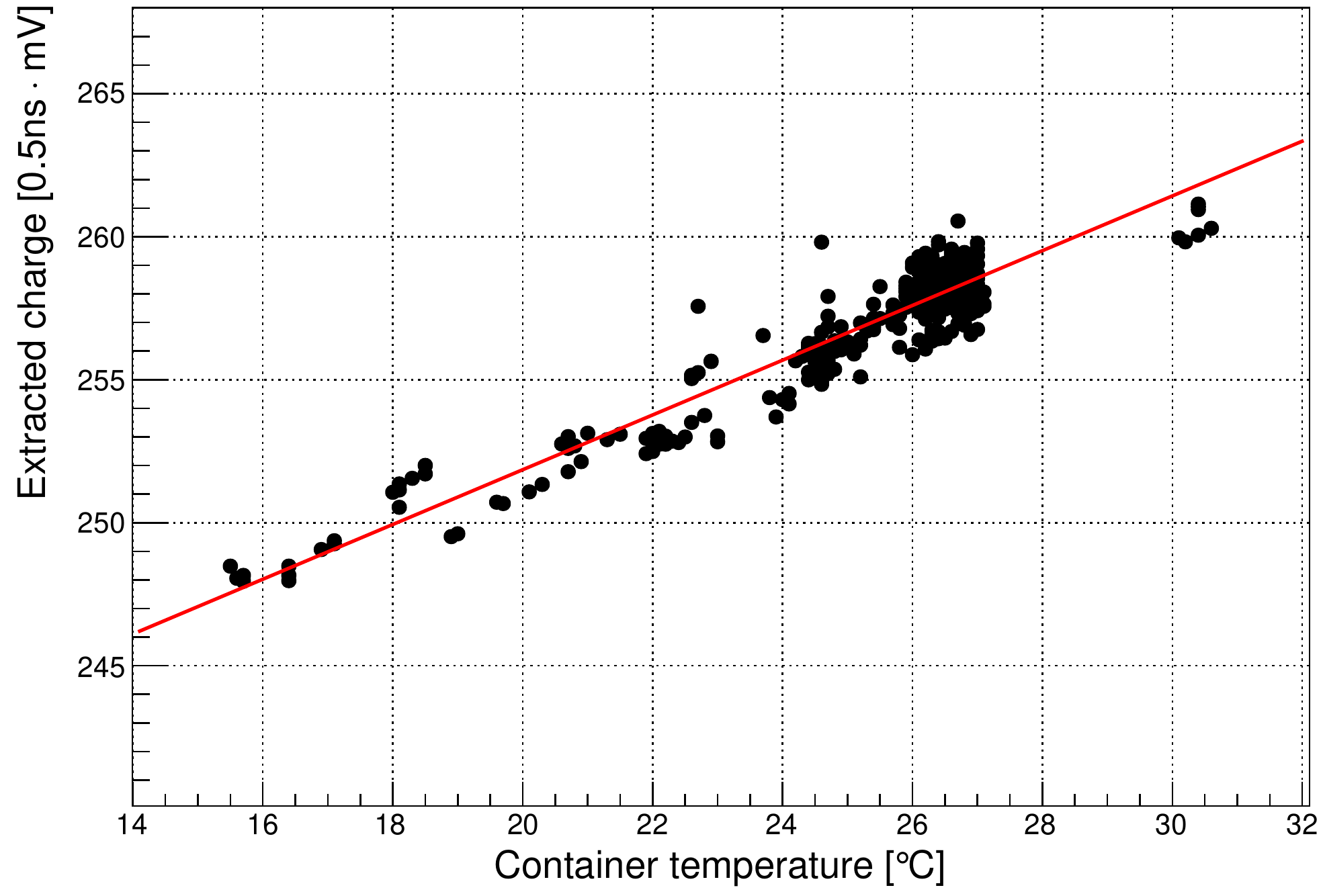}
\caption{The average of the extracted charge obtained from all channels
versus the air temperature in the container hosting the bias power
supply. A clear linear dependency can be seen corresponding to less
than 1\textperthousand{}/\textdegree{}C  in absolute voltage. The
effect most probably originates from the calibration resistor or the
op-amp driving the power supply circuit.}  

\label{fig:container} 
\end{figure} 

Surprisingly, a dependence of the gain on the air temperature of the
container in which the bias power supply is hosted has been found, as
shown in figure~\ref{fig:container}. Investigating the data further, it
turns out that only a fraction of the bias channels show this effect
while channels hosted on two boards out of eleven are not affected.
This makes an individual calibration of every channel
necessary. Although this is doable, the missing control over the
temperature renders a calibration very difficult. The most reasonable
explanation is that two different charges of either the calibration
resistor or the op-amp have been used in assembly and one of them has a
higher temperature gradient. The observed dependency is about 4\% over
10\,\textdegree{}C corresponding to less than
1\textperthousand{} in absolute voltage which is on the order of the
typical temperature gradient of some commercial resistors. Fortunately,
the container is air conditioned and heat waste from the electronics
generates a stable temperature most of the time. In special conditions
like strong winds, clouds or fog on exceptionally cold days, the
temperature can still drop which creates the observed effect. Although
this affects the gain only on a few days a year, it is planned to
either replace the boards or stabilized the temperature. To get an
unbiased result, runs taken at measured container temperatures below
25.5\,\textdegree{}C and above 27\,\textdegree{}C have been rejected so that for the analysis the whole
range  is restricted to roughly 1\,\textdegree{}C. This range contains
the majority of the data points, as it can be seen in
figure~\ref{fig:container}.

\begin{figure}[p]
 \centering
 \includegraphics*[width=\textwidth,angle=0,clip]{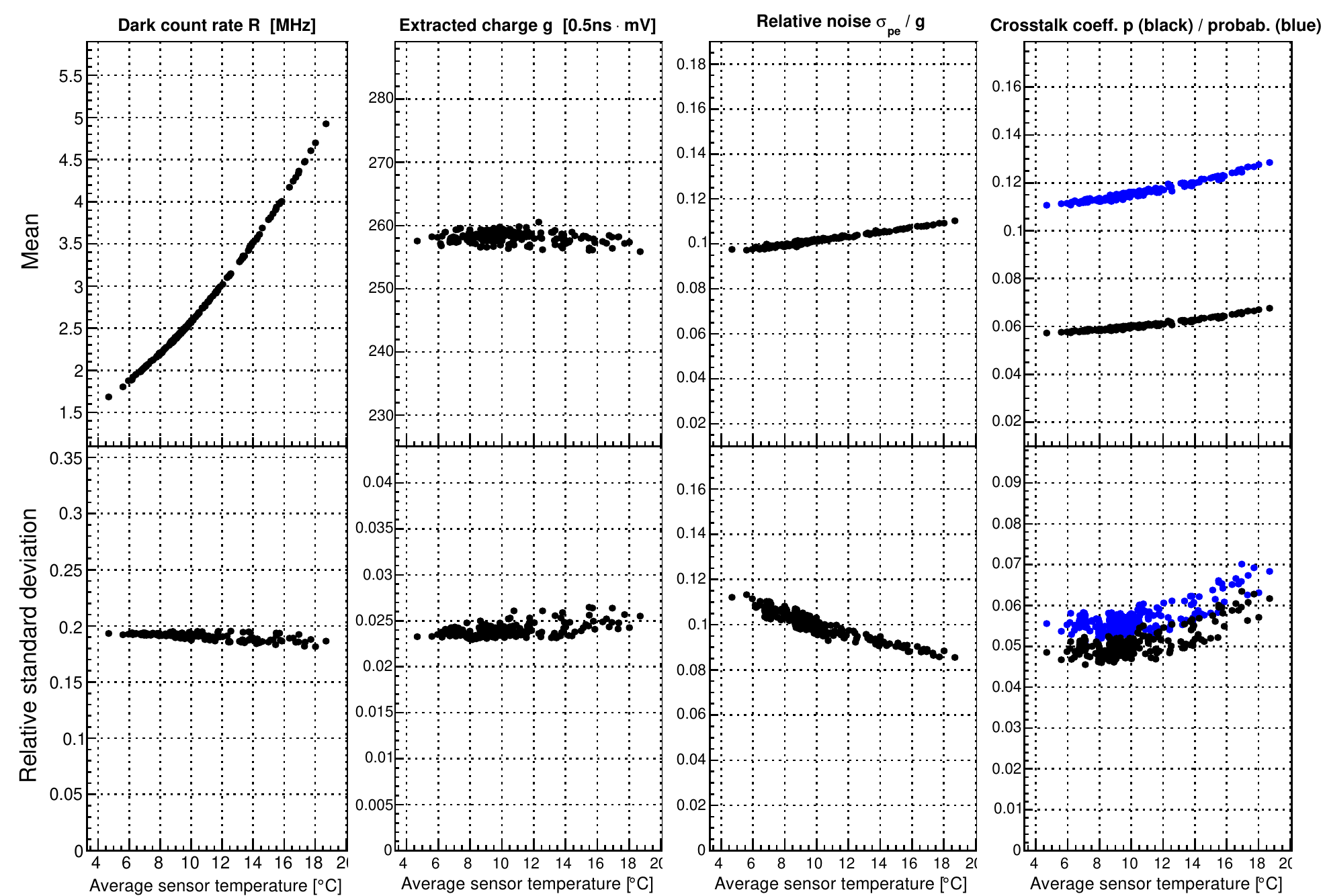}\\[1ex]
 \includegraphics*[width=\textwidth,angle=0,clip]{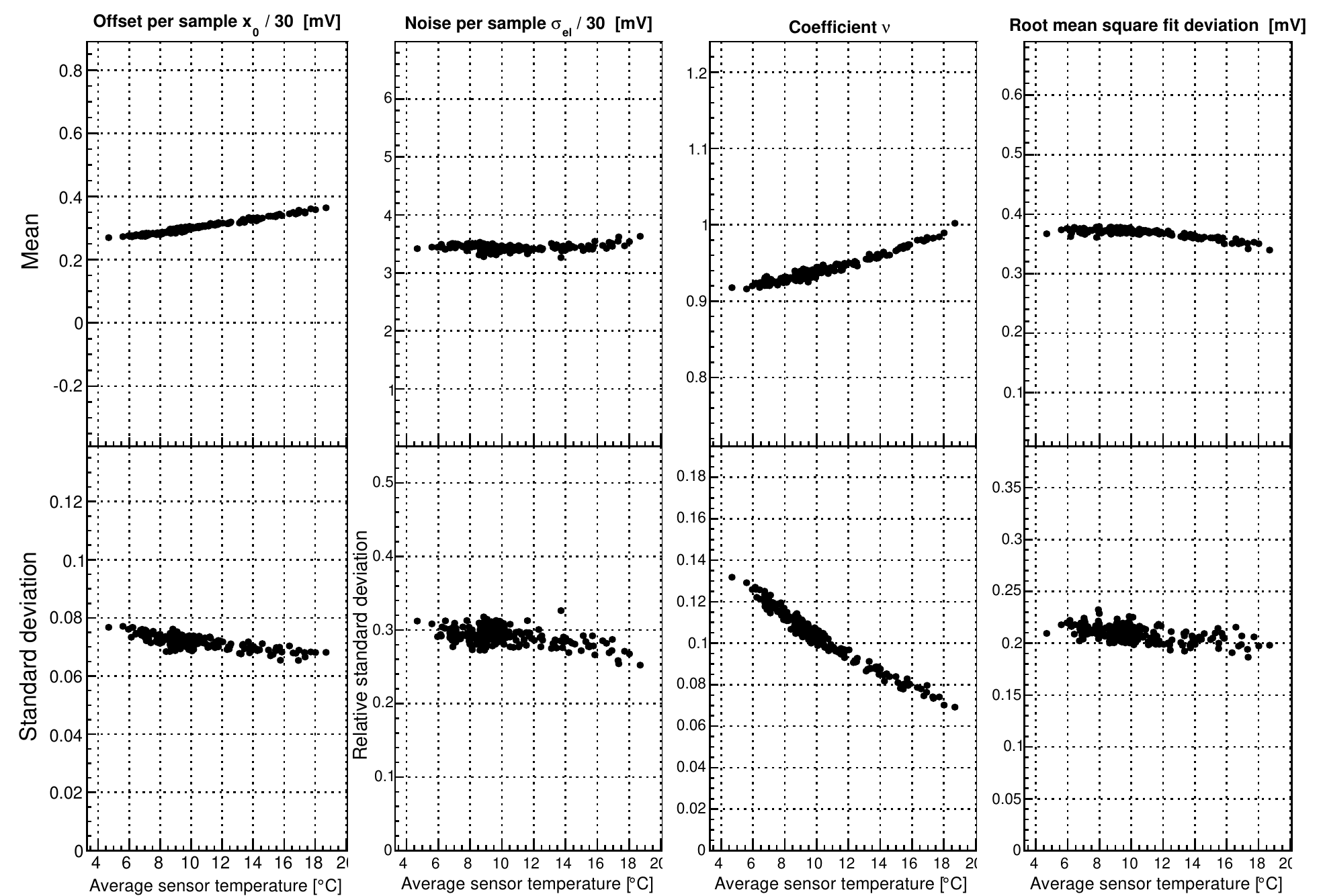}
\caption{Result from the fit to the recorded dark count spectra at 
different sensor temperatures. Shown is the average and standard
deviation of the values obtained from all channels for one run. For all
values except the offset \(x_0\) the standard deviation is expressed
relative to the mean.}  \label{fig:resultvalues}

\end{figure}

An overview of the results of the measurement of the temperature
dependecy is shown in figure~\ref{fig:resultvalues}.
The crosstalk shows a small residual temperature dependency. This can 
be interpreted as a dependency on the absolute voltage. In this study,
both effects can not be disentangled. The dark count rate shows the
dependency on temperature expected from the data-sheet. The distribution
coefficient \(\nu\) shows a dependency, although this affects only
multiplicities larger than \(N\sim 7\). It might as well be related to
small changes in the efficiency of the charge extraction algorithm for
\(N=1\).

Fitting a polynomial of second order to the dark count rate and the
crosstalk coefficient yields the results given in
table~\ref{tab:polynomial}.

\begin{table}[h]
\centering
\begin{tabular}{|l||c|c|c|}\hline
                      & \(c_0\) & \(c_1\) & \(c_2\) \\\hline\hline
Dark count rate/MHz   & 1.263\,\txtpm\,0.006   & 0.0587\,\txtpm\,0.0012 & 0.00730\,\txtpm 0.00005    \\\hline
Crosstalk coefficient & 0.0558\,\txtpm\,0.0003 & (1.7\,\txtpm\,0.5)\(\cdot 10^{-4}\) & (2.362\,\txtpm 0.022)\(\cdot 10^{-5}\) \\\hline
\end{tabular}
\caption{Coeffients of a fit of a second oder polynomial to the temperature
dependence of dark count rate \(R\) and crosstalk coefficient \(p\) as
shown in figure~\protect\ref{fig:resultvalues}.}
\label{tab:polynomial}
\end{table}

The distribution of most fit values in the camera scales nicely
with the value itself so that the relative standard deviation
is mainly temperature independent. The distribution of the relative
noise \(\sigma_{pe}\) and the coeffiecient \(\nu\) becomes smaller with
increasing temperature which can be attributed to the increased number
of available freedoms to fit, more precisely the higher multiplicities
available in the distributions. This is an effect of the increased
crosstalk probability. It is also supported by the slightly increasing
fit quality expressed by the root mean deviation between the distribution
and the fit function.

\subsubsection{Gain stability}

\begin{figure}[t]
 \centering
 \includegraphics*[width=\textwidth,angle=0,clip]{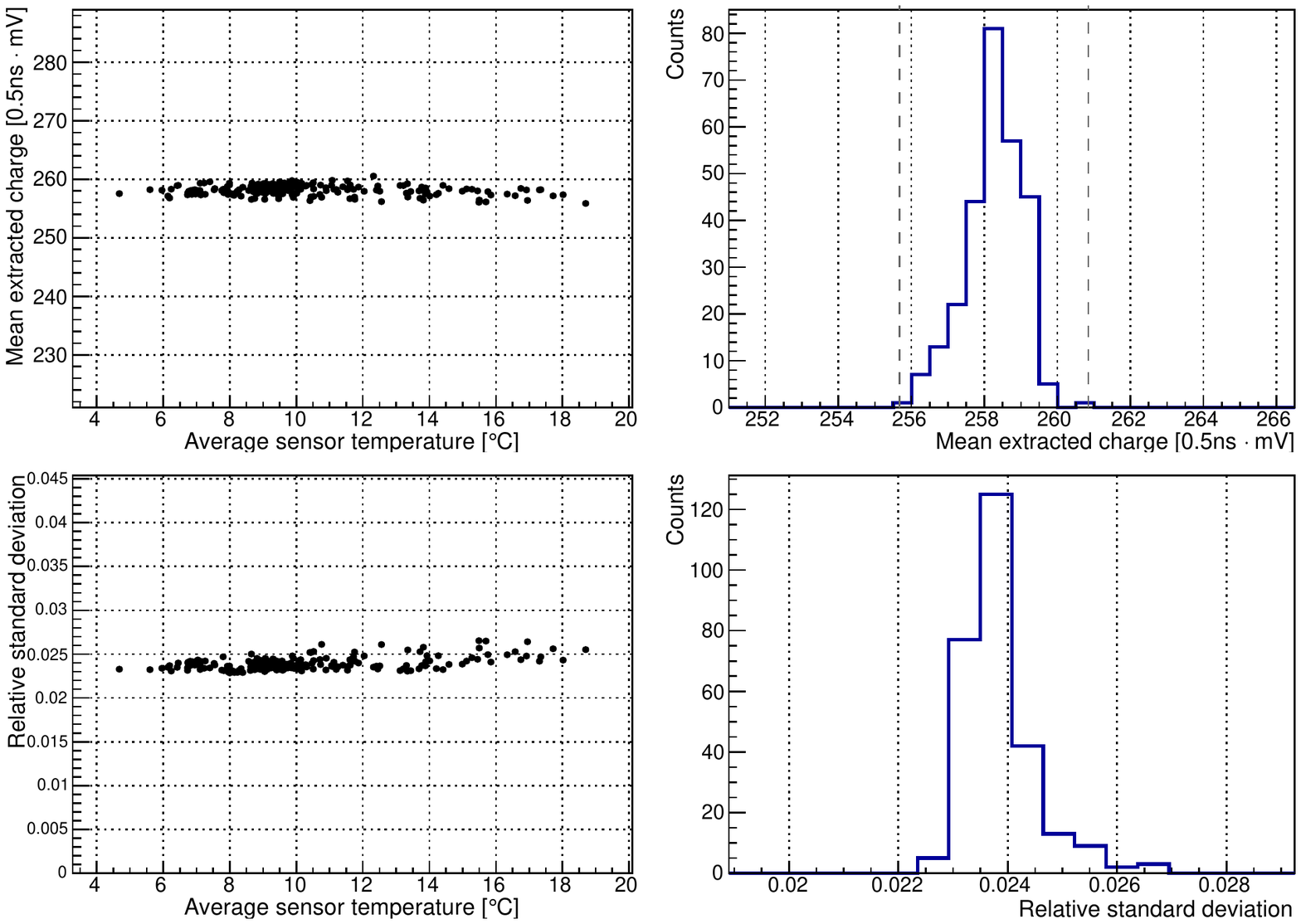}
\caption{Run-wise mean (top) and standard deviation (bottom) of the
extracted charge \(g\) of all channels versus average sensor
compartment temperature (left) and as distributions (right). No
significant dependency on temperature is visible. The width of the
distribution of means is \(\sim\)3\textperthousand{}. Two gray dashed
lines denote the mean \(\pm\)1\%. The average
standard deviation is around 2.4\%.}
\label{fig:gainresult}
\end{figure}

\begin{figure}[t]
 \centering
 \vspace{1em}
 \includegraphics*[width=\textwidth,angle=0,clip,trim=0 7.0cm 0 0]{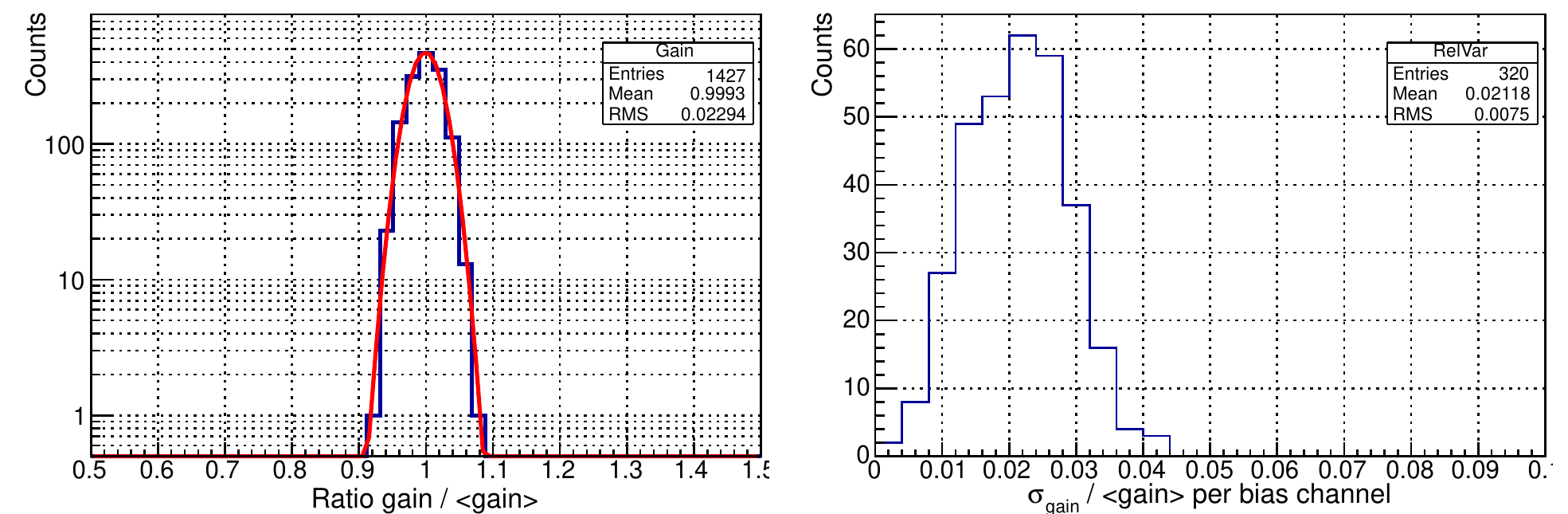}
\caption{Randomly selected example distributions. The left plot
shows the distribution of the gain of all individual channels
normalized to the average gain. The standard deviation of this
distribution is \(\sim\)2.4\%. The right distribution shows the
standard deviation of the gain per bias channel relative to its
average. For an operation voltage which corresponds to an overvoltage
of 1.4\,V, its average of 2.1\% corresponds to a voltage of
\(\sim\)29\,mV.} \label{fig:result-dist} 

\end{figure}

One of the most important results for this project and for future
detectors is the stability obtained for the gain using G-APDs under
real environmental conditions. In figure~\ref{fig:gainresult}, the mean
and standard deviation of the average extracted charge over all channels
is shown versus average sensor temperature together with their
corresponding distributions. 
The standard deviation of the distribution of the measured average gain
values is \(\sim\)3\textperthousand{}. The average of the standard
deviation of the distribution of the gain values in the camera 2.4\%.
Although this standard deviation is in the order of the expectation
from the precision of the hardware and feedback values, it still means
that the maximum deviating channels reach out to almost
\(\pm\)10\% due to the high number of channels. The average standard
deviation within single patches is on the order of 2.1\% and
independent of the temperature. For an assumed overvoltage of 1.4\,V
this corresponds to 29\,mV. This value fits reasonably well with the
precision with which the operation voltages was specified by the
manufacturer used to sort the sensors into patches and the precision of
the individual serial resistors. An example distribution for the gain
of all channels relative to the average gain and an example
distribution for the patch standard deviation is shown in
figure~\ref{fig:result-dist}. For single sensors, this result means
that they stay within the achievable limit defined by the hardware
voltage setting, otherwise the standard deviation would be increased.
Apart from the large number of sensors, the full width of the
distribution can be explained with the standard deviation of values
within patches of up to 4\% plus the 2.2\% shift introduced from a
single voltage step.


\subsubsection{Sum spectra}

\begin{figure}[hp]
 \centering
 \includegraphics*[width=0.98\textwidth,angle=0,clip]{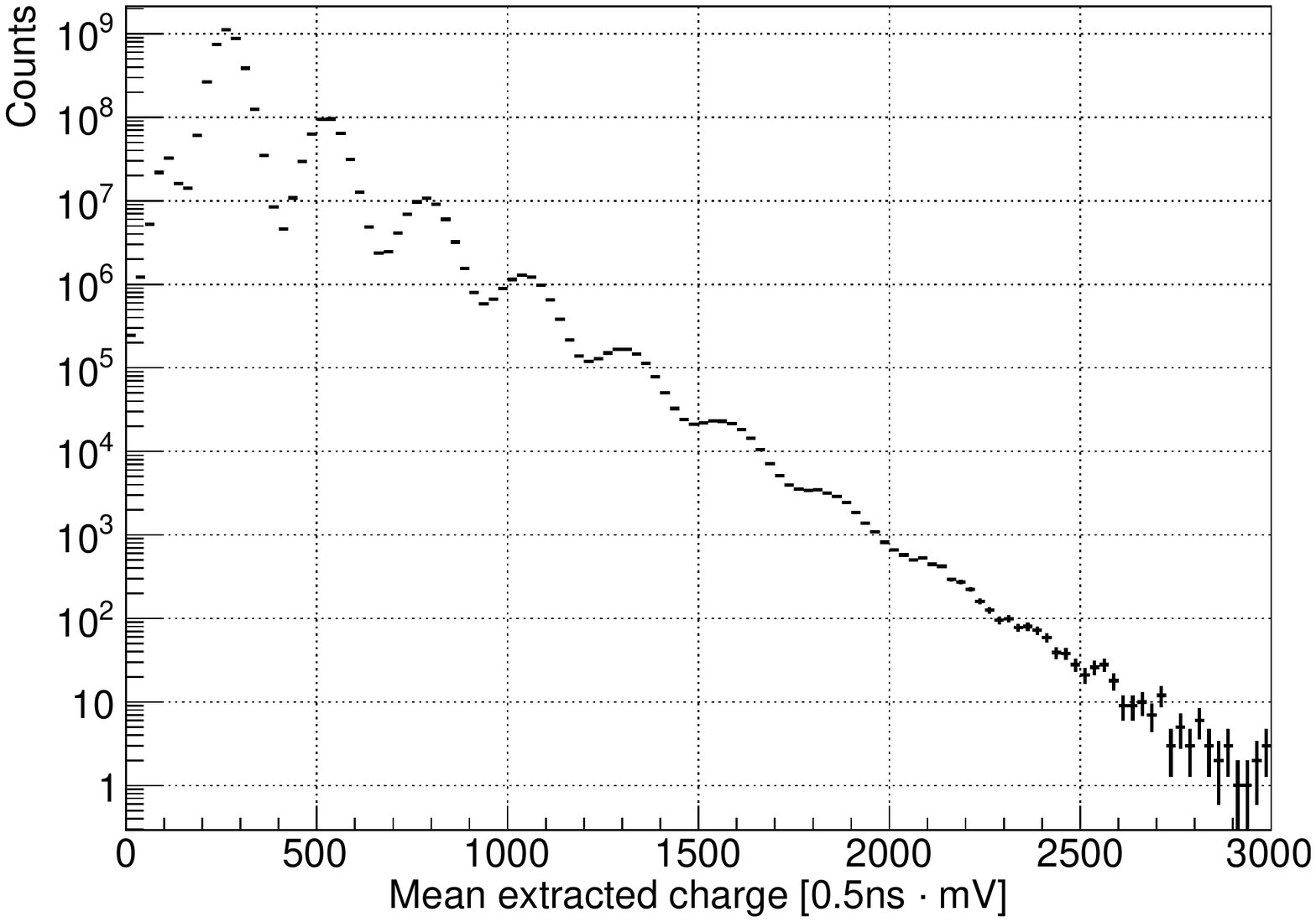}\\[2em]
 \includegraphics*[width=0.98\textwidth,angle=0,clip]{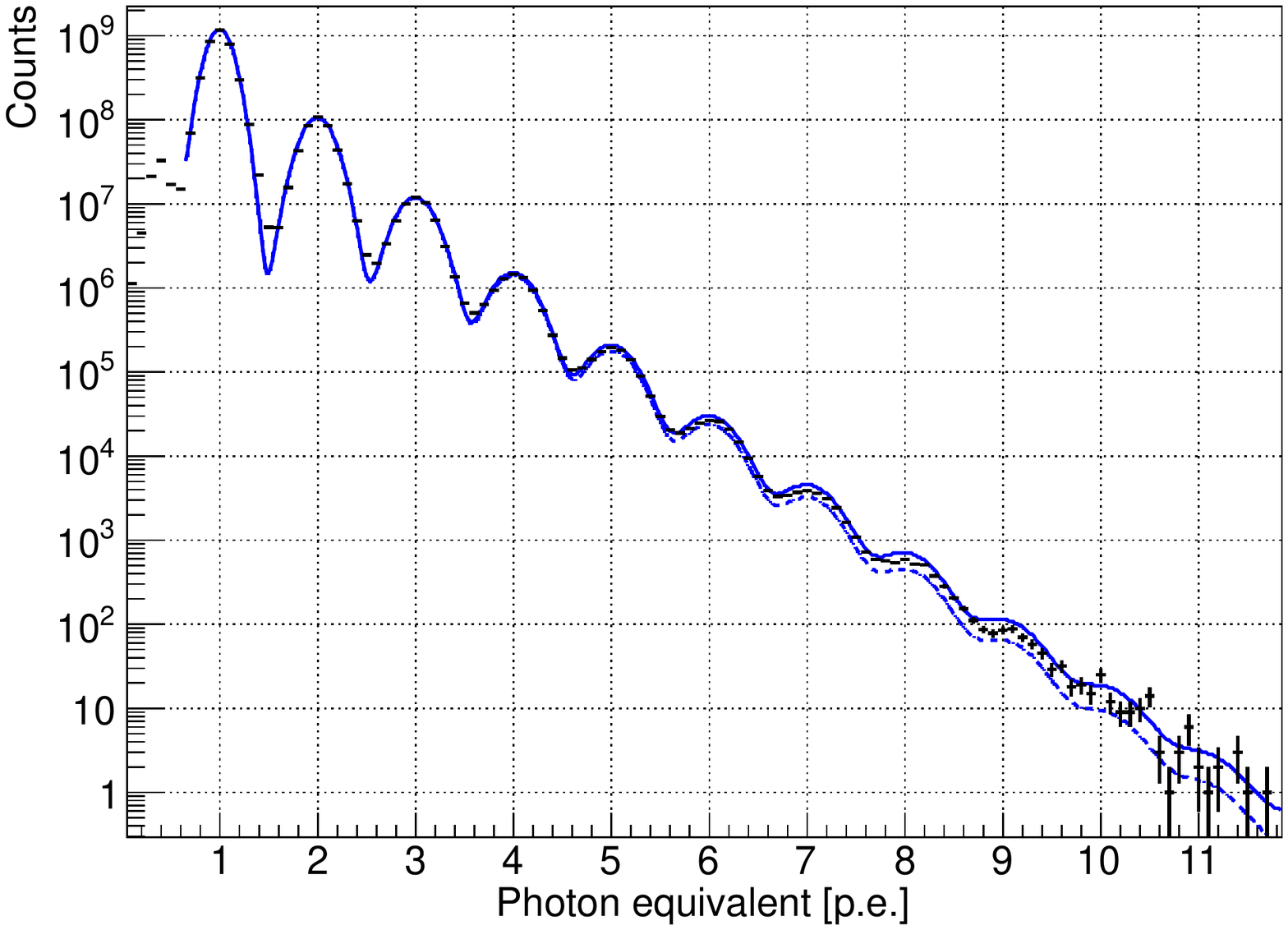}
\caption{Top: The sum of all dark count spectra. Different peaks can easily
be distinguished up to \(N\sim7\). Bottom:
Sum of the same spectra, but each one normalized
individually with the extracted offset and gain. Overlayed is a fit
of a modified Erlang distribution (solid line, top) and an Erlang distribution
(dashed line, bottom). More detail given in the text.}
\label{fig:spectrum}
\end{figure}

Due to the very small variation of the gain, all measured spectra can
be combined into a single spectrum, as shown in
figure~\ref{fig:spectrum} (top). The peaks from individual breakdowns
can easily be distinguished up to a multiplicity of \(N\sim 7\). This
result improves further, when all individual spectra are offset (\(x_0\))
subtracted and normalized with the determined gain \(g\). This result
is shown in figure~\ref{fig:spectrum} (bottom). Here, multiplicities up
to at least \(N=10\) can be distinguished. The high level of similarity
between the scaled and unscaled distributions proves not only the
stability of the system and the successful operation of the feedback
system but also the precision of the applied charge extraction. The
overlaid fit of a modified Erlang distribution (solid) and a standard
Erlang distribution (dashed) proves the applicability of these
distributions up to very high multiplicities. A difference between both
is only visible for multiplicities \(N>5\).


The fit results for the modified Erlang distribution with \(g=1\) and
\(x_0=0\) are summarized in table~\ref{tab:sumspectrum}. The relative
errors of all parameters are in the order of \(10^{-4}\) or smaller.
The corresponding crosstalk probability \(p_{xt}\) according to 
equation~\ref{eq:crosstalk} is 11.7\%.

\begin{table}[h]
\centering
\begin{tabular}{|c||c|c|c|c|c|}\hline
Parameter & \(A_1\)                & \(\sigma_{pe}\) & \(\sigma_{el}\) & \(p\)    & \(\nu\)  \\\hline
Value     & \( 1.2282\cdot 10^9 \) & 0.10469         & 0.060770        & 0.070314 & 0.93715  \\\hline
\end {tabular}
\caption{Resulting coefficients from fitting a modified Erlang distribution
to the sum spectrum shown in~\protect\ref{fig:spectrum}. All significant digits
according to the obtained statistical errors are provided.
}
\label{tab:sumspectrum}
\end{table}

If in a camera with 1440 sensors, a rate of avalanches per channel
between 50\,MHz and 2\,GHz induced by night-sky background
photons is expected, this gives a total rate between
72\,GHz and 2.9\,THz. Taking the fit result for the spectrum in
figure~\ref{fig:spectrum}, this yields roughly one event per second
with a multiplicity of \(N\)\,=\,15 in low light conditions and one
with \(N\)\,=\,18 for bright light conditions excluding possible pile-up from
the night sky itself. Including fluctuations,
it can be assumed that with a trigger threshold of at least 20\,\pe
(\(N\)\,=\,20) not more than one trigger per second is induced by
optical crosstalk events.

\afterpage{\clearpage}

\subsubsection{Pulse shape}

\begin{figure}[tp]
 \centering
 \includegraphics*[width=0.49\textwidth,angle=0,clip]{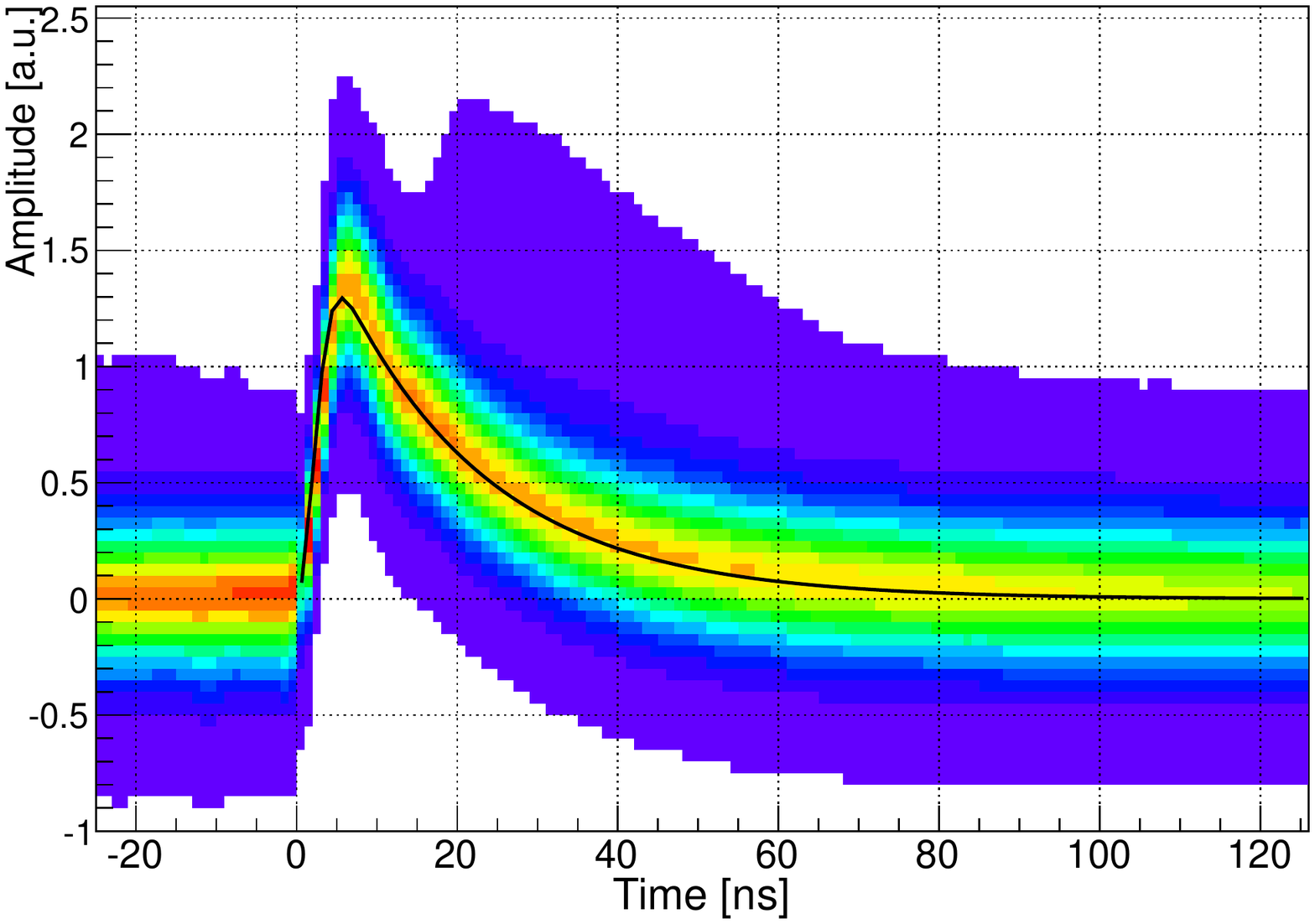}
 \includegraphics*[width=0.49\textwidth,angle=0,clip]{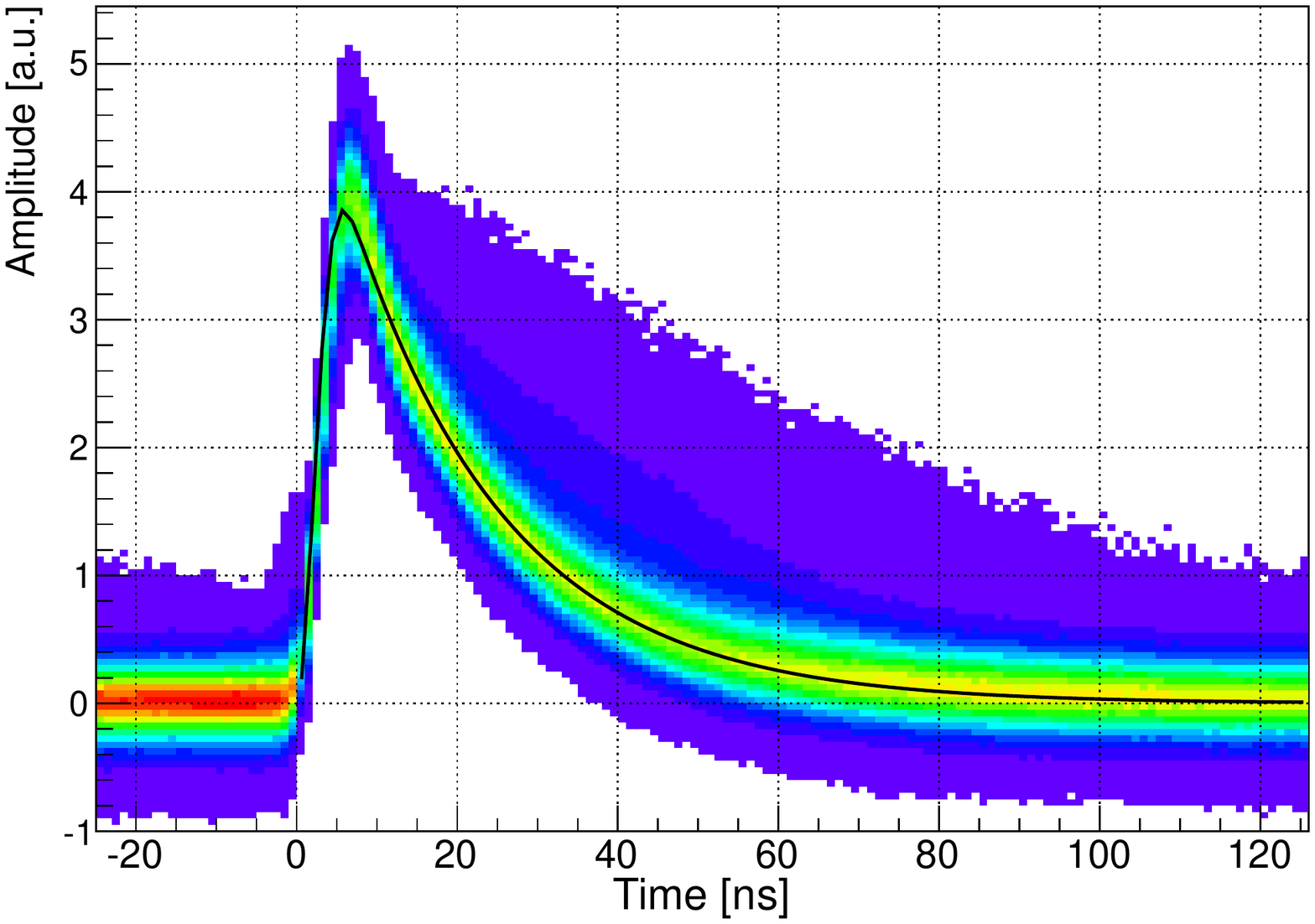}\\
 \includegraphics*[width=0.49\textwidth,angle=0,clip]{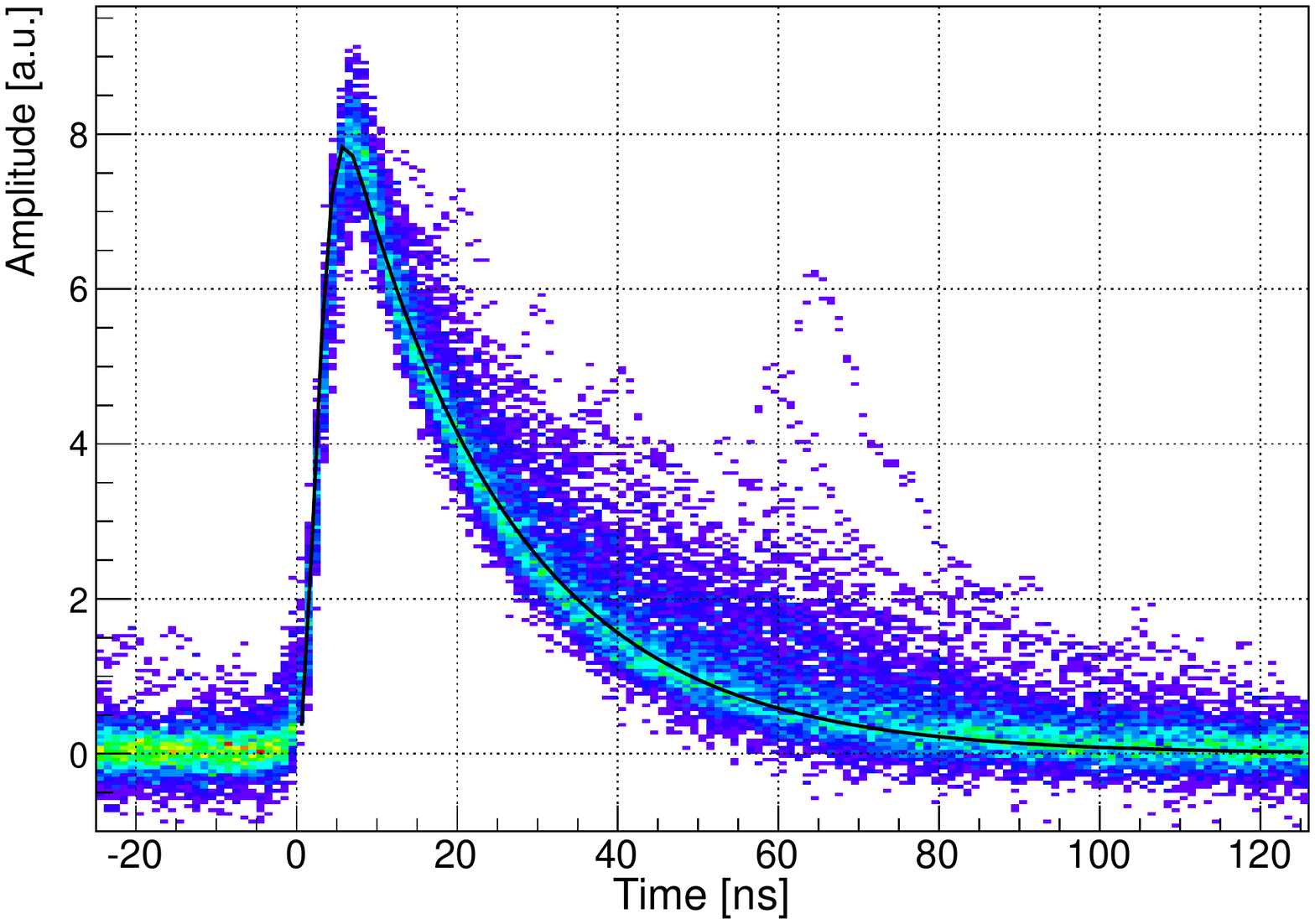}
 \includegraphics*[width=0.49\textwidth,angle=0,clip]{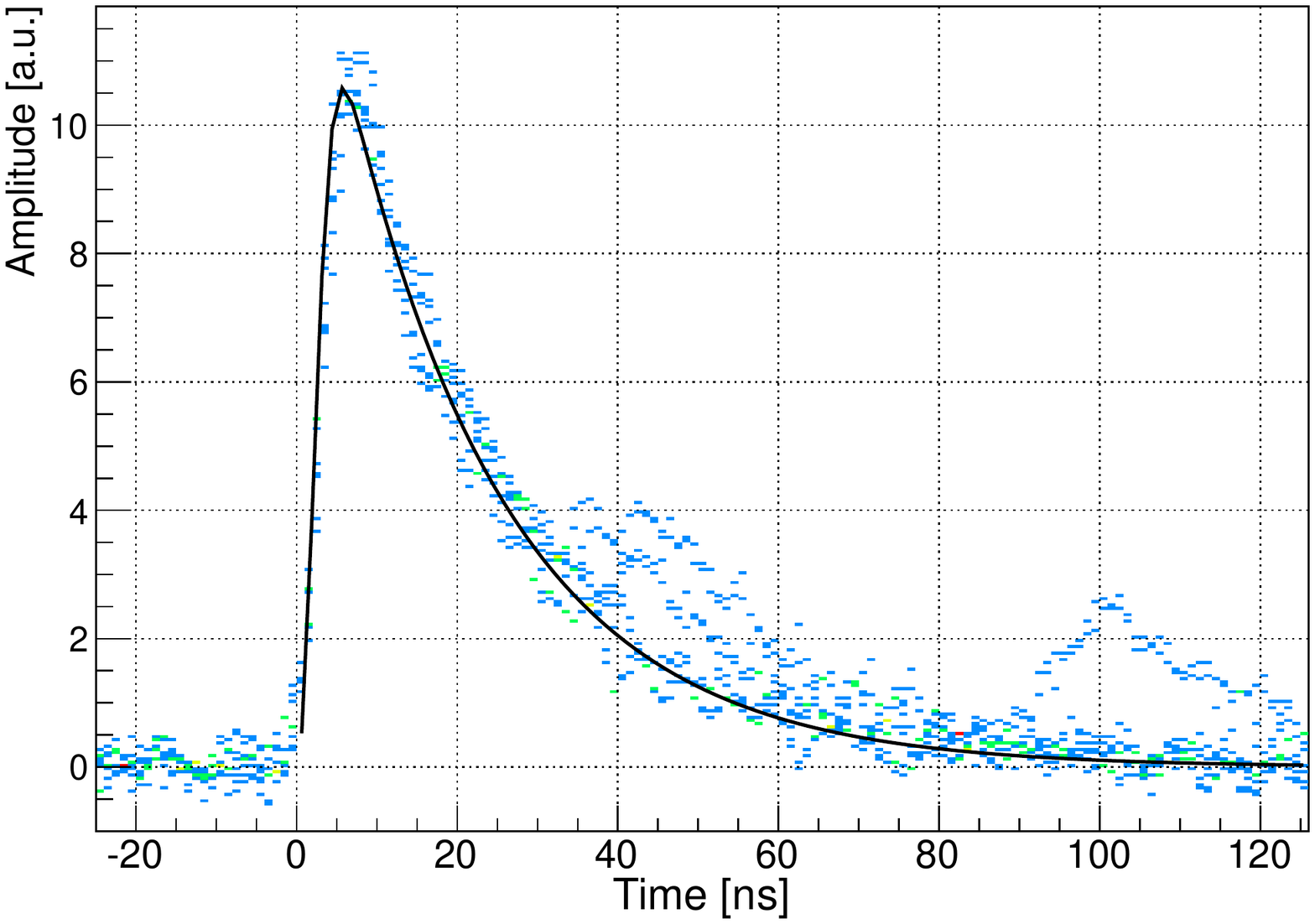}\\[2em]
 \includegraphics*[width=0.98\textwidth,angle=0,clip]{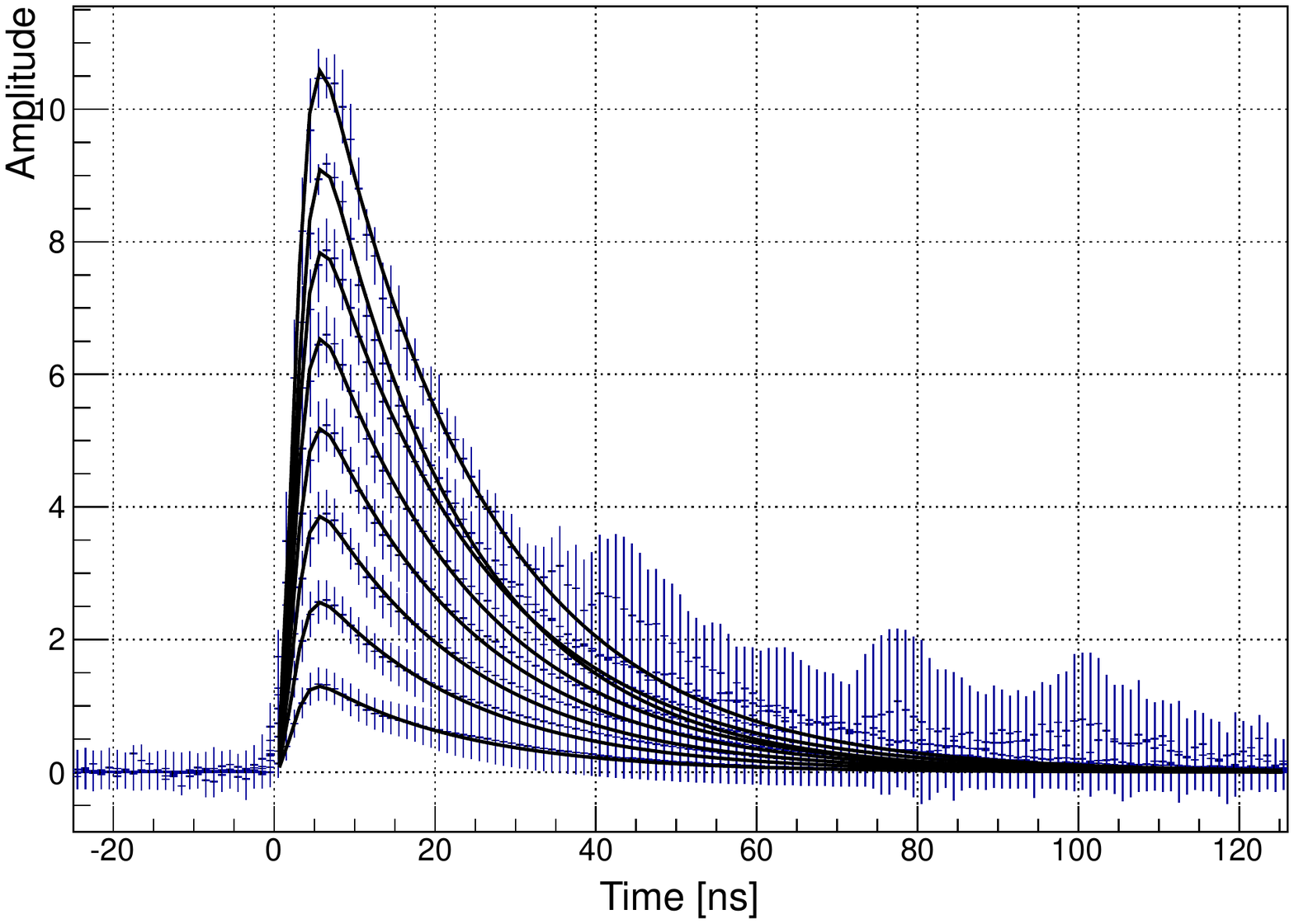}  
\caption{
Two dimensional histograms (top) of the sampled waveform. The color
scale starts at 0.5\% of the maximum bin. The shown multiplicities are
1, 3, 6 and 8 (top left to bottom right). For all multiplicities between
one and eight, a profile histogram has been filled (bottom). Fits
to the profile histogram are shown as black lines. The inlay shows all
eight fits re-normalized with \(c_0\)\,=\,1, \(c_1\)\,=\,0 and
\(N\)\,=\,1. A good agreement is visible.}

\label{fig:pulse}
\end{figure}

The average gain of each channel can be used to scale the extracted
pulses and overlay them shifted by their arrival time.  From this, two
dimensional histograms were filled and profiles calculated, separated
for different multiplicities, for example, for multiplicity \(N\)\,=\,1
for extracted signals between \(0.5 g\) and \(1.5 g\). 
The result for a randomly selected run at nominal operation voltage
can be seen in figure~\ref{fig:pulse}. A very good match of individual
pulses is evident. Also visible is the influence from afterpulses.
Since their amplitude is attenuated by the amount of remaining charge
in the cell and at the same time their probability is exponentially
decreasing, they show an activity maximum a few nanoseconds after the
primary pulse. If the range in which a pulse is integrated is chosen
carefully, the bias from afterpulses can be suppressed completely. 

To all profiles, the following function has been fit between 1\,ns
and 25\,ns:

\[A(t/\mbox{ns}, N) = c\,N \cdot (1-\frac{1}{1+e^{\frac{t-t_0}{\tau}}})\cdot e^{-\frac{t-t_0}{\lambda}}
\label{eq:pulseshape}\]

The shift of function with respect to the extracted arrival time is a
priori unknown, therefore a free parameter for the time shift \(t_0\)
is introduced. If the charge scales with the multiplicity \(N\) as
expected, the normalization \(c\) should be independent of the
multiplicity. The rise and fall times are described by \(\tau\) and
\(\lambda\) respectively.  Its maximum is reached at
\(t=t_0+\tau\log(\lambda/\tau-1)\). The fit range was chosen to
suppress the effect of afterpulses and optical crosstalk events induced
from afterpulses. The results of the fits for different multiplicities
are summarized in the table~\ref{tab:pulseshape}.

\begin{table}[h]
\centering
\begin{tabular}{|c||c|c|c|c|}\hline
N  & \(c\)              & \(t_0\)/ns         & \(\tau\)/ns         & \(\lambda\)/ns    \\\hline\hline
 1 & 1.57\,\txtpm\,0.20 & 2.7 \,\txtpm\,0.5  & 0.9 \,\txtpm\,0.3   &   19\,\txtpm\,5   \\\hline
 2 & 1.57\,\txtpm\,0.13 & 2.8 \,\txtpm\,0.3  & 1.05\,\txtpm\,0.19  &   19\,\txtpm\,3   \\\hline
 3 & 1.59\,\txtpm\,0.10 & 2.8 \,\txtpm\,0.3  & 1.07\,\txtpm\,0.16  & 19.3\,\txtpm\,2.6 \\\hline
 4 & 1.60\,\txtpm\,0.08 & 2.8 \,\txtpm\,0.3  & 1.07\,\txtpm\,0.16  & 19.4\,\txtpm\,2.2 \\\hline 
 5 & 1.62\,\txtpm\,0.07 & 2.86\,\txtpm\,0.23 & 1.09\,\txtpm\,0.14  & 19.4\,\txtpm\,1.9 \\\hline 
 6 & 1.62\,\txtpm\,0.07 & 2.91\,\txtpm\,0.23 & 1.14\,\txtpm\,0.14  & 19.9\,\txtpm\,1.9 \\\hline 
 7 & 1.63\,\txtpm\,0.03 & 3.00\,\txtpm\,0.14 & 1.04\,\txtpm\,0.08  & 18.2\,\txtpm\,0.7 \\\hline
 8 & 1.62\,\txtpm\,0.04 & 2.74\,\txtpm\,0.20 & 1.07\,\txtpm\,0.10  &   20\,\txtpm\,1.0 \\\hline
\end{tabular} 
\caption{Resulting coefficients of fitting the pulse shape described by equation~\protect\ref{eq:pulseshape}
to the pulse profiles as shown in figure~\protect\ref{fig:pulse}.}
\label{tab:pulseshape}
\end{table}

Applying the signal extraction used for the dark count spectrum to the
fit functions, or integrating the pulse from half-height-maximum for
15\,ns, leads to a charge consistent with the expected linear behavior
within errors. This match is illustrated by the good match of all
eight pulses re-normalized with \(c_0\)\,=\,1, \(c_1\)\,=\,0 and
\(N\)\,=\,1. This is shown in the inlay in figure~\ref{fig:pulse}
(bottom). This is consistent with the expectations taking the band width
of the readout chain into account.


\afterpage{\clearpage}

\section{Light pulser measurements}\label{sec:lightpulser}


The evaluation of dark count spectra taken with closed lids cannot
measure the dependence of the gain from the current, i.e.\ the ambient
light condition. Instead, the measured amplitude of an external light
pulser is used. The light pulser is installed in the center of the
reflector dish. Its light yield is temperature stabilized. 

For each measurement, the light pulser is flashed one thousand times
with a rate of 25\,Hz. The readout is self-triggered by the camera's
trigger system. To avoid triggers on showers, the time window for each
trigger is only 12\,ns and a trigger signal from at least 25 out of 40
trigger boards is required.

Light pulser runs are currently taken roughly every 20 minutes during
standard data taking. The data from the 650 analyzed runs
presented here were taken between 31/12/2013 and 21/03/2014. During
all measurements the feedback system was in operation.


\subsection{Method}

\paragraph{Signal extraction}
After the digital values have been converted to physical units, the
signal is extracted by a peak-search and an integration of 5\,ns before
and 80\,ns after the half-height leading edge. To determine the maximum
and the position of the half-height leading edge, a \nth{3} order
spline interpolation is applied. For each channel, the average
amplitude of all events is calculated.

The emitted pulses have a typical length of 50\,ns with an average
amplitude of \(\sim\)30\,\pe per pixel. Although the event-to-event
fluctuation of the amplitude is comparably large, it is stable on average.




\paragraph{Light pulser properties}
Laboratory
measurements have shown that a small temperature dependency of the light yield still
remains.
As the correlation of the measured amplitude on the ambient temperature
is stronger than on the sensor temperature, it can be assumed that this
is an effect of the light pulser exposed to ambient temperature rather
than of the gain of the sensors. This is also justified because the
amplitude increases with increasing temperature. In the case that this
would be an effect of the gain of the sensors, the measured temperature
had to be higher than the real temperature of the sensor so that the
feedback system would apply a too high voltage. As the main heat source
in the sensor compartment are the sensors themselves, there is no
reason to assume that with increasing current a temperature measured
close to the sensors is overestimated. Therefore, the residual
temperature dependency has been fitted with a line with a slope
corresponding to \(\sim\)5\%/\(100\,\mu\)A and the data has been corrected
accordingly.

The light emitted by the pulser is spatially inhomogeneous on the
camera surface to within a few percent. The light distribution is approximated
with an average amplitude distribution from all data taken under 
moonless conditions with \(I<7\,\mu\)A for which the smallest influence is
expected. For this study, it has then been subtracted patch-wise from
all runs. 

Another bias on the measurement is the tidiness of the light-emitting
diode. Small jumps of the measured amplitude are observed between
consecutive nights, especially after rain, fog or snow.

\subsection{Results}


\begin{figure}[t]
 \centering
 \includegraphics*[width=\textwidth,angle=0,clip]{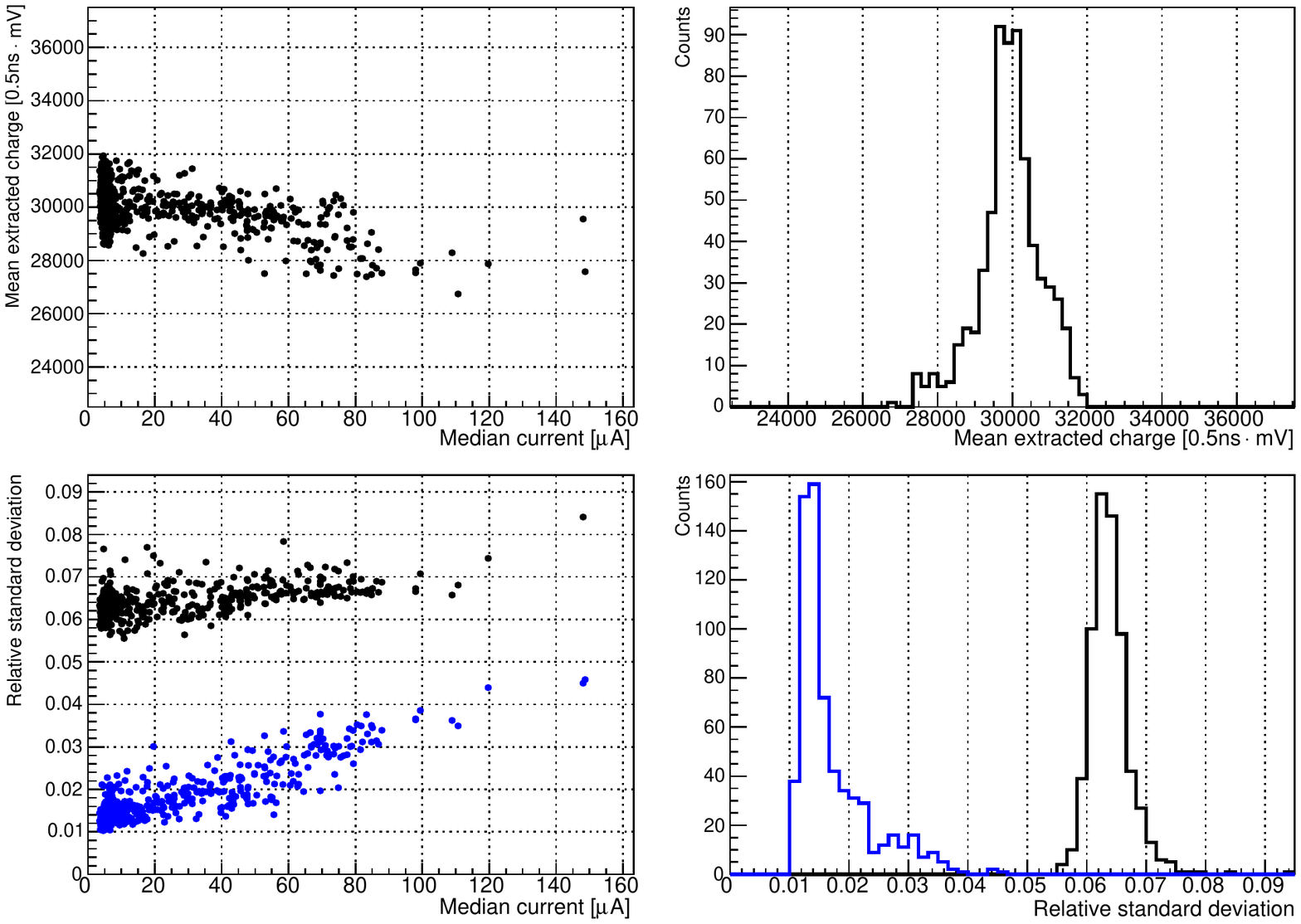}
\caption{Run-wise mean (top) and standard deviation (bottom) of the
light pulser amplitude from 1428 channels versus median per-sensor
current (left) and as distribution (right). Blue denotes the standard
deviation of the amplitude distribution, corrected for the
inhomogeneous illumination of the camera surface. }
\label{fig:light-pulser} \end{figure}

The results of the measurements are shown in
figure~\ref{fig:light-pulser}. Each entry represents the mean (top) or
sigma (bottom) of a Gaussian fitted to the distribution of the average
amplitude of all 1428 channels per run. Black denotes the uncorrected
distribution, blue denotes the distribution corrected for the average
light distribution. In the following, the indicated current is the
average of the currents determined for the individual sensors to avoid
the ambiguity from a different number of pixels served by different
voltage channels. A current of 100\,\(\mu\)A therefore gives a current
in the bias channel of 400\,\(\mu\)A or 500\,\(\mu\)A depending on
whether four or five sensors are served by this channel. While the mean
(top left) shows a decrease between new moon (\(\sim\)5\,\(\mu\)A) and
bright moon light conditions (\(\sim\)100\,\(\mu\)A), the sigma
(bottom, left) shows an increase.

The decrease of the average amplitude between new moon and bright moon
light conditions is on the order of 5\%. This trend is independent of
the applied temperature correction. The standard deviation of the
distribution is smaller than 3\%.

Both trends are due to inhomogeneities in the amplitude distribution
appearing in the camera with increasing current. Looking into the data
in more detail, it turns out that while the majority of the pixels
still shows the nominal expected value, large scale  patterns which
show lower amplitude exist. Runs taken at different pointing directions
and different moon positions show different patterns.

The reasonable explanation found for the decrease with increasing
current, could be that the temperature of the sensor
compartment is dominated by the waste heat of the sensors themselves.
At a voltage of \(\sim\)70\,V and a current of 100\,\(\mu\)A per
sensor, the sensors dissipate a total power of \(\sim\)10\,W.
Generally, waste heat is dissipated through the solid cones, the photo
sensor's carrier and the air.  Especially at high currents, a fraction
of the waste heat is lost through the solid cones. Therefore the
temperature at the temperature sensors slightly smaller than of the
photo sensors themselves which in turn facilitates a too low applied
voltage. While an 8\% decrease can be explained with the temperature
difference of only 2\,\textdegree{}C, a mismatch in resistor value or
current measurement of 8\% is unlikely within the small range of
operation compared to the full available range. Former studies did not
show this trend hidden by the larger spread of the data due to the
lower resolution of the applied analysis.

Another possible explanation could be the decrease of detection
efficiency at bright light conditions because on average a finite
number of G-APD cells recently suffered a breakdown and are recharged.
Taking the spectrum of the diffuse night-sky background and the
spectral response of the instrument into account, the rate of induced
avalanches per sensor under new moon conditions does not exceed 50\,MHz,
corresponding to \(\sim\)5\,\(\mu\)A. The brightest
conditions, corresponding to a \(\sim\)98\% illumination of the moon
disk, produced a current of not more than 200\,\(\mu\)A. Which is the
maximum at which regular observations have been carried out so far.
For a dark count rate of 5\,MHz and a dark current of 0.5\,\(\mu\)A
a corresponding rate of \(\sim\)2\,GHz for breakdowns per sensor can be
deduced. Taking the determined pulse shape as reference with a
half-value time of 20\,ns, this yields on average 1\% of cells which
show no or a significantly reduced response. Taking 100\,ns until the
cell is fully recharged, on average another 4\% of all cells will show
an attenuated signal. That means that even during the brightest
observations ever recorded, the detector efficiency was not reduced by
more than 5\%. Taking the average of the released charge, the decrease
will not be more than 2\%. Consequently, at 100\,\(\mu\)A the effect is
negligible and cannot explain the measured 5\% decrease.\\



The uncorrected standard deviation in the camera is between 6.5\% and 7\%.
Correcting for the average light distribution during new moon nights, it
increases from \(\sim\)1\% to a maximum of 4\% during bright moon light
conditions. This increase is expected for several reasons: The brighter
the ambient light gets the more inhomogeneous the background light
yield in the camera becomes due to direct moon light, reflections and
shadowing. While the purpose of the feedback is to correct this effect,
it can only be corrected on average per bias voltage channel, but not
for the individual sensors connected to each channel. Consequently,
the distribution is expected to get broader, the more inhomogeneous the
background light becomes. 
In addition, the noise in the readout
increases with the square-root of the background light flux which
directly affects the variation of the reconstructed amplitude
especially at bright light conditions. This is supported by the fact
that at low currents the event-to-event variation correlates with the
average amplitude. This effect vanishes towards higher currents. A
possible underestimation of the sensor temperature has been discussed
previously. 


\section{Ratescans}\label{sec:ratescans}

With the measurement of the dark count spectrum and the light pulser
amplitude, the stability of the system in terms of temperature and
ambient light conditions has been demonstrated. In both cases, only the
response of a fraction of the whole system is taken into account and
the light pulser measurement shows influence from the properties
of the light pulser itself. Instead, {\em ratescans} offer the
possibility to directly measure the detector response on ambient light
and air showers. Ratescans determine the detector's trigger rate as a
function of the trigger threshold. At low thresholds, this rate is
completely dominated by triggers induced randomly from night-sky
background photons. At high thresholds, triggers by air showers or
their secondary particles dominate. At low thresholds, the total
trigger rate is the sum of the triggers of all trigger channels because
triggers of individual channels are independent. At high thresholds,
the total trigger rate is lower than the sum of the individual  trigger
channels because in most cases several patches trigger in coincidence.

To understand this in more detail, the trigger hardware is outlined
hereafter. A  more detailed description is available in~\cite{Design}.

\subsection{Method}
\paragraph{Trigger system}

First, an analog sum of the nine signals of the channels
corresponding to one four- and one five-pixel bias voltage channel is
performed. Clipping is used to shorten the length of the output signal.
Next in line is a comparator whose threshold is set by a 12\,bit
digital-to-analog converter. Roughly 15 counts correspond to the
amplitude of a single photon equivalent. As the conversion is not
precise but depends on many factors, as for example the gain of the
sensors, in the following only counts are used, although this number
allows for an estimate of the amplitude in units of photon equivalents.
The comparator signal from four patches is summed serving as input for
a 1-out-of-4 discriminator logic. Requiring a minimum length of its
input signal, this logic suppresses electronics noise. The final
trigger decision is a simple OR of the incoming discriminator signals,
more precisely the result of a N-out-of-40 logic realized in a Field
Programmable Gate Array (FPGA) with N\,=\,1.

To monitor the rate of all trigger channels, a counter for all
comparator and discriminator outputs as well as a counter for the final
trigger decision is implemented in the controlling FPGAs. Their readout
time is set to once every second during ratescans and once every five
seconds during data taking.

\paragraph{Measurement}

Under normal data taking conditions, one or two ratescans are performed
every night. They typically last about five to ten minutes. For a 
pointing direction, a sky region without bright stars is chosen with a
zenith distance not more than 25\textdegree{}. This avoids bias from
bright stars in the field-of-view and increased attenuation towards the
horizon. Ratescans taken under verifiable non-ideal weather conditions
have been discarded. Each measured point contains at least 400 triggers
so that a statistical error of at least 5\% is achieved for the total
trigger rate. Ratescans are started at a threshold of 100 and stopped 
when the time needed to collect enough data to achieve the 5\%
statistical error exceeds 2.5 minutes. 

\subsection{Results}

\begin{figure}[t]
 \centering
 \includegraphics*[width=0.49\textwidth,angle=0,clip]{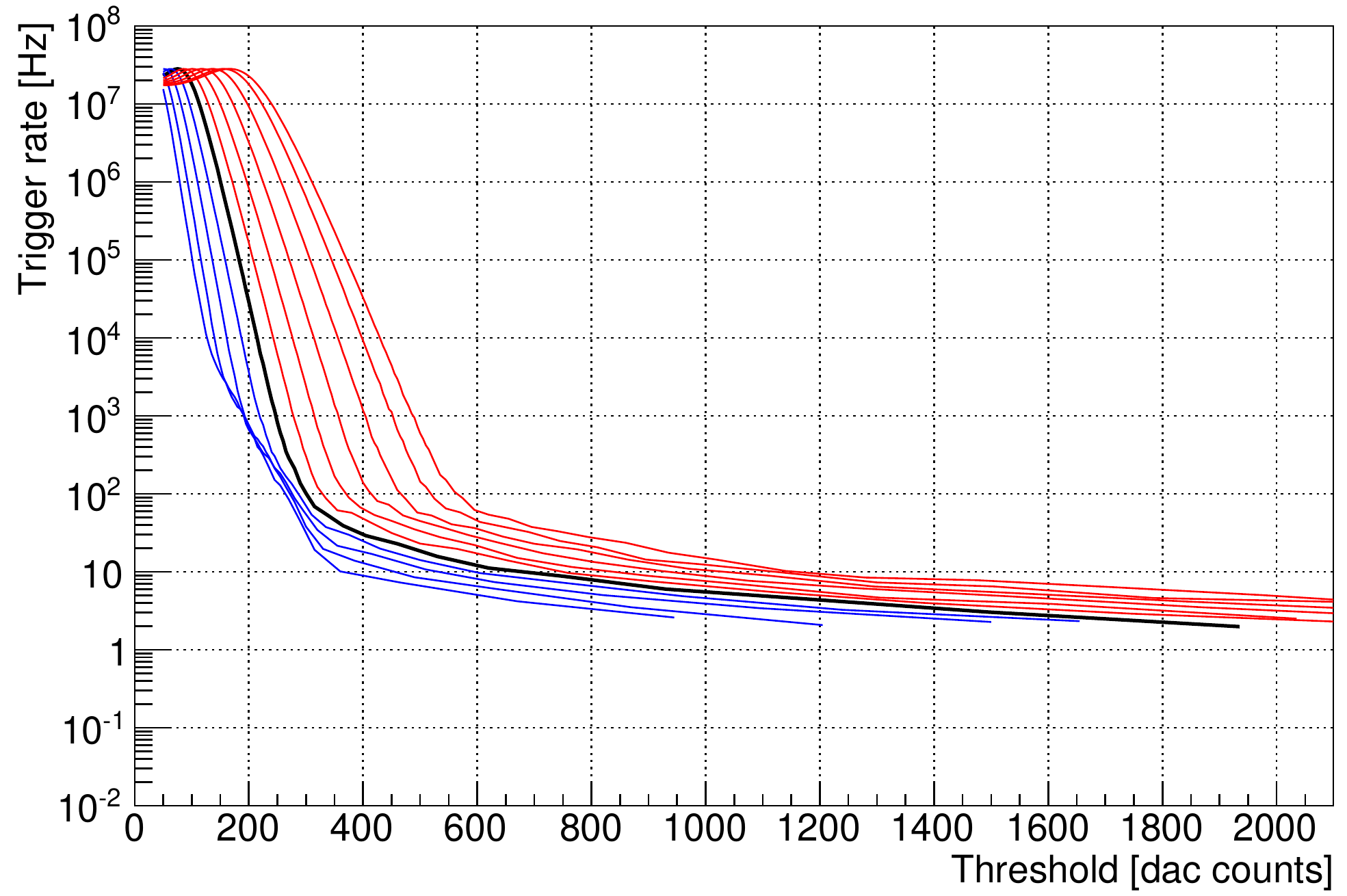}\hfill
 \includegraphics*[width=0.49\textwidth,angle=0,clip]{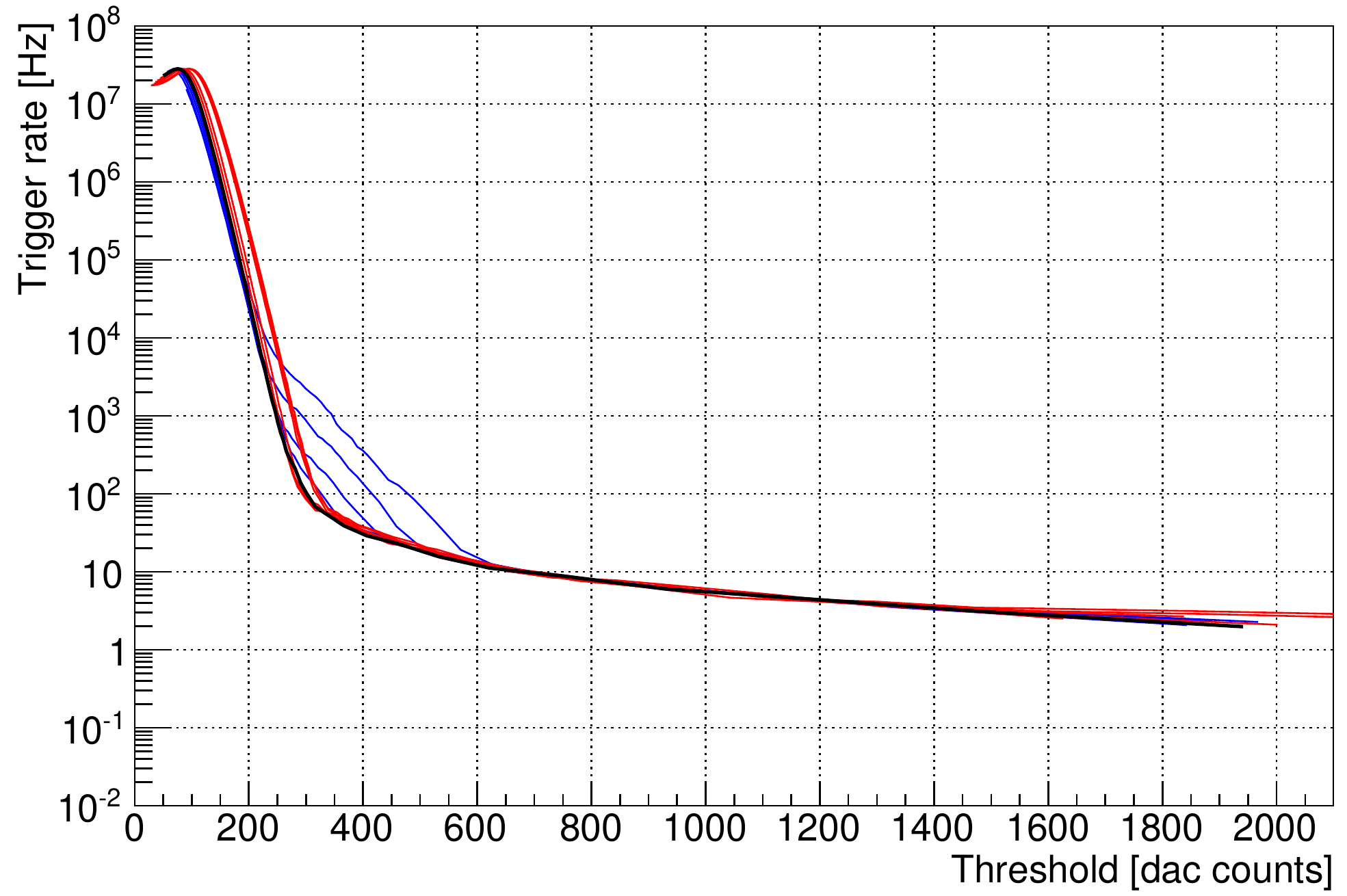}\\
 \includegraphics*[width=0.49\textwidth,angle=0,clip]{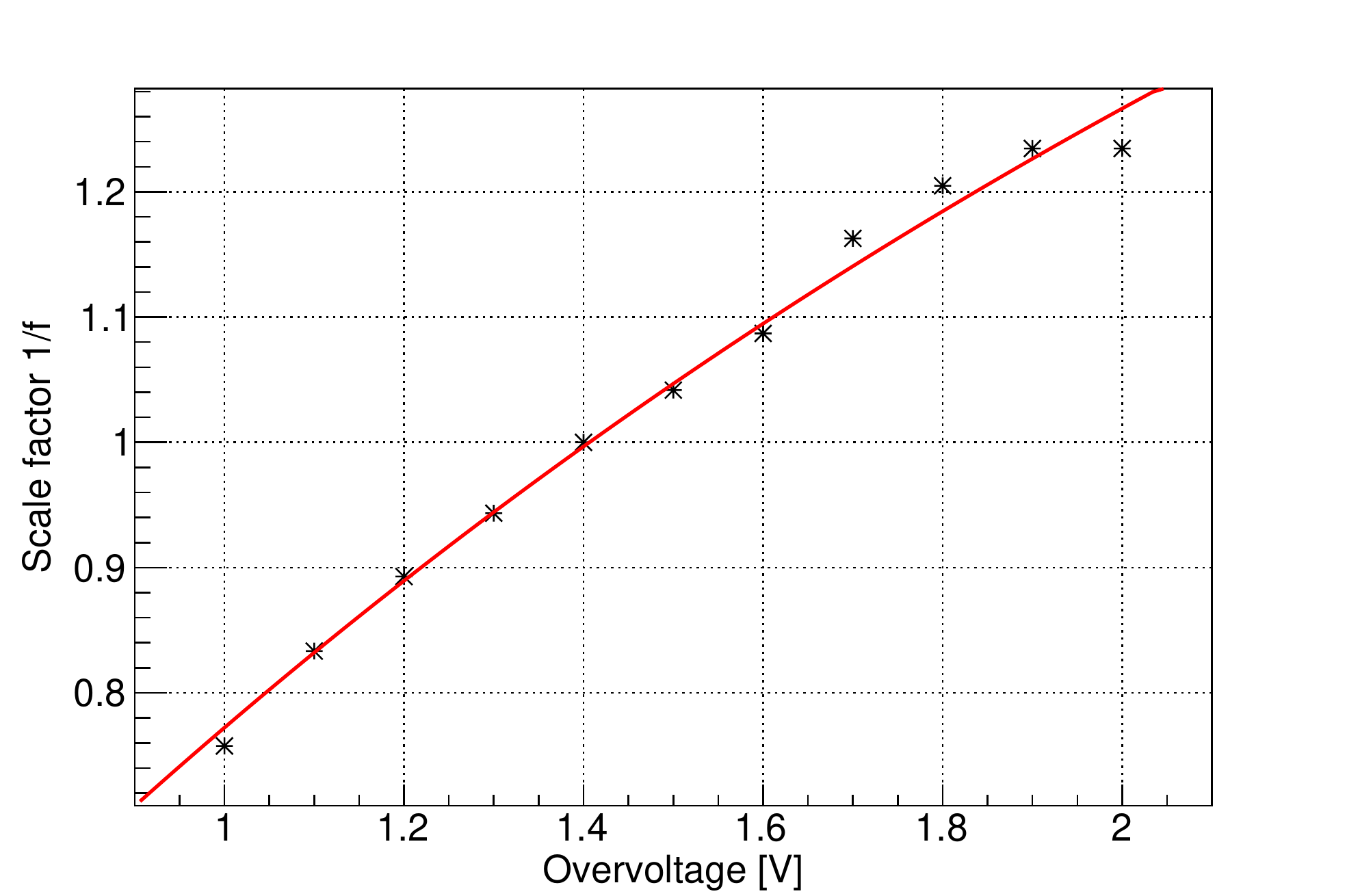}
\caption{Ratescans taken in one night (01/02/2014) during good
atmospheric conditions with different applied voltages. The applied
voltage has an offset from the nominal voltage between -0.4\,V and
0.6\,V in steps of 0.1\,V. Curves with voltages lower than the nominal
voltage are shown in blue, curves with higher voltages in red. The
curve at nominal voltage is shown in black. The left plot shows the
curves as recorded, the right after applying a scale factor. The scale
factor was determined manually for best fit and then fitted (bottom). A
good match of the shower tail is observed. More visible details are
explained in the text.}
\label{fig:ratescans1} 
\end{figure}

To better understand the effect of different gains on the rates, ratescans
with different bias voltage settings were taken. The measurements in
figure~\ref{fig:ratescans1} (left) were recorded during a single dark
night at voltage offsets from the operation voltage between -0.4\,V and
+0.6\,V in steps of 0.1\,V corresponding to an overvoltage between
1.0\,V and 2.0\,V. These ratescans have been taken at zenith distances
smaller than 20\textdegree{}. Each ratescan typically shows three
different regions. At very low thresholds, a saturation of the counter
is visible, followed by a steep slope dominated by triggers induced
from ambient photons. After a transition, a long and flat tail is
visible representing air showers induced from cosmic rays.\\

As the change in voltage implies a change in gain and respectively in
threshold, it should be possible to rescale the threshold
value according to the applied voltage to match one another. To
obtain this match, an additional artificial scale factor \(f\) is necessary,
to correct for changes in photo detection efficiency and crosstalk
probability. Such an artificial factor has been applied manually until
all curves matched the reference curve at nominal voltage. The derived
scale factors are shown in figure~\ref{fig:ratescans1} (bottom). They
have been fitted with \(1/f=(2.15\pm0.13)[1-e^{(0.45\pm0.04)\cdot U}]\)
and the corresponding resulting factors applied. The scaled curves are 
shown in figure~\ref{fig:ratescans1} (right). The cavity visible
around the kink for low voltages originates from noise of the digital
electronics and is related to the readout of the counters and
the switching of the comparators and discriminators.

The good match of the scaled curves proves the stability of the
system. The unscaled curves show that a change in gain would
generate a rescaling of the threshold value appearing as a shift of the
shower tail.


The small gap in the falling edges is due to the change of pointing
direction which induced a small change in ambient light level.
Generally, a small shift towards higher thresholds and a steepening of
the slope is visible with increasing voltage. This is most probably a direct
effect of the increase of photo detection efficiency and crosstalk
probability.

The curves with the highest overvoltages show a saturation at around
2\,Hz for high thresholds. It is assumed that this originates from two
faulty channels. In these channels, the serial resistor is too low and
therefore the voltage at the G-APDs far higher than nominal.
Consequently, they show a significant increase in crosstalk probability
and a finite probability that a single breakdown eventually induces a
discharge of all cells in the sensor. Although the trigger input for
these two channels is disabled, such a high signal leads to electronic
crosstalk in neighboring channels. Under normal data taking conditions,
this can be neglected because the induced random triggers are well
below normal data taking rates. With the additional increase of
voltage, it becomes significant and is therefore visible in the plots
as a saturation at high  thresholds.\\


\begin{figure}[t]
 \centering
 \includegraphics*[width=0.49\textwidth,angle=0,clip]{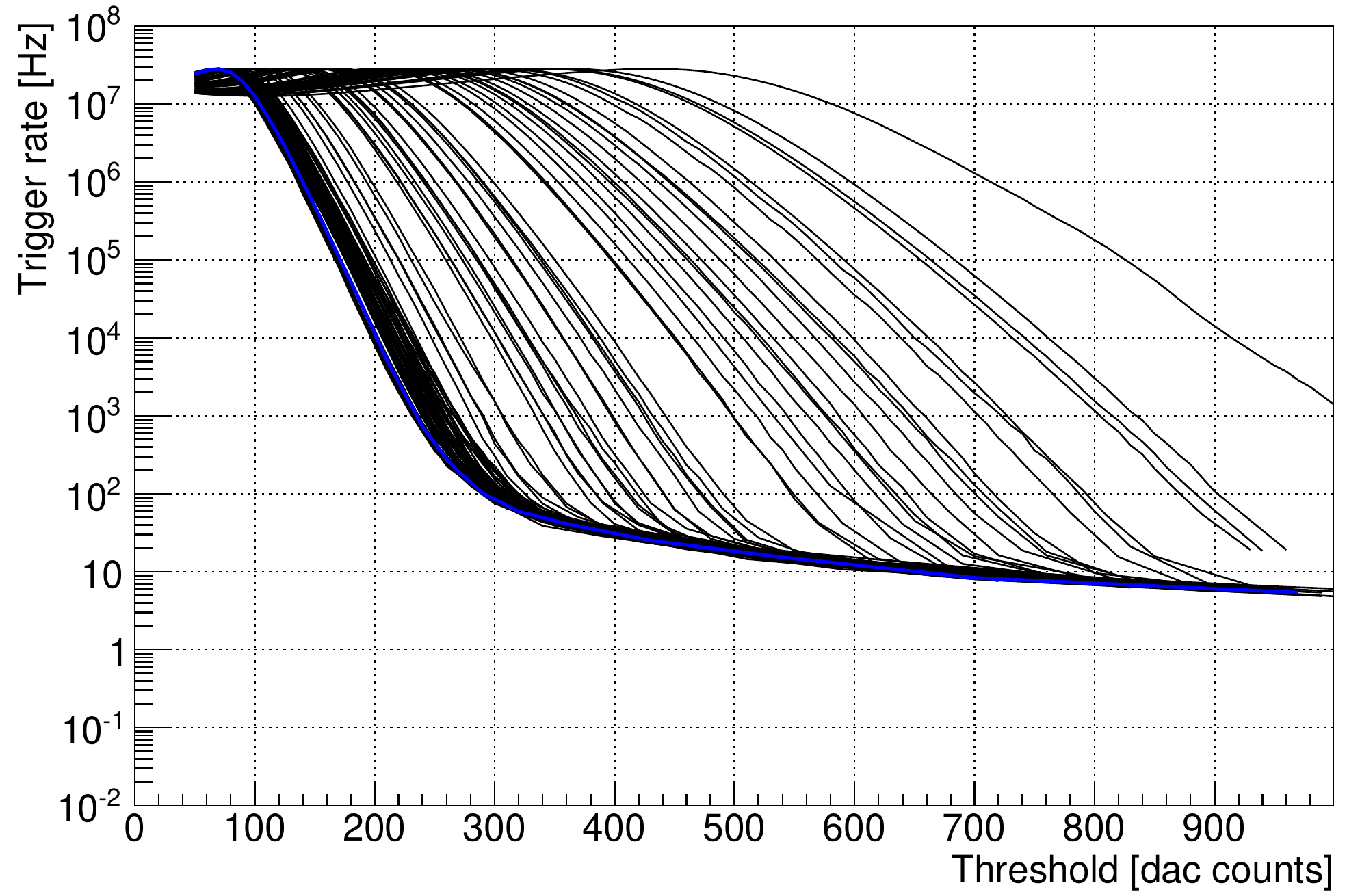}\hfill
 \includegraphics*[width=0.49\textwidth,angle=0,clip]{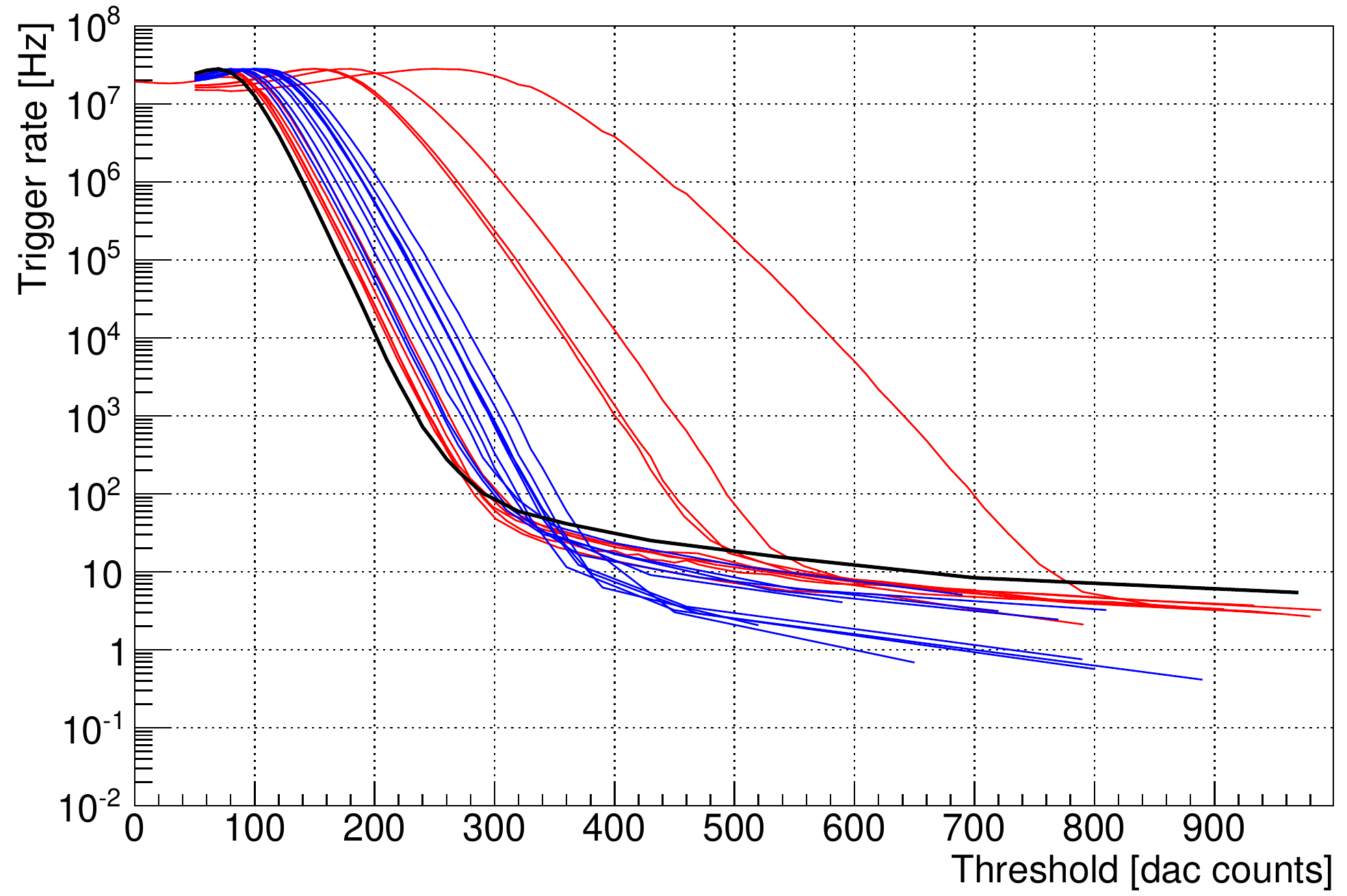}\\
 \includegraphics*[width=0.49\textwidth,angle=0,clip]{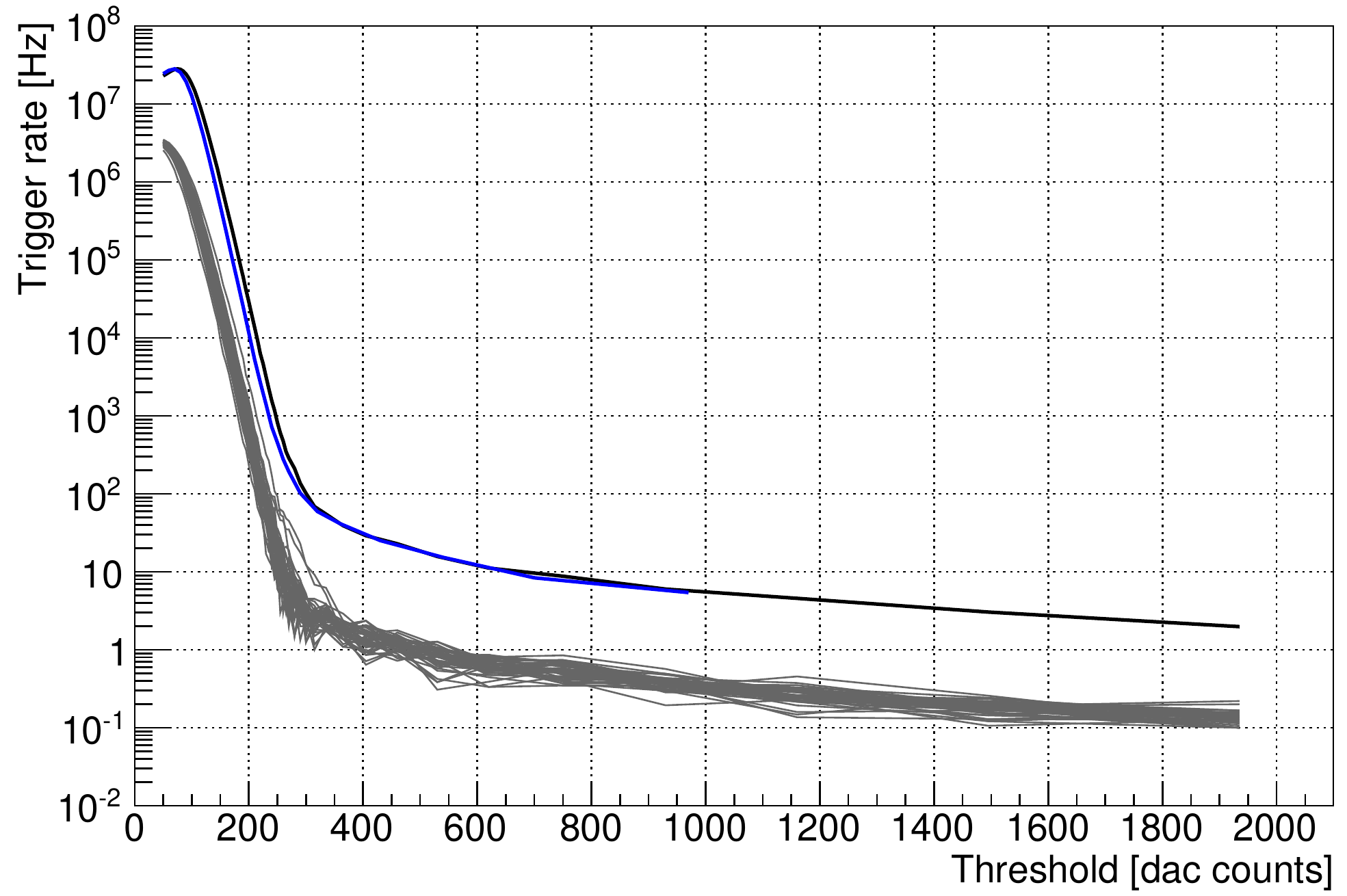}
\caption{Ratescans taken close to zenith at good atmospheric conditions
(left) but at ambient light levels ranging from new moon to almost
full moon. The blue line is a selected reference ratescan. A good
agreement of the shower tail independent of the ambient light level can
be seen. The right plot shows ratescans taken under known poor
atmospheric conditions as clouds (blue) and Calima (red). For
comparison, the reference ratescan is shown  as well (black). The effect
of non ideal conditions is apparent. For easier comparison, the bottom
plot shows two ratescans taken  under comparable conditions at two
different nights: Jan.\ \nth{1} 2014 (blue) and Feb.\ \nth{1} 2014 (black). The gray
curves show the 40 individual board rates. An almost perfect match is
visible. The small difference in the falling edge is due to a slightly
different background light level.
}
\label{fig:ratescans3} 
\end{figure}

To prove a stable shower tail independent of light conditions,
ratescans have been recorded between 08/10/2013 and 21/07/2014 at
different light levels. They are shown in figure~\ref{fig:ratescans3}
(left). The light conditions range from new moon to more than 90\% moon
disk.  A very good agreement of the shower tail of all curves is
visible. At the same time, the changing light levels produce a shift of
the falling edge. They all agree well within their
statistical error which is in the order of 15\% for low count rates.
As an example for the good match, two ratescans
taken in two different nights one month apart are shown (bottom)
including the 40 discriminator rates. 
For comparison, the right plot shows selected rate scans taken under
poor weather conditions such as thin cloud layers (blue) or Calima
(red), a dust layer of Saharan sand in the atmosphere, c.f.~\cite{SAL},
and the reference ratescan (black) from the left plot. Clearly visible
is the strong influence of increased absorption in the atmosphere on
the shower tail. Consequently, comparing the measured shower rate at a
given threshold within the shower tail with the expected shower rate
can reveal valuable information about data quality and for data
analysis, c.f.~\cite{Hildebrand}.

\section{Threshold parametrization}\label{sec:threshold}

To better understand the relation between the kink of the ratescans and
the ambient light level, all ratescans were parametrized and fitted. 
The parameters obtained from the fits were used to determine a relation
between the measured current and the threshold.

\paragraph{Method}

A function of the rate \(R\) versus threshold \(t\) has been
fitted to all ratescans:

\[R(t)/\mbox{Hz} = R_0(t) + R_1(t)\]
with
\[R_0(t) = e^{m\cdot(t-t_0)}\qquad\mbox{and}\qquad R_1(t) = 1.8\cdot 10^9\ t^{-3} + 3.5\]

\begin{figure}[t]
 \centering
 \includegraphics*[width=0.495\textwidth,angle=0,clip,trim=0 0 0.9cm 0.6cm]{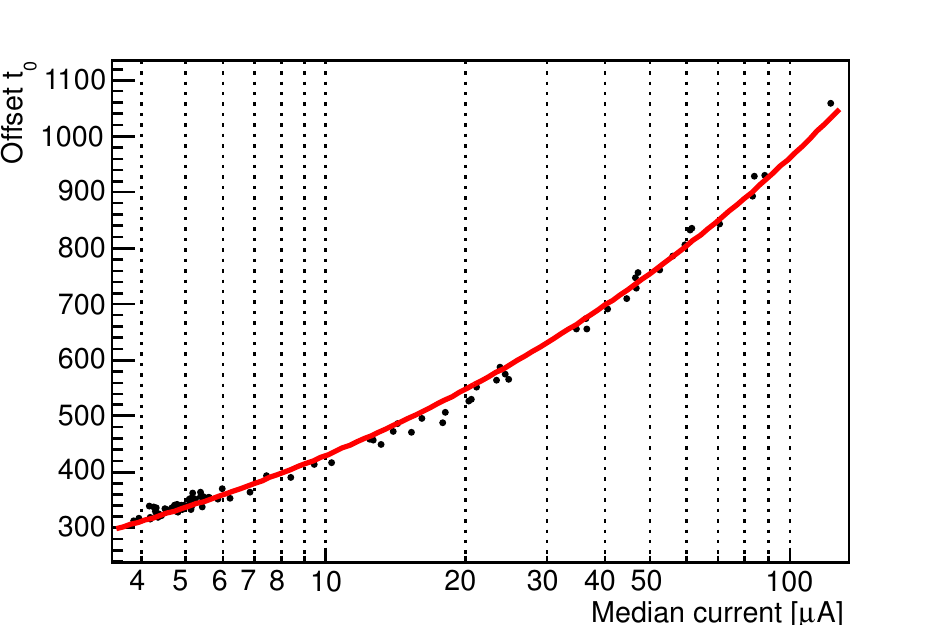}
 \hfill
 \includegraphics*[width=0.495\textwidth,angle=0,clip,trim=0 0 0.9cm 0.6cm]{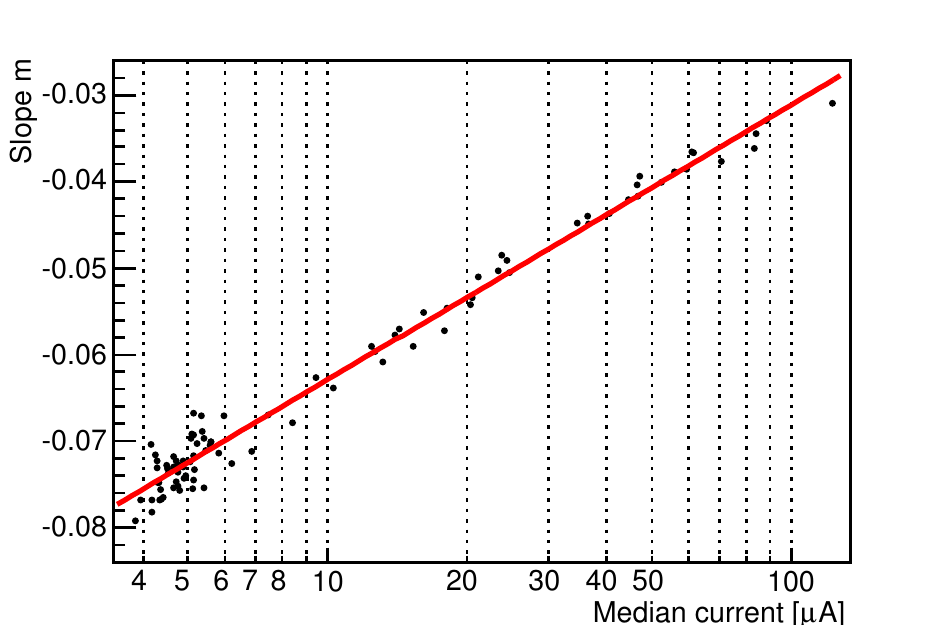}
\caption{Coefficients from the fit to several ratescans at different
ambient light level. Details and the result of the fits (red) can be
found in the text. } 
\label{fig:coefficients} 
\end{figure}

The function \(R_0\) denotes the rate of triggers from
random photons and \(R_1\) the rate of triggers from coincident photons,
or ambient light and showers, resp. The function \(R_1\) was fit
independently for \(t\)\,>\,350 on the reference ratescan shown in
figure~\ref{fig:ratescans3}. The starting point of all fits is chosen such that only the 
mainly exponential part of the falling edge just before the kink is
considered. 


\paragraph{Result}

The resulting coefficients \(m\) and \(t_0\) are shown in
figure~\ref{fig:coefficients} as a function of the median current
of all bias voltage channels. Fits to the data yield

\[m = (-0.0947\pm0.0005) + (0.0318\pm 0.0005)\cdot\mbox{ln}\left(\frac{I}{\mu\mbox{A}}\right)\]

and

\[t_0 = (192.0\pm 1.6)\left(\frac{I}{\mu\mbox{A}}\right)^{0.3500\pm0.0024}\,.\]

\begin{figure}[t]
 \centering
 \includegraphics*[width=0.49\textwidth,angle=0,clip,trim=0 0 0.9cm 0.6cm]{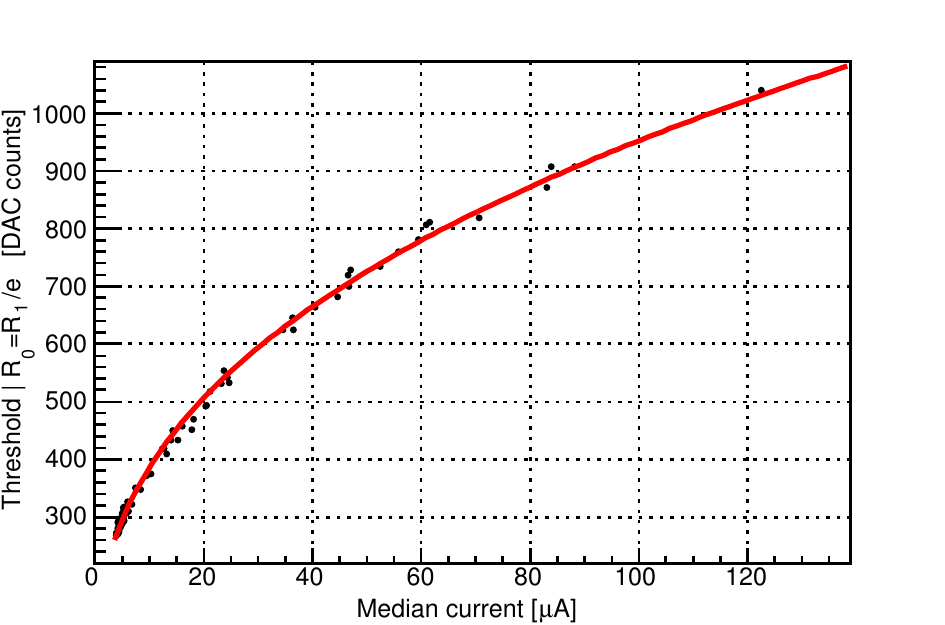}
\caption{From the fits to all ratescans, the point at which the
trigger rate from background photons falls below \(1/e\) of the
rate from the shower tail is determined. The result for the corresponding thresholds
is shown versus the median current in the camera. Details
on the function which was fit to the data (red) can be found in the text.}
\label{fig:prediction}
\end{figure}

For each fit, the threshold at which the
condition \(R_1=R_0/e\) is fulfilled is determined. The result
is shown in figure~\ref{fig:prediction}. A fit to the data provides
the following relation for the threshold \(t\):

\[t = (156.3\pm 1.2)\left(\frac{I}{\mu\mbox{A}}\right)^{0.3925\pm0.0022}\]

This result does not describe the attenuation of the shower tail by changing
atmospheric conditions or with increasing zenith distance. Both effects are
neglected in view of the very steep edge induced by triggers from
background photons.

\paragraph{Application}

For more than a year, the obtained relation between threshold and
current is used during data taking to set the trigger threshold of all
comparators. Just before a run is started, the median current is
determined and the threshold level set accordingly. To avoid a bias on
the analysis due to changing thresholds, the threshold level is kept
constant during runs. Only single channels with high rates due to direct star light 
are continuously regulated. Due to the fast changing ambient light level
during twilight, runs with only one minute instead of the
usual five minutes are taken. Under normal circumstances, all other
fluctuations during the night are slow enough to not result in too high
rates within this five minute interval. In rare cases of sudden rise of
brightness, like overclouding or direct light from  cars passing by on
the nearby road, it can happen that the data acquisition saturates, but
in these cases data are not suited for further analysis anyway. 

The application of the derived threshold has lead to very
stable rates and observation conditions. Currently, exceptionally high
rates from single patches, like bright stars, are still suppressed with
an algorithm based on the measured patch  rates, but efforts are
ongoing to base that on their individual currents as well. The lowest
threshold achieved during clear new moon nights is on the order of 300
counts which corresponds to roughly 20\,\pe per trigger patch or
\(\sim\)2.2\,\pe per pixel.



\section{Current prediction}\label{sec:prediction}

The measured currents in the camera are directly correlated with the
light flux detected by the sensors. Consequently, the currents are an
ideal measure for the sky brightness. As shown in the previous section~\ref{sec:threshold},
they directly define the trigger threshold which itself is 
linked with the energy threshold of each observation. If it is possible
to estimate the expected currents from environmental conditions, the
comparison with the measured currents can directly be used as a quality
monitor for the data and as weather monitor. Being able to predict the
energy threshold of each observation in advance also allows for a
further improved scheduling optimizing sensitivity on all
targets.

\paragraph{Method}

The main influence on the sky brightness are moon and sun properties.
Effects which can easily be included, because they are predictable, are:
moon phase, moon and sun position as well as pointing direction.
Effects which are difficult to include, because they are difficult to
predict or unpredictable, are: zodiacal light, backlighting, Albedo, or
the different increase of light towards the horizon depending on
weather conditions. Especially, during sunrise and sunset atmospheric
scattering can significantly increase the light yield.








To derive a prediction of the median current in the camera, all physics
triggered data taken between 01/09/2013 and 22/07/2014 have been used.
Only  a single night was excluded due to snow on the ground which
significantly increased the background light yield. From these data,
relations from the measured current have been derived. Generally, all
effects can be described by physical formulas. The disadvantage of such
an approach is that instrumental effects like the angular acceptance of
the light guides are not included and need additional components. It is
unclear, if an easy formula can be derived this way. Formulas to
describe the brightness of moonlight have been suggested in literature,
e.g.~\cite{Krisciunas}, but describe the measured current less accurate
than the presented empirical model, c.f.~\cite{MaxICRC}. To derive such
an empirical model, the residual between the predicted current and the
measured current has been used versus the Sun's zenith distance
\(\Theta\), the illuminated fraction \(f\) of the moon disk, the Moon's
altitude \(\alpha\), the Moon's distance \(d\) from the earth relative
to its semi-major axis \(d_0\)\,=\,384,400\,km, and the angular
separation \(\delta\) between the moon position and the pointing
direction. The following fit was achieved:

\newcommand{\ma}{\,\mu A}

\[I_{est}/\ma = 5.7 + 95.8 \cdot 
\frac{f^{2.73}\cdot \sin^{0.70}\alpha}{(\frac{d}{d_0})^2} 
e^{0.77\cos^4\delta} + 
e^{-97.9+105.9\sin^2\Theta}\]



\begin{figure}[t]
 \centering
 \includegraphics*[width=0.49\textwidth,angle=0,clip]{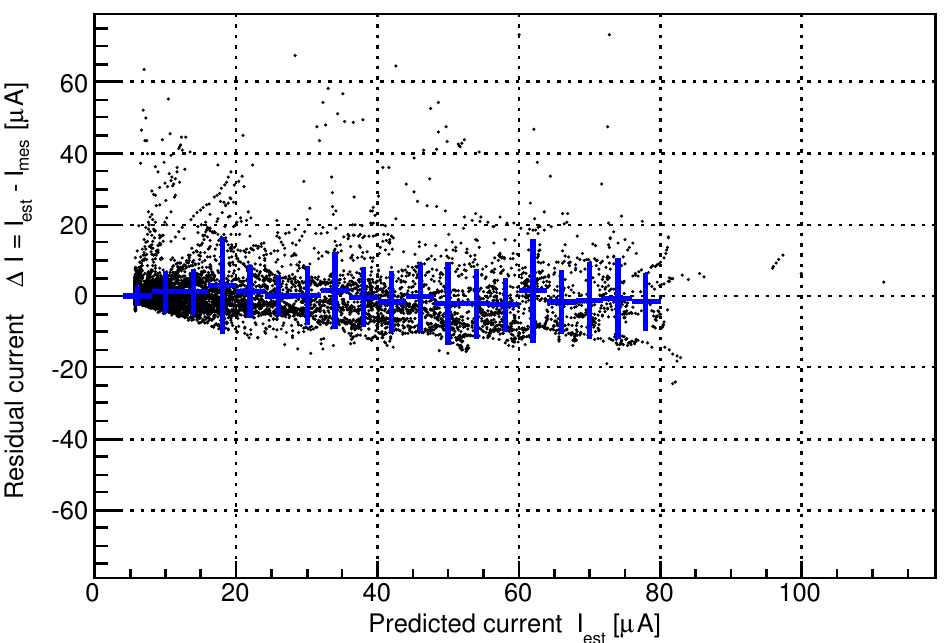}\hfill
 \includegraphics*[width=0.49\textwidth,angle=0,clip]{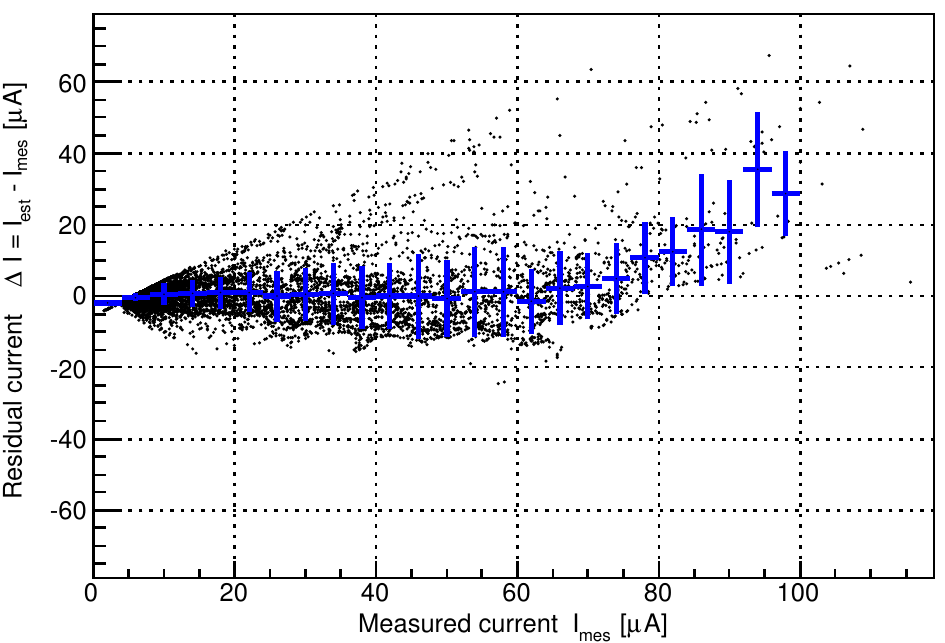}\\[1ex]
 \includegraphics*[width=0.49\textwidth,angle=0,clip]{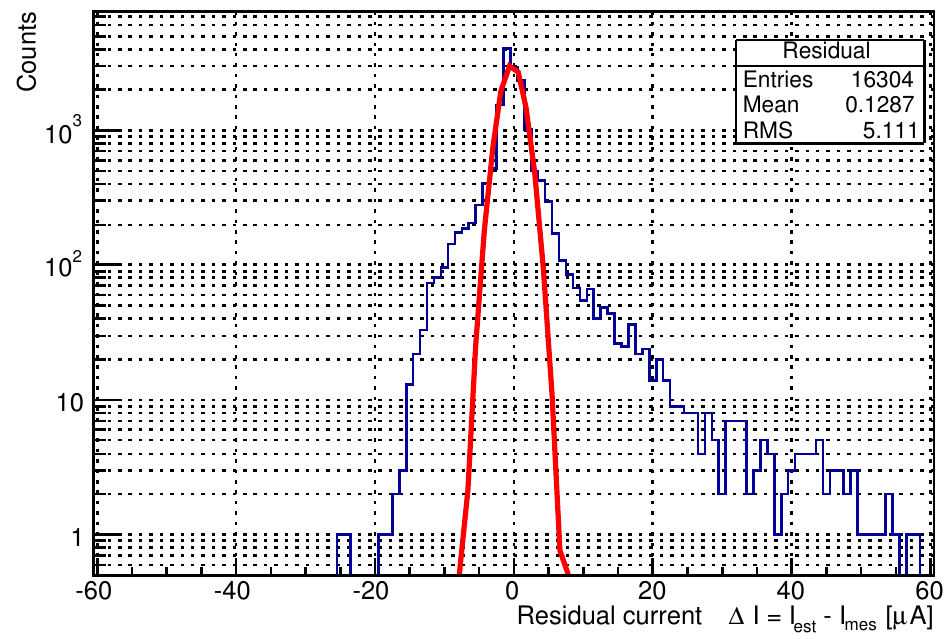}
\caption{All three plots show the residual between the predicted
current and the measured current. The left plot shows the distribution.
The majority of the data is well predictable, presumably the data with
good quality. The tails on both sides correspond to non ideal
observations conditions. Enhanced current is due to scattering and reflection,
while reduced currents are due to attenuation. The fit of a
Gaussian yields a sigma of \(\sim\)1.6\,\(\mu\)A. The central plot
shows the residual as a function of the predicted current overlaid with
a profile (blue) which error bars denote the standard deviation of the
distribution. No significant structure is visible. The right plot is
compiled as the central plot but shows the residual versus the measured
current. An overestimation of high currents is visible.} 
\label{fig:prediction-residual} 
\end{figure}

\paragraph{Result}
In figure~\ref{fig:prediction-residual}, the distribution of residuals
between the resulting prediction and the measured currents are shown.
While the majority of the data shows a residual of less than
\(\pm\)5\,\(\mu\)A, higher residuals are visible originating from
attenuation, increased scattering or reflections. For high currents,
the prediction is usually overestimating the measured current which is 
an effect of the exponential increase of light when the Sun is just below
the horizon. 

While in the above formula some dependencies are easy to fit, such as
the dependency of the moon brightness, others like the dependency on sun
brightness and angular separation are difficult because of very limited
statistics. An additional complication is that in case of the Sun,
already small changes in atmospheric scattering imply significant
changes in absolute light flux. Due to the exponential behavior,
this limits the quality of a simple prediction. This can be
verified with the data which shows a significant increase of
background light towards sunset and sunrise even during dark
time, so called zodiacal light.





\begin{figure}[t]
 \centering
 \includegraphics*[width=\textwidth,angle=0,clip,trim=0 9.7cm 0 2.3cm]{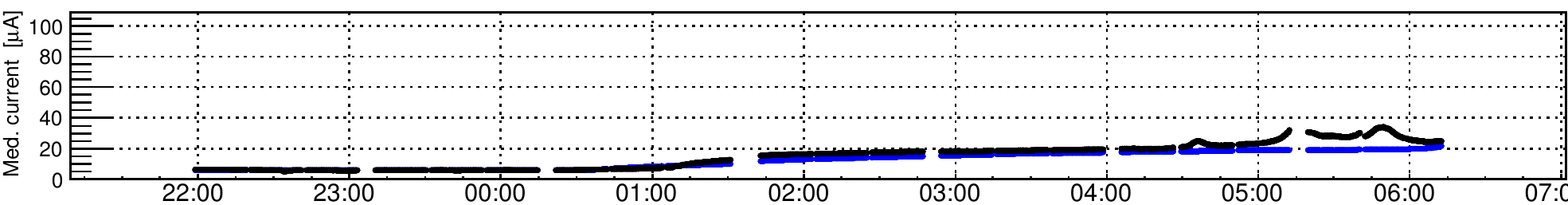}\\
 \includegraphics*[width=\textwidth,angle=0,clip,trim=0 9.7cm 0 2.3cm]{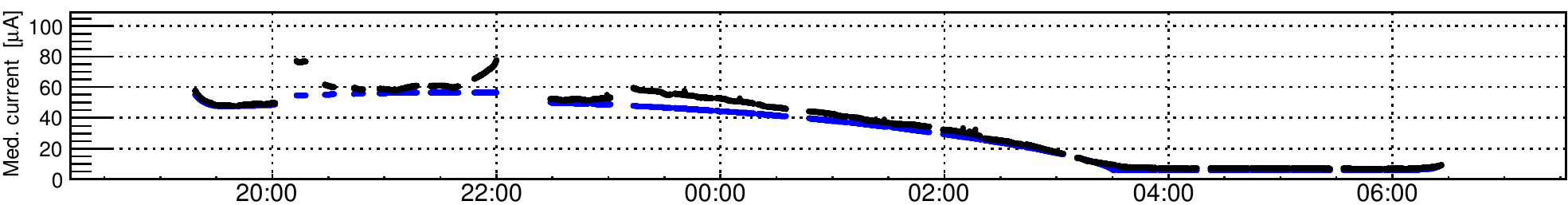}\\
 \includegraphics*[width=\textwidth,angle=0,clip,trim=0 9.7cm 0 2.3cm]{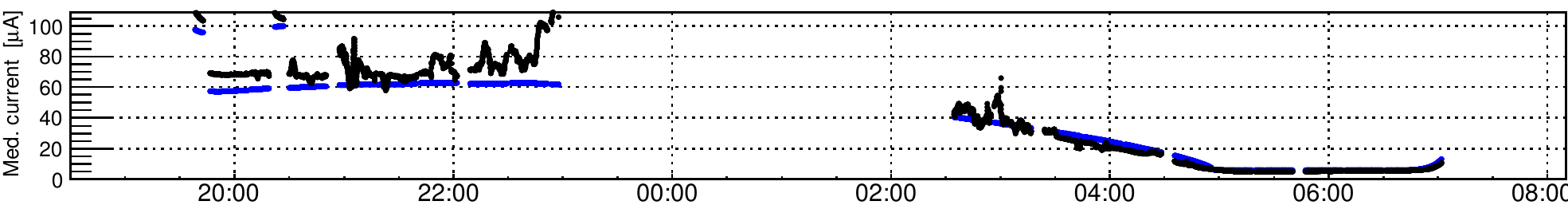}\\
 \includegraphics*[width=\textwidth,angle=0,clip,trim=0 9.7cm 0 2.3cm]{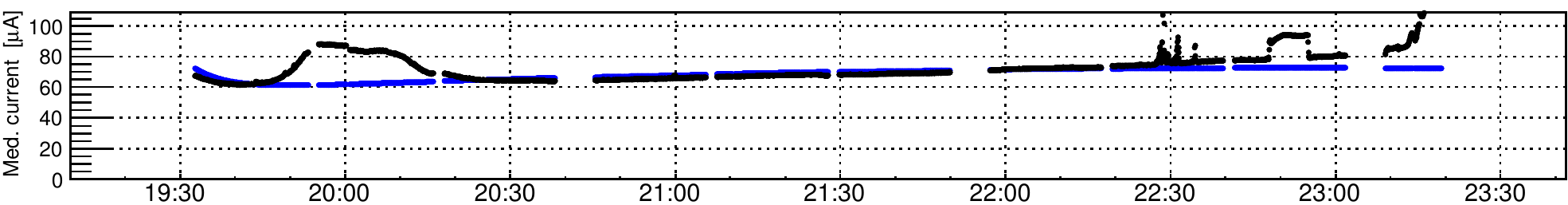}\\
 \includegraphics*[width=\textwidth,angle=0,clip,trim=0 9.7cm 0 2.3cm]{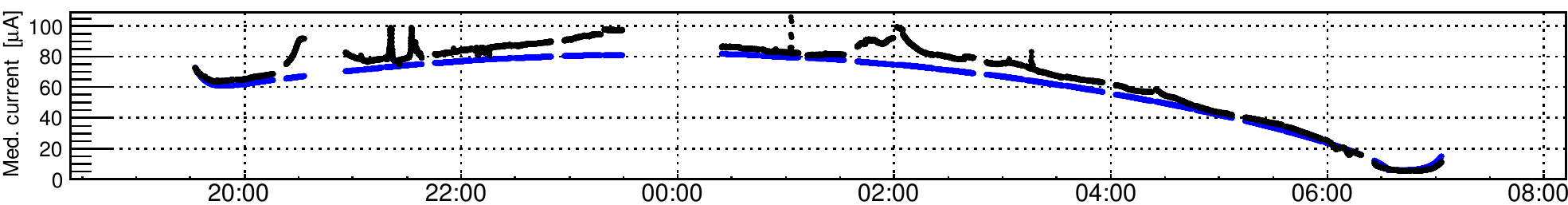}
\caption{The five plots show the average median current (black) per
run (5\,min) versus time (UTC) compared to the corresponding current
prediction (blue) for the nights 26/10/2013, 12/11/2013, 11/01/2014,
12/01/2014 and 13/01/2014 (top to bottom) with a maximum illuminated fraction of the
moon disk of roughly 47\%, 73\%, 84\%, 91\% and 96\% respectively. 
Poor weather conditions like high  humidity and clouds can be seen as
short time variation. Smooth variations on longer time scales are
assumed to be an effect from reflections. Generally, a good agreement
is visible. }
\label{fig:predictions}
\end{figure}


In figure~\ref{fig:predictions}, some examples are shown, comparing 
measured and predicted current. The measured current is the median
over the camera. The maximum illuminated fraction of the Moon
ranges from 47\% to 96\%. The Moon culminates at a zenith distance of
\(\sim\)10\textdegree. Pointing directions are between
10\textdegree{} and 70\textdegree{} zenith distance. The smallest
angular distance to the Moon is around 10\textdegree{}.

It is apparent that unexpected events are easy to recognize. While
static reflections moving into or out from the field-of-view 
usually introduce a slow change, reflections from clouds introduce
a noise like shape. In other cases, clouds or additional attenuation in
the atmosphere show a current drop compared to the prediction.
Evaluating this not only visually but also with statistical 
methods are an ideal tool for data quality checks needed for
a stable long term monitoring.

Furthermore, the predicted current is directly related to the energy
threshold which can be derived from the trigger threshols,
c.f.~\cite{ThresholdPrediction}. This can be used to optimize
observation schedule.

\section{Conclusions and Outlook}


The results presented in this article prove that the FACT camera can
be operated under low light conditions with a gain variation of its
sensors better than 0.5\% on average while providing a homogenous
response over the camera of better than 2.5\% at the same time. The
achieved homogeneity is close to the limit provided by the power 
supply. This result was derived from measurements of the dark count
spectra of the sensors. A special achievement is that for the data
analysis, a distribution function, the modified Erlang distribution,
was introduced which fits the dark count spectrum even at high
multiplicities  extremely well. A simple Monte Carlo simulation of the
dark count spectrum shows matching results allowing for a proper
detector simulation in the data analysis chain. Investigations of the
temperature and voltage dependence of the obtained parameters gave a
very valuable overview of the behavior of the sensors under varying
conditions. Measurements with an external light pulser proved that this
stability is maintained on a few percent level up to the brightest
light conditions.

It was demonstrated that measurements during bright moon light (see
also~\cite{Hildebrand,MaxICRC,ThresholdPrediction}) are possible under
stable trigger conditions.

For these successes, no external calibration device was needed.
Even for monitoring purpose the existing light pulser can be omitted
because a decrease of sensor properties taking place only at bright
light conditions, is not very likely and can therefore be monitored
with the dark count spectra. It turned out that a very detailed
characterization of the sensors in a laboratory is not necessary and
can be done measuring their dark count spectra under varying
conditions in the field.


The excellent stability of the gain in space and time also demonstrates
that for the existing system the precision of the gain is ultimately
limited only by the resolution of the voltage setting. It can be
assumed that a more precise voltage supply and knowledge of the voltage
applied to the sensor itself can further improve the stability. On the other hand, the
presented results show that a power supply which needs to achieve a
precision better than 0.5\textperthousand{} at voltages above 70\,V
needs a very careful design or detailed characterization especially to
avoid any temperature gradients. When several sensors share the same
support voltage, it is essential that they are carefully selected in
advance, even more if thousands are used. In any case a power supply must
be designed to provide currents in the order of a hundred to several
hundred \(\mu\)A per channel if the sensors should be operated during
bright light conditions.

In the new generation of G-APDs appearing on the market, most of their
features are improved. A new material for the internal quenching
resistor reduces the temperature gradient which lowers the requirement
for the precision of the temperature measurement and voltage setting.
Higher purity of the silicon facilitates a reduction of dark counts and
afterpulses. The application of trenches between G-APD cells allows to
order sensors with reduced optical crosstalk with the drawback of a
reduced active area. Allowing for the same level of crosstalk as in
recent sensors, the applied voltage can be increased operating the
sensor in a regime in which photo-detection efficiency shows a weaker
dependence on temperature and is close to saturation. Through-Silicon
Via technology allows to tile several sensors together with almost no
gap in between, but also for the price of a small decrease in active
area.


This study has impressively proven that even sensors which have been on
the market already for several years can be used with high precision
for photo detection in cameras even when operated under varying
environmental conditions. Regular operations during bright moon light
without any hint for decreasing performance since assembly have proven
the durability of G-APDs. Although, several hardware failures happened
during 2013, as a leaking pump, a blown professional power supply and a
broken lid actuator, not a single problem was related to the sensors
themselves.

Future projects will strongly benefit from this experience. Already
today, several sub projects within the Cherenkov Telescope Array (CTA)
project are developing new G-APD based cameras. For a project which is going to
operate about one hundred telescopes, especially their
stability will lower maintanance costs. The discussed improvements in
technology will simplify their application further and at the same
improve the achieved performance and consequently simpify data analysis.
The very stable operation facilitates an analysis threshold closer to
the trigger.

\newpage

\appendix 
\section{The distribution function for the dark count spectrum}\label{sec:appendix}

\subsection{The distribution function}

To fit the dark count spectrum measured from a G-APD sensor, a proper
distribution function has to be used. The distribution function describes
the probability to measure a charge which corresponds to \(N\) times
the charge released by a single breakdown as a function of \(N\). In
the following, several distribution functions which come into question
are presented and discussed. Some of them have been rearranged to
emphasize their common basis. 

\begin{figure}[tb]
\includegraphics[width=\textwidth]{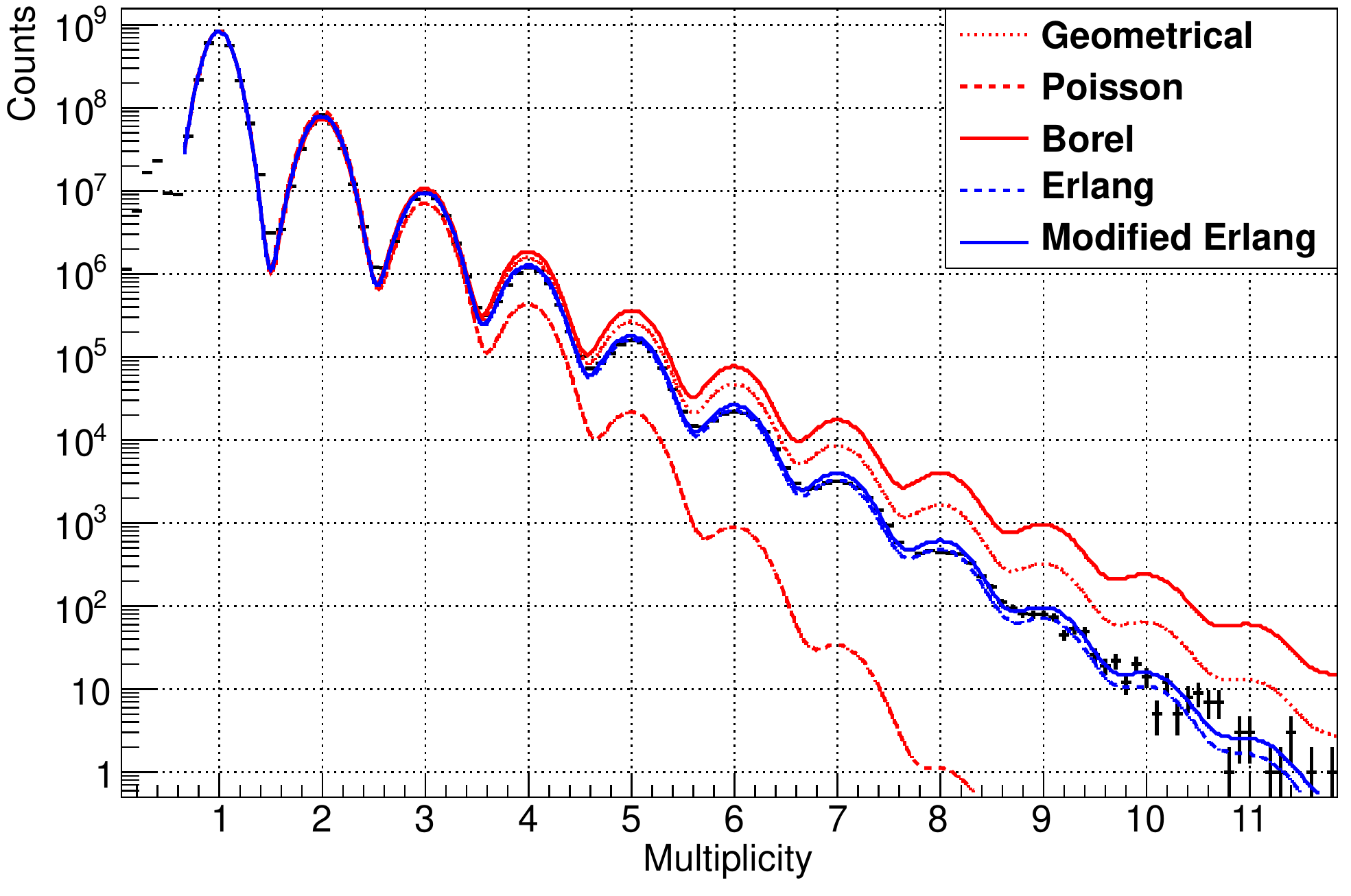}
\caption{A measured single-\pe spectrum overlayed with several
different distribution function fitted to the data. The pure Poisson
distribution significantly underestimates the data, while the Borel
distribution clearly overestimates the data. The geometrical
distribution by chance gets very close to the data. A good fit is
obtained by the Erlang and modified Erlang distribution.}
\label{fig:example}
\end{figure}

In figure~\ref{fig:example}, example fits to existing data for the
distributions discussed in the following are shown.

\paragraph{Geometric distribution}

The most simple distribution function is the geometric distribution. It
describes the probability to measure \(N\) breakdowns in total, if
every breakdown introduces a maximum of one additional breakdown, each
with a probability \(p\) 

\[P(N) = p^N (1-p) \propto p^{N-1} \]
\[\Rightarrow \frac{P(N+1)}{P(N)} = p \]

\paragraph{Poisson distribution}

In reality, every avalanche emits several photons which can initiate an
additional number \(p\) of breakdowns on average. Such a process is
described by a Poisson distribution.

\[P(N) = \frac{p^N}{N!} e^{-p} \propto \frac{p^{N-1}}{N!}\]
\[\Rightarrow \frac{P(N+1)}{P(N)} = \frac{p}{N} \]

\paragraph{Conway-Maxwell-Poisson distribution (CMP)}
A generalization of the geometrical and the Poisson distribution is the
Conway-Maxwell-Poisson distribution which has the Poisson distribution
and the Geometric distribution as special cases. An additional
parameter \(\nu\) describes a decrease in the probability for
successive breakdowns.

\[P(N) \propto \frac{p^{N-1}}{(N!)^\nu} \]
\[\Rightarrow \frac{P(N+1)}{P(N)} = \frac{p}{N^\nu} \]

\paragraph{Borel distribution}
In all previous cases, the fact that every secondary avalanche can
induce further avalanches is neglected. The absence of branching in the
process results in a significant underestimation of the
probability for high multiplicities \(N\). Including the possibility of
branching yields the so-called Borel distribution. It describes the
probability to have exactly \(N\) breakdowns if every induced
avalanche induces a number of independent additional breakdowns and
each of these processes is Poisson distributed. If the process is
started with more than a single synchronous breakdown, the Borel
distribution can be extended to the Borel-Tanner distribution, which is
mentioned here only for completeness.

\[P(N) = \frac{1}{N}\frac{(pN)^{N-1}}{(N-1)!}e^{-pN}
\propto\frac{1}{N}\frac{(qN)^{N-1}}{(N-1)!} \qquad\mbox{with}\qquad q=pe^{-p}\]
\[\Rightarrow \frac{P(N+1)}{P(N)} = q \left(\frac{N+1}{N}\right)^{N-1} \]

For high multiplicities \(N\) this ratio converges to \(qe\).\\


For sensors, in which each generation is nearly independent from
the previous generation, this
description is already fully sufficient as shown in~\cite{Vinogradov},
although here, a mismatch at higher orders is visible.
In the case of the sensors applied in the FACT camera, it turns out
that the Borel distribution overestimates the probability for high
multiplicities. Empirically, it has been found that by adding an arbitrary
factor \(N\), a much  better agreement between the probability function
and the measured data is obtained.

For geometrical reasons, in general, consecutive generation are not
indepent and the number of newly triggered cells is decreasing on
average. In every sensor, each cell has only a limited number of
neighboring cells which can be triggered and in addition, each cell can
be triggered only once. Therefore, it is not expected that the Borel
distribution is matching generally.


The emperically obtained distribution is known in the literature as
Erlang distribution.

\paragraph{Erlang distribution}

The Erlang distribution describes the probability to have exactly \(N\)
breakdowns in case each breakdown produces \(k\) independent breakdowns,
each exponentially distributed.

\[P(N,k) = \frac{p^kN^{k-1}}{(k-1)!} e^{-pN} \]

for \(k=N\) this yields

\[P(N) \propto \frac{(qN)^{N-1}}{(N-1)!}  \qquad\mbox{with}\qquad q=pe^{-p} \]
\[\Rightarrow \frac{P(N+1)}{P(N)} = q \left(\frac{N+1}{N}\right)^N \]

In the Erlang distribution the ratio between two consecutive
probabilities is larger by a factor proportional to \(1+1/N\) as compared
to the Borel distribution. For high multiplicities \(N\), this ratio
also converges towards \(qe\).

All mentioned distributions describe either a single process, or an
infinite chain of processes. Due to the fact that the number of cells
in a sensor which can discharge is limited, it turns out that a
minor correction for a proper description for multiplicities
above \(N\approx 7\) is necessary.

\paragraph{The modified Erlang distribution}

In analogy to the Conway-Maxwell-Poisson distribution a modified Erlang
distribution is introduced, which allows a change of probability with
an increasing number of breakdowns. A possible explanation for this 
small correction is the limited size of the sensor or different
crosstalk probabilities of different cells, \cf~\cite{Eckert,Otte}.

\[P(N) \propto \frac{(qN)^{N-1}}{[(N-1)!]^\nu}  \qquad\mbox{with}\qquad q=pe^{-p}\label{eq:Erlang}\]
\[\Rightarrow \frac{P(N+1)}{P(N)} = \frac{q}{N^{\nu-1}} \left(\frac{N+1}{N}\right)^N \]

For \(\nu=1\) this distribution transforms into the standard Erlang distribution.

\paragraph{Approximation}

For an easy comparison, the Sterling approximation \(N!\approx
\sqrt{2\pi N} (N/e)^N\) can be applied. For the modified Erlang
distribution this yields

\[P(N) \propto \frac{q^{N-1}}{ N^{0.5\nu + (N-1)(1-\nu)}}
\qquad\mbox{with}\qquad q=pe^{-p\nu} \]

For \(\nu\) very close to unity, the function can be further simplified into

\[P(N) \propto \frac{q^{N-1}}{\left.\sqrt{N}\right.^\nu}\]
\[ \frac{P(N+1)}{P(N)} = q \left.\sqrt{\frac{N}{N+1}}\right.^\nu\]

The similarity with the Conway-Maxwell-Poisson distribution is
immediately apparent, which fits the dark count spectrum equally well
with \(\nu\approx 0.5\).

\paragraph{Discussion}
In the limit of high multiplicities \(N\), the ratio between two
consecutive peaks converges to \(qe\) for the Borel distribution and
the Erlang distribution. At the same time, the ratio of the first two
peaks in the Borel distribution is \(q\) and in the Erlang distribution
\(2q\). While in the case of the Borel distribution the deviation from
a geometric distribution is obvious, the change in slope for
consecutive multiplicities \(N\) and \(N+1\) decreases from 10\% for
\(N\)\,=\,1 to only 3\% for \(N\)\,=\,3. Therefore, the resulting
distribution can easily be confused with the geometrical distribution
if the measurement is not sufficiently sensitive,
c.f.~figure~\ref{fig:example}.

%
%
%
%
%

\subsection{Simulation}

\begin{figure}[tb]
\includegraphics[width=.49\textwidth]{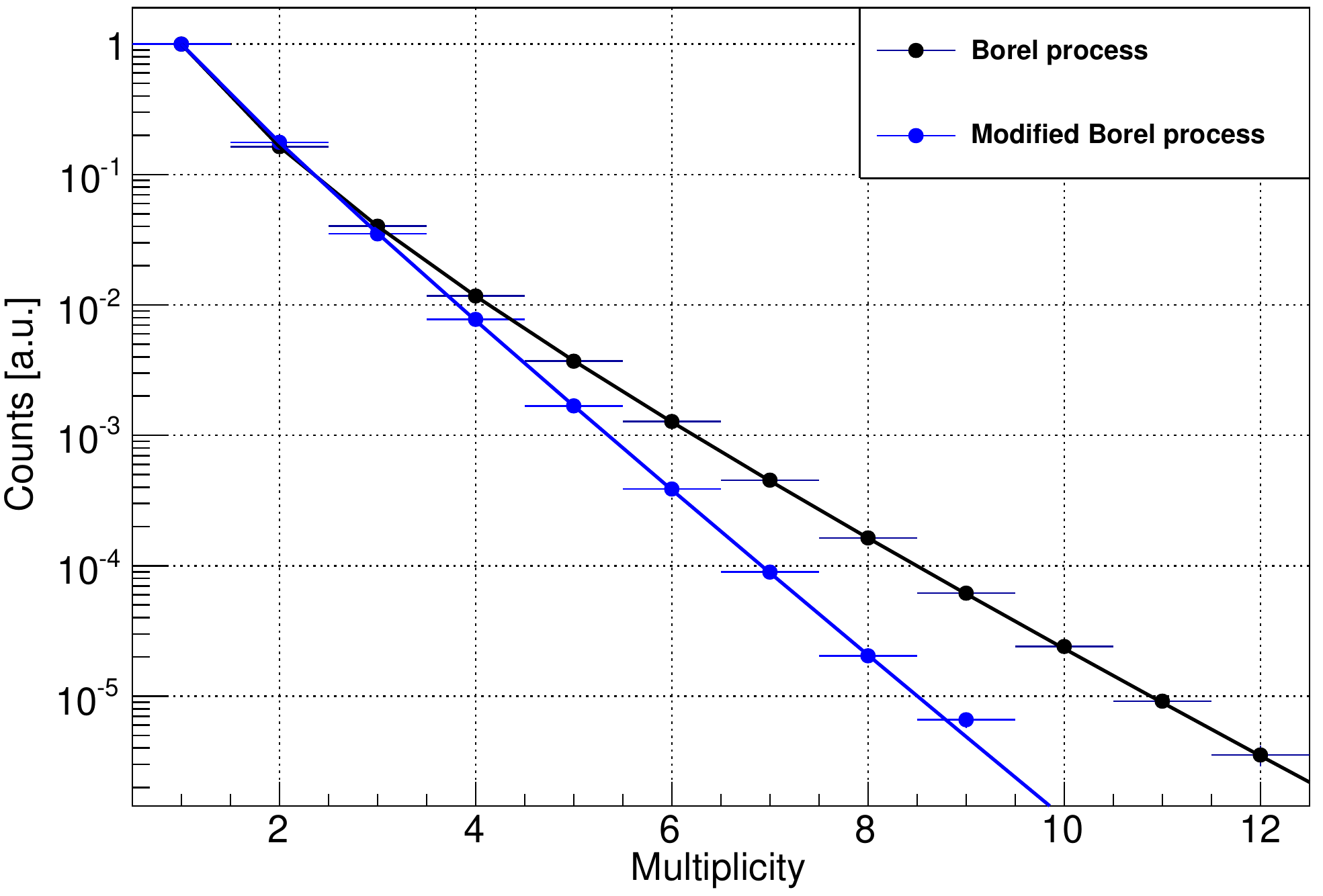}\hfill
\includegraphics[width=.49\textwidth]{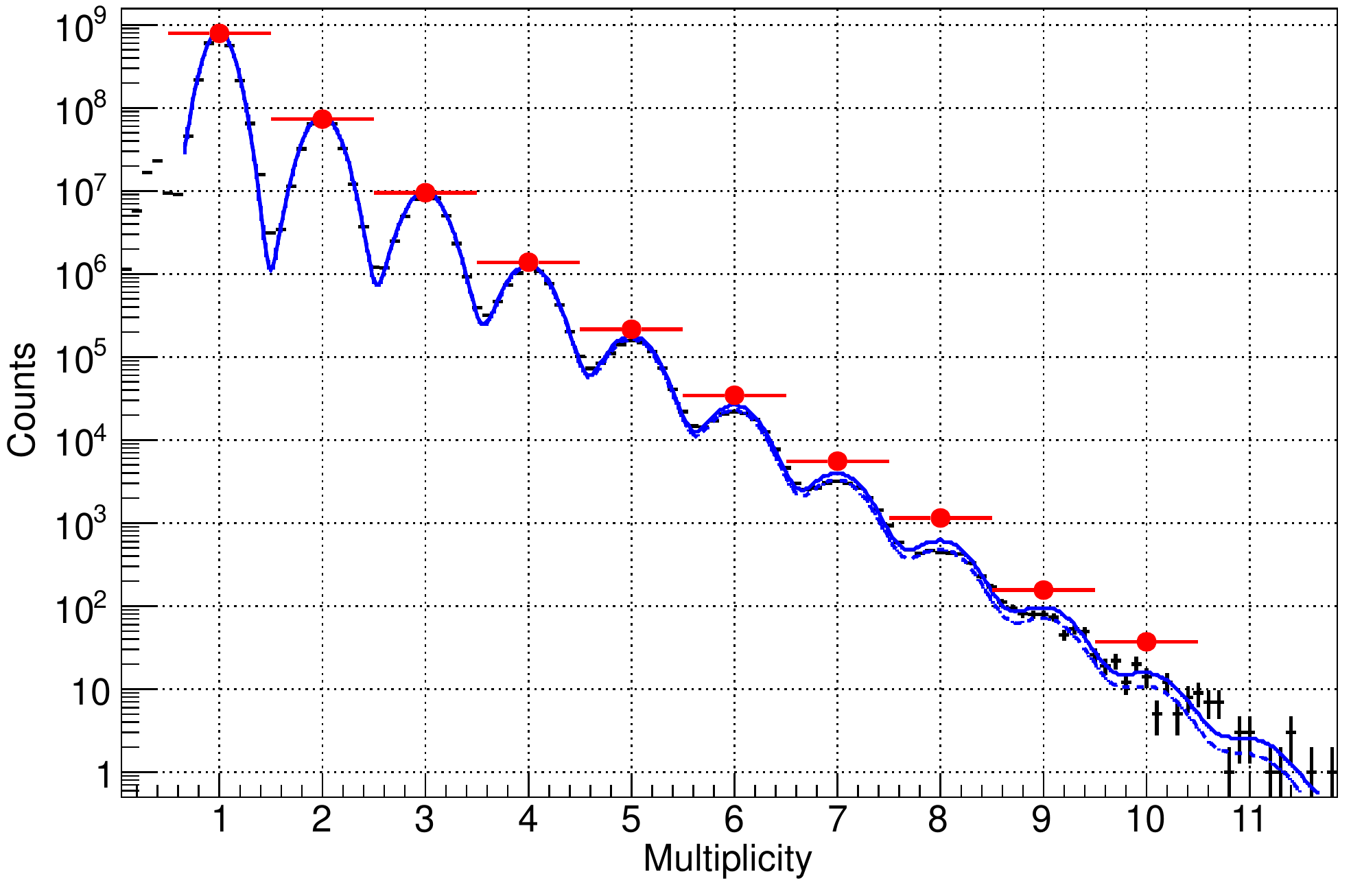}
\caption{Left: Simulation of a simple branching Poisson process
(black), i.e.\ the probability for a successful crosstalk induced
breakdown does not change and a fit of the corresponding Borel
distribution; and the simulation of a Poisson process which takes the
geometry of the sensor into account (blue) and a fit of the
corresponding modified Erlang distirbution. Right: Data fitted with an
Erlang distribution (dashed line) and a modified Erlang distribution
(solid line). Superimposed is a geometry aware simulation using the
properly converted fit parameters as input. Each red dot corresponds to
the amplitude of a Gaussian with noise parameters as obtained from the
fit. A good match between the simulated process and the
original distribution is visible.} \label{fig:simulation} \end{figure}

\begin{figure}[tb]
\centering
\includegraphics[width=.49\textwidth]{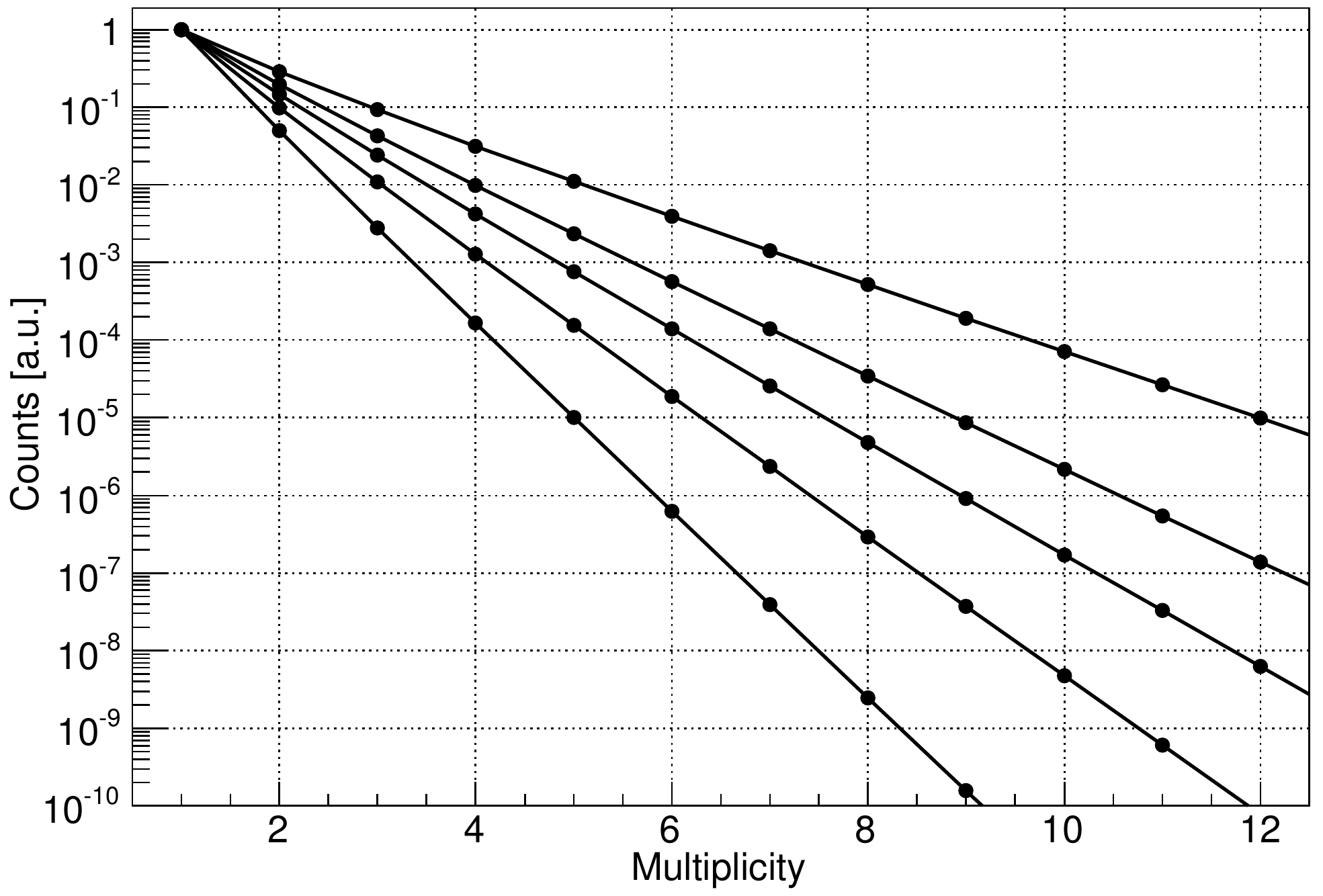}
\caption{Erlang distribution for crosstalk probabilities of 5\%, 10\%, 15\%, 20\% and
30\%. The crosstalk probability is the number of of breakdowns with \(N>1\) divided
by the total number of breakdowns.
}
\label{fig:plot_ct}
\end{figure}

Although a distribution function has been found empirically which fits
the measured dark count spectra extremely well, its understanding  is
essential for a proper detector simulation.

To initially understand the empirical modification on the
Borel-distribution, at first, an ideal Borel distribution has been
simulated.

\paragraph{Branching Poisson process (Borel-distribution)}

In the Borel process, each breakdown creates further breakdowns with 
identical Poisson distributions. The following recursive code snippet
returns a random number of total breakdowns including the primary
breakdown. 

{\small
\begin{verbatim}
int hit()
{
    int n = random_poisson(probability);
    int counter=1;
    for (int i=0; i<n; i++)
        counter += hit();
    return counter;
}
\end{verbatim}
}

The result for a Poisson probability of 20\% is shown in 
figure~\ref{fig:simulation} (left) in black. Superimposed is a fit of a
Borel distribution. As expected, simulation and fit show a good agreement.
The fit yields a probability \(p\) within errors consistent with 0.2.

\paragraph{Branching Poisson process with geometry awareness}

For the simulated Borel process it is assumed that each branching process
takes place with identical probabilities. Considering a real crosstalk
process on a sensor, it would mean that the emitted photons always have
exactly the  same probability to induce further breakdowns. While the
number of emitted photons is Poisson distributed, the number of charged
cells in the direct vicinity of the emitting cell is decreasing with
each discharge. The requirement that only direct neighboring cells can
suffer a breakdown effectively limits the number of crosstalk induced
discharges to four. This geometrical effect has to be taken into
account in the simulation. In addition, it must be considered that a
real sensor has only a finite number of available cells in total.
Therefore, a realistic simulation has to take the geometry into account
and memorize discharged cells. An implementation for a quadratic device
with 3600 cells is shown in the following code snippet, assuming that
crosstalk only affects direct neighbors. The returned value is the
random number of total breakdowns induced including the primary
breakdown. 




{\small
\begin{verbatim}
int hit(int x, int y)
{
    if (not_inside(x, y) || is_discharged(x, y))
        return 0;
    discharge(x, y);
    int n = random_poisson(probability);
    int counter = 1;
    for (int i=0; i<n; i++)
    {
        switch (random_direction())
        {
        case 0: counter += hit(x+1, y); break;
        case 1: counter += hit(x-1, y); break;
        case 2: counter += hit(x, y+1); break;
        case 3: counter += hit(x, y-1); break;
        }
    }
    return counter;
}
\end{verbatim}
}

The result of this simulation for a probability of 20\% is shown in
figure~\ref{fig:simulation} (left) in blue. Superimposed is a fit of a
modified Erlang distribution as introduced earlier. Simulation and fit
show a good match. The fit yields the following results:

\[ p  = 0.2196\pm0.0002\]
\[\nu = 0.978\pm0.002\]

\begin{figure}[tb]
\centering
\includegraphics[width=.49\textwidth]{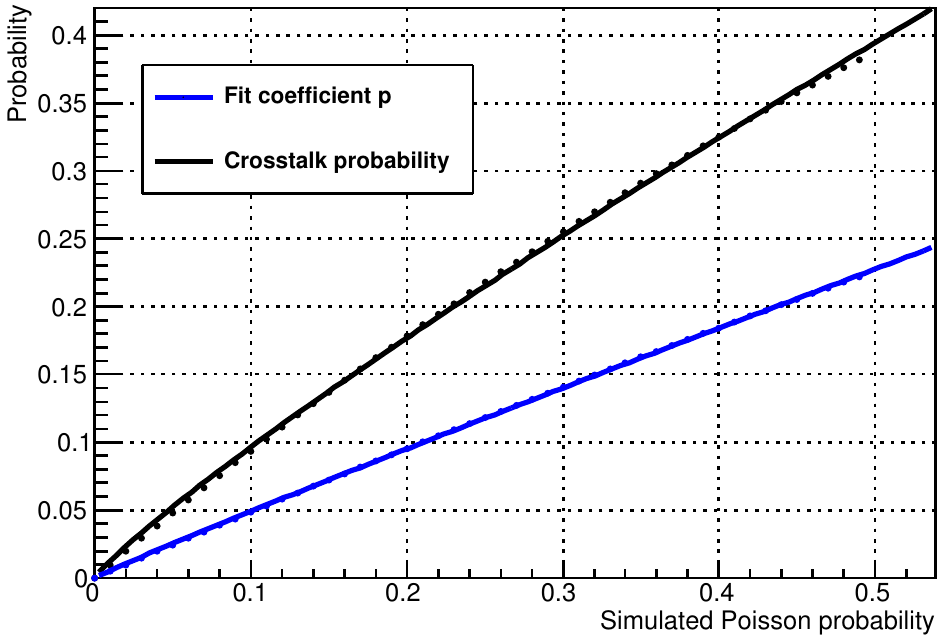}\hfill
\includegraphics[width=.49\textwidth]{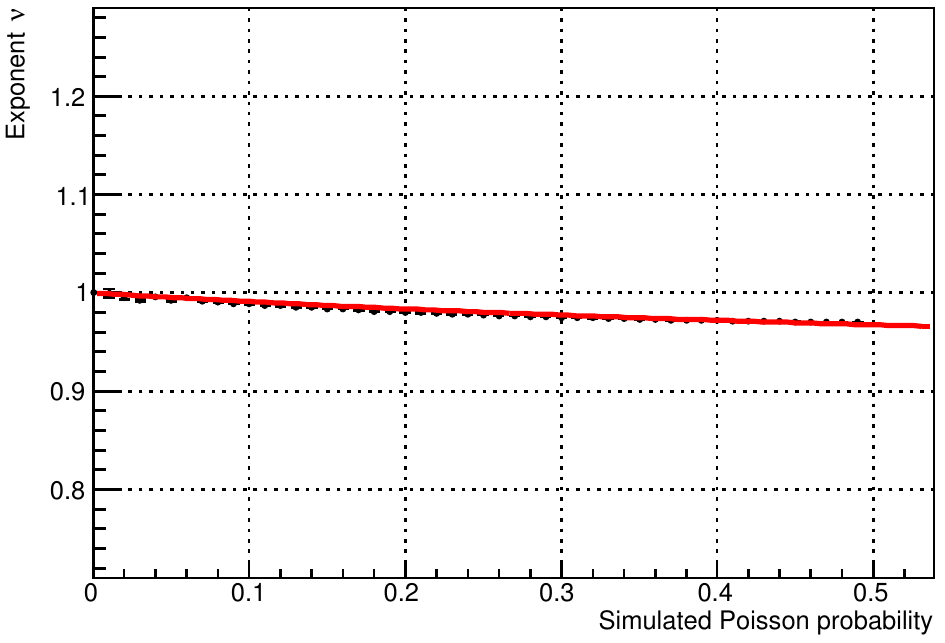}
\caption{The probability \(p\) (left, blue) and the exponent \(\nu\) (right) 
of the a fit of a modified Erlang distribution versus the Poisson probability \(p\)
of a simulation taking the geometry into account. The black dots show the
corresponding crosstalk probability \(p_{xt}\). Both probabilities are superimposed
with a fit of the form \(c_0p^{c_1}\). The right curve was fitted with
\((cp+1)/(p+1)\).} 
\label{fig:fit_ct} 
\end{figure}

By changing the simulated probability, no decrease in the fit quality
is observed. It also allows to find a relation between the Poisson
probability \(p\) in the simulation and the coefficient \(p_{fit}\) of
the fit of an Erlang distribution, the crosstalk probability \(p_{xt}\)
and the exponent \(\nu\). The result is shown in
figure~\ref{fig:fit_ct}. The fits yield

\[p_{fit} = (0.440106\pm 0.00010)\cdot p^{0.9515\pm 0.0020}\]
\[p_{xt}  = (0.723\pm 0.004)\cdot p^{0.875\pm 0.005}\]
\[\nu = \frac{(0.902387\pm 0.00006)\cdot p+1}{p+1}\]

These relations can be used to estimate the probability
\(p\) needed for the simulation from a fit to the data. More precise
fits can be found, but for the simulation a higher precision is
not necessary.\\ 

So far, the possibility that a crosstalk photon induces a breakdown in
another cell other than a direct neighbor has still been neglected.
This can be taken into account if photons are simulated with a
distribution which is exponentially falling from the center of the
emitting cell and uniform in direction. Since this effect is only
visible for \(N>10\), it is negligible for image reconstruction in
Cherenkov astronomy and only relevant if exact rate for high
thresholds need to be calculated.



%
%
%

\subsection{Application}  

In figure~\ref{fig:simulation} (right) the measured data superimposed
with a fit of an ideal Erlang distribution (solid line) and a modified
Erlang distribution (dashed line) are shown. Superimposed (red dots) is
a simulation with Poisson probability \(p\) obtained from  the
parameter \(p_{fit}\) from a fit of a modified Erlang distirbution. It can be
seen that a good match with the ideal Erlang distribution is obtained.

Knowing the proper distribution function also allows to compare the
influence of the crosstalk probability on the distribution.
Figure~\ref{fig:plot_ct} shows the expected distribution for different
crosstalk probabilities.

\subsection{Result} 

The spatial compactness of successive breakdowns in crosstalk events
has a strong influence on the distribution. This influence is described
well by a slightly modified Erlang distribution. The additional factor
\(N\) by which the Erlang distribution deviates from the Borel
distribution can be interpreted as the loss of charged cells in the
vicinity of the avalanche emitting photons. The exponent \(\nu\) can be
interpreted as a further fine tuning of the geometrical behavior, for
example, the probability that non direct neighbors are triggered. That
the Erlang distribution yields reasonable results can be interpreted
such that each breakdown induces additional breakdowns with an
exponential distribution due to the loss of available charged cells
with increasing number.

\newpage

\acknowledgments
The important contributions from ETH Zurich grants ETH-10.08-2 and
ETH-27.12-1 as well as the funding from the Swiss SNF, the German BMBF
(Verbundforschung Astro- und Astroteilchenphysik) and the DFG
(collaborative research center SFB 876/C3) are gratefully acknowledged.
We are thankful for the very valuable contributions from E.\ Lorenz,
D.\ Renker and G.\ Viertel during the early phase of the project. We
also thank the Instituto de Astrofisica de Canarias for allowing us to
operate the telescope at the Observatorio Roque de los Muchachos in La
Palma, the Max-Planck-Institut f\"ur Physik for providing us with the
mount of the former HEGRA CT\,3 telescope, and the MAGIC Collaboration
for their support. We further thank the group of M. Tose from the
College of Engineering and Technology at Western Mindanao State
University, Philippines, for providing us with the scheduling
web-interface.

\end{document}